\documentclass[utf8]{aa}  

\usepackage{graphicx}
\usepackage[varg]{txfonts}
\usepackage{tabularx}
\usepackage{lscape}
\usepackage{multirow}
\usepackage[Large]{subfigure}

\usepackage[colorlinks,
linkcolor={blue},
citecolor={blue},
urlcolor={blue}]{hyperref}
\begin{document}

   \title{VIBES: VIsual Binary Exoplanet survey with SPHERE}

   \subtitle{Upper limits on wide S-planet and S-BD frequencies, triple system discovery, and astrometric confirmation of 20 stellar binaries and three triple systems\thanks{Based on observations collected at the European Southern
   		Observatory, Chile (ESO Open Time 096.C-0835, 097.C-0826, 098.C-0643, 0100.C-0543, 0101.C-0405).}}

   \author{J. Hagelberg
          \inst{1,2}
          \and
            N. Engler\inst{1}
          \and
          C. Fontanive\inst{3}
          \and
          S. Daemgen\inst{1}
          \and
          S. P. Quanz\inst{1}
          \and
          J. Kühn\inst{1,3}
		 \and
		 M. Reggiani\inst{4}
          \and
          M. Meyer\inst{5}
          \and
          R. Jayawardhana\inst{6}
          \and
          V. Kostov\inst{7}
        }

   \institute{Institute for Particle Physics and Astrophysics, ETH Zurich,
   	Wolfgang-Pauli-Strasse 27, 8093 Zurich, Switzerland\\
              \email{janis.hagelberg@unige.ch}
    \and
    Geneva Observatory, University of Geneva, Chemin des Mailettes
51, 1290 Versoix, Switzerland
   \and
   Center for Space and Habitability, University of Bern, Bern, Switzerland
   \and
Institute of astrophysics, KU Leuven, Celestijnlaan 200D, 3001 Leuven, Belgium
   \and
   University of Michigan, Astronomy Department, USA
   \and
   Department of Astronomy, Cornell University, Ithaca, NY 14853, USA
      \and
   NASA/GSFC, USA
             }

   \date{Received September 18, 2020; accepted March 16, 1997}

  \abstract
   {Recent surveys indicate that planets in binary systems are more abundant than previously thought, which is in agreement with 
   	theoretical work on disc dynamics and planet formation in binaries. So far, most observational surveys, however, have focused on short-period planets in binaries,
   	thus little is known about
   the occurrence rates of planets on longer periods ($\geq 10$ au).}
   {In order to measure the abundance and physical characteristics of wide-orbit giant exoplanets in binary systems, we have designed the 'VIsual Binary Exoplanet survey with Sphere' (VIBES) to search for planets in visual binaries. It uses the SPHERE instrument at VLT
   	to search for planets in 23 visual binary and four visual triple systems with ages of <145 Myr and distances of <150 pc.}
   {We used the IRDIS dual-band imager on SPHERE to acquire high-contrast images of the sample targets. For each binary, the
   	two components were observed at the same time with a coronagraph masking only the primary star. For the triple star, the tight components were treated as a single star for data reduction. This enabled us to effectively search for companions around 50 individual stars in binaries and four binaries in triples.
   }
   {We derived
   upper limits of $<$13.7\% for the frequency of sub-stellar companions
around primaries in visual binaries, $<$26.5\% for the fraction of sub-stellar companions around secondaries
in visual binaries, and an occurrence rate of $<$9.0\% for giant planets and brown dwarfs around either
component of visual binaries.
   We have combined our observations with literature measurements to astrometrically confirm, for the first time, that 20 binaries and two triple systems, which were previously known, are indeed physically bound. Finally, we discovered a third component of the binary HD~121336. }
   {The upper limits we derived are compatible with planet formation through the core accretion and the gravitational instability processes in binaries. These limits are also in line with limits found for single star and circumbinary planet search surveys.}


   \maketitle

\section{Introduction}

One key statistical outcome from the more than 4000 planets that have been detected so far is that almost every Sun-like star harbours a planet
\citep{fressin_false_2013, udry_statistical_2007, dressing_occurrence_2013, howard_planet_2012}.
Given that almost half of all stars in our Milky Way are bound in multiple systems \citep{raghavan_survey_2010}, one could expect a large fraction
of the detected exoplanets to be in binaries.
Yet, less than 200 planets are known to reside in multiple stellar systems\footnote{http://www.univie.ac.at/adg/schwarz/multiple.html} \citep{schwarz_new_2016} , either as circumstellar planets orbiting one of the two stars in the binary 
or as circumbinary planets orbiting the centre of mass of both stars (also known as S- and P-planets, respectively, \citet{dvorak_planetenbahnen_1982, dvorak_numerical_1984}).

Early theoretical work predicted a lower occurrence rate of circumstellar planets compared to single star planets, 
due to the fact that the gravitational potential of a multi-star system and truncated protoplanetary discs would 
hinder planet formation \citep{artymowicz_dynamics_1994} and also render the planet dynamically unstable for long-term survival \citep{holman_long-term_1999}.
Large exoplanet search surveys thus often avoided or gave low priority to binary stars in order to enhance the planet
detection yield. Furthermore, radial velocity, transit, and direct imaging techniques all tend to have lower detection
sensitivities for most binary configurations, thus further enhancing the observational bias against planets in binaries.

Various multiplicity surveys, mostly searching with imaging for stellar companions to known radial velocity (RV) or transit detected planet hosting stars, 
have shown that the occurrence rate of circumstellar planets is in 
the $\approx$ 10--30\% range  (e.g. review of various surveys in \citet{wang_influence_2014}).
However for very close binaries, which have a separation smaller than 47 au, planet occurrence is only 0.34 
times the one of single stars or wider binaries
 \citep{kraus_impact_2016}.  
 For very close binaries tighter than 20~au, disc truncation and the high velocities of gas and dust induced by the secondary star 
 render planet formation almost impossible \citep{zsom_first_2011}.
 The existence of a handful of such planets, such as $\gamma$ Cephei~Ab \citep{hatzes_planetary_2003}, HD~196885~Ab
 \citep{correia_elodie_2008, chauvin_planetary_2011}, and Kepler~420~Ab \citep{santerne_sophie_2014}, may actually 
 be the results of stellar
  scattering \citep{marti_stellar_2012} instead of having formed within the binary.

 In contrast to the earlier theoretical studies mentioned above, recent theoretical and observational results
 show that there is a possibility that planet formation in multiple stars is in fact enhanced compared
to single stars as long as the binary separation is larger than 50 au.
This enhancement can be caused by the secondary inducing spiral density waves in the protoplanetary disc, which can potentially
stimulate gravitational instability (GI)
\citep{batygin_formation_2011, rafikov_planet_2013}.
This is in line with  observations that
suggest a 3-fold increase in hot Jupiter occurrence in binary stars compared to singles 
\citep{wang_influence_2015, ngo_friends_2016, evans_high-resolution_2018, fontanive_high_2019}.
Binary stars should thus be taken into consideration when analysing planet formation and
 evolution, 
not only because binary stars represent $\sim 1/3$ of stars in the Milky Way (M: $26.8\%\pm 1.4\%$ \citet{winters_solar_2019}, FGK: $33\%\pm 2\%$  \citet{raghavan_survey_2010}, 
A: $32.1^{+3.9}_{-3.5}\%$ \citet{de_rosa_vast_2014}), but also to 
understand the
robustness and diversity of planet formation.

In order to probe the population of wide companions to binary stars we started the 
'VIsual Binary Exoplanet survey with Sphere' (VIBES) which combines SPHERE’s planet
discovery and characterisation potential with its ability to simultaneously target 
all components of a stellar multiple, at the cost of a degraded sensitivity around the secondary star. 
Our survey searching for wide circumstellar planets also fills the gap between the SHINE survey targeting planets orbiting single
 stars 
and the 'Search for Planets Orbiting Two Stars' \citep[SPOTS,][]{thalmann_spots:_2014}) survey which looks for wide circumbinary
 planets. By combining the result on S- and P-type planet populations probed by the VIBES and SPOTS surveys respectively, we have a
  census of the overall population of wide giant planets in binaries. 
Given that a majority of our targets are in the Scorpius-Centaurus association we are probing younger but farther stars than 
other surveys like the NACO-LP \citep{chauvin_vlt/naco_2015, desidera_vlt/naco_2015}, the IDPS \citep{vigan_international_2012}, and
 the NICI Campaign \citep{liu_gemini_2010, biller_gemini/nici_2013} which were more sensitive to higher-mass planets at smaller
  orbital separations than our survey. The smaller inner working angle achieved with SPHERE and the fact that we are probing a 
  population at wider separation mitigates the effect of larger distance.

The survey is described in the
 following sections starting with the target sample definition (Sect. \ref{sec:sample}). 
 The strategy adopted for the observations and data reduction are described in Sects.
   \ref{sec:obs-strategy} and \ref{sec:data-reduction} respectively.
The confirmation of 26 binaries is described in
Sect. \ref{sec:obs-results}, followed by the statistical analysis of the complete survey in Sect. \ref{sec:analysis}, and a discussion of the
survey outcome in Sec. \ref{sec:discussion}.

\section{Target sample}\label{sec:sample}

We compiled a sample of 26 multiple systems that are members of young clusters and associations
  based on the catalogues listed in Table \ref{tab:astro_error}. 
 We set selection criteria on age and distance to be younger than 50 Myr and closer than 150 pc.
  Considered regions were Taurus  \citep[1--2 Myr, d$\simeq$145 pc,][]{torres_vlba_2009, kraus_coevality_2009}, 
 Scorpius Centaurus  
 \citep[5--15 Myr, d$\simeq$140 pc,][]{de_zeeuw_hipparcos_1999,song_new_2012, pecaut_revised_2012}, the $\beta$ Pic moving group  \citep[$22\pm 6$\ Myr, d$\simeq$30 pc,][]{binks_lithium_2014,bell_self-consistent_2015,shkolnik_all-sky_2017},
  Tucana--Horologium \citep[$45 \pm 4$\ Myr, d$\simeq$46 pc,][]{bell_self-consistent_2015}, and Columba
\citep[$42^{+6}_{-4}$\ Myr, d$\simeq$50 pc,][]{bell_self-consistent_2015}.
The heterogeneity induced by the wide range of age and distance in the associations 
has less impact on the survey sensitivity than the 
large variation in observing conditions, also because the final sample is dominated by Sco-Cen stars. 
Nonetheless this heterogeneity is taken into account in the final statistical analysis.

\begin{table}
	\caption{Astrometric and photometric uncertainties for this work and for measurements sourced in the literature.}             %
	\label{tab:astro_error}
	\centering                          %
	\begin{tabular}{c c c c}        %
		\hline\hline                 %
		 Ref & Error Sep & Error PA & mag\\    %
			&	(\arcsec) & (deg)&\\
		\hline                       %
        \noalign{\smallskip}
		This work & 0.005 & 0.3 & 0.5\\
		\citet{chauvin_adaptive_2003} & 0.03 & 2.5 & 0.16 \\      %
		\citet{daemgen_sub-stellar_2015} & 0.001    & 0.1 & 0.01 \\
		\citet{elliott_search_2015} & 0.01   & 0.29 & 0.03 \\
		\citet{fabricius_tycho_2002} & 0.14 & 1 & -- \\
		\citet{gaia_collaboration_gaia_2018} & 0.0001 & 0.07 & -- \\
		\citet{hartkopf_iccd_1996} & 0.003 & 0.2& -- \\
		\citet{herschel_catalogue_1874} & -- & -- & -- \\
		\citet{janson_multiplicity_2013} & 0.008  & 0.4 & 0.2 \\
		\citet{kouwenhoven_primordial_2005} & 0.0015 & 0.03 & --  \\
		\citet{kouwenhoven_primordial_2006} & 0.0015 & 0.03 & 0.12  \\
		\citet{kohler_multiplicity_1998} &  0.003  & 0.4 & -- \\ 
		\citet{mcalister_iccd_1990} & 0.003 & 0.2 & --\\
		\citet{tokovinin_msc_1997} &  --  &  -- & --\\
		\hline                                   %
	\end{tabular}
\end{table}

\begin{table*}
	\caption{Associations and the age adopted for its members.}             %
	\label{tab:assoc_ages}
	\centering                          %
	\begin{tabular}{c c c c}        %
		\hline\hline                 %
		Acronym & Name & Age & Ref \\    %
		\hline                        %
        \noalign{\smallskip}
		ABD & AB Doradus 					& $145^{+51}_{-19}$	& \citet{bell_self-consistent_2015}	\\
		BPMG & $\beta$ Pictoris MG 				& $22\pm 6$ &	\citet{shkolnik_all-sky_2017}\\
		LCC	& Lower Centaurus Crux 			& $17\pm 1$ &	\citet{pecaut_revised_2012}	\\
		ROPH & $\rho$ Ophiuchi				& 2--5	& \citet{wilking_star_2008}\\
		TAU	& Taurus Molecular Cloud		& 1--2	&	\citet{kenyon_pre-main-sequence_1995}\\
		THA	& Tucana-Horologium Association	& $45 \pm 4$ & \citet{bell_self-consistent_2015}	\\
		UCL	& Upper Centaurus Lupus			& $16\pm 1$ &	\citet{pecaut_revised_2012}\\
		USCO& Upper Scorpius				& $11\pm 3$ &	\citet{pecaut_revised_2012}\\
		\hline                                   %
	\end{tabular}
\end{table*}

Our targets were selected in order to be bright enough (R < 11 mag) to provide good adaptive optics correction, with a binary separation in the 0.8\arcsec -- 5\arcsec range to prevent AO wavefront sensing instabilities while still having both stars simultaneously in the IRDIS field of view.
The targets' primary spectral types range between B6 and M0 with a median mass
of 2.1 $M_\sun$. The selected projected binary separations translate to $\sim 30$--600 au at their respective
distances. Considering that dynamical interaction limits the maximum orbital separation of potential planetary
 companions on S-type orbits to $\sim0.3$ times the binary separation on average, we should expect these planetary companions to be in the  $\sim10$--170 au range \citep{holman_long-term_1999}. 
The age range of 1--145 Myr is chosen in order to optimise our planet detection sensitivity as planets are brightest when they are young  
\citep[independent of the exact formation process, e.g. hot- or cold-start;][]{fortney_synthetic_2008, marley_luminosity_2007}. 

The publications out of which our sample was sourced are listed in Table \ref{tab:astro_error}
along with the uncertainties on the measured separation, position angle, and magnitude differences
given by the authors.
To determine the age of our targets we used the BANYAN $\Sigma$ tool  \citep{gagne_banyan_2018}. This Bayesian analysis algorithm uses galactic coordinates (XYZ) and space velocities (UVW) of the star to compute the membership probability to nearby young associations listed in Table~\ref{tab:assoc_ages}. The results of this analysis are presented in Table~\ref{tab:MG} giving the identified moving group to which
the star is most probably associated and the membership probability. The
table also lists the spectral types, coordinates, and distance given by Gaia \citep{gaia_collaboration_gaia_2018}. 
Finally, for HD~102026 which was identified as a field star we used the age determined 
photometrically by \citet{tetzlaff_catalogue_2011} of $15.8 \pm 7.3$ Myr.

\begin{figure}
	\centering
	\includegraphics[width=9cm,clip]{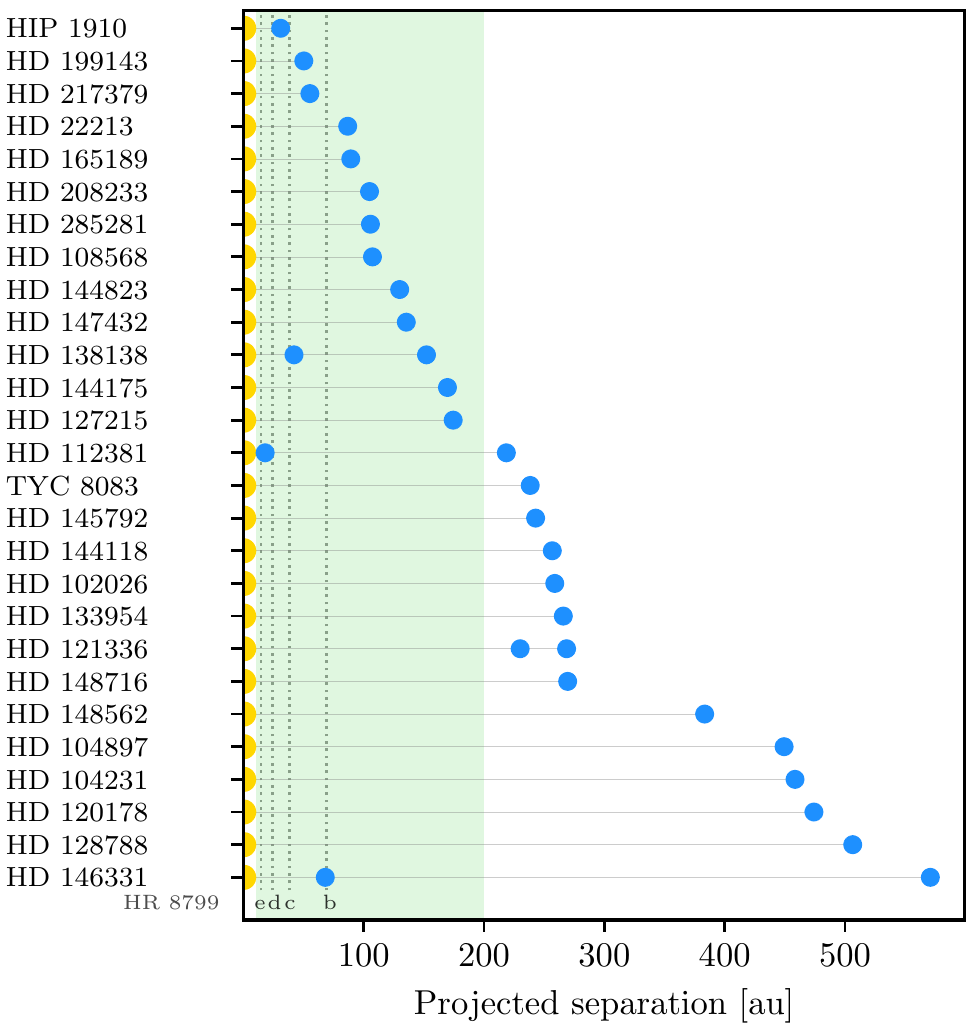}
	\caption{Projected separation of the binaries in the observing sample. The primary and secondary stars are represented by the yellow and blue dots respectively. The green area represents the 10--200~au
	considered by the statistical analysis.
	HR~8799 bcde are represented by vertical dotted lines to illustrate the position of this wide giant planet system.}%
	\label{fig:loliplot}
\end{figure}

\begin{figure}
	\centering
	\includegraphics[width=9cm,clip]{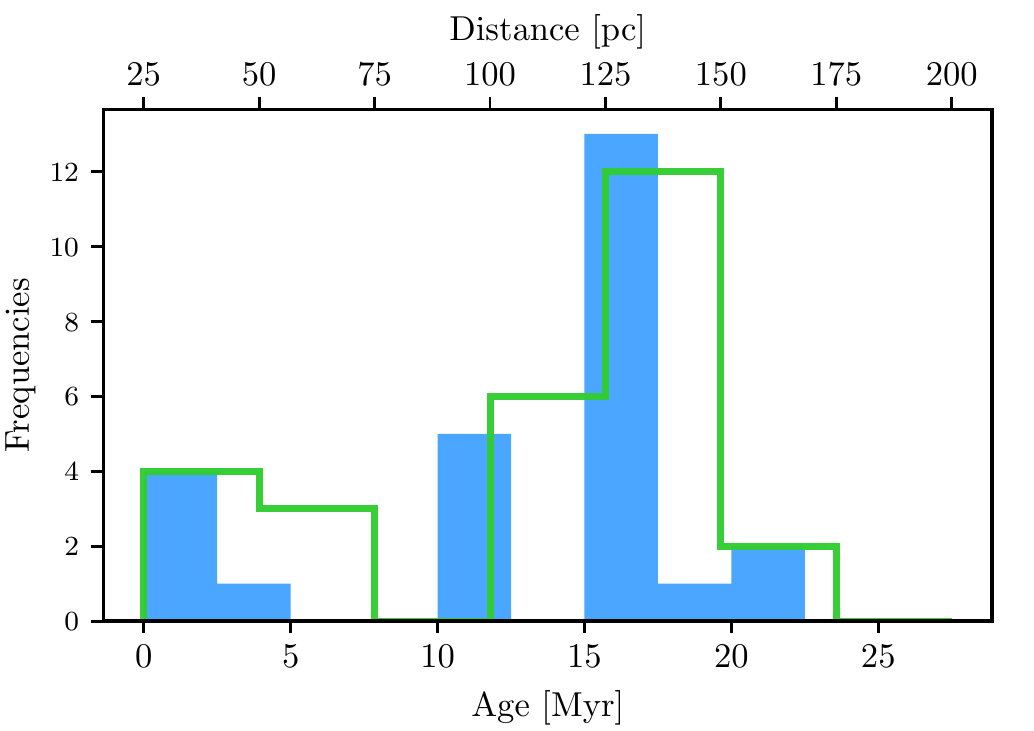}
	\caption{Histogram of age and distance of the VIBES sample, blue bars and green line respectively. HD~217379 and HD~285281 (145 and 45 Myr respectively) have been left out in the age histogram for better readability.}%
	\label{fig:sample_histogram}
\end{figure}

The sample we assembled based on these criteria is listed in Table \ref{tab:MG} including 
the existing astrometric measurements of the binaries. The spread in binary separations is
illustrated in Fig. \ref{fig:loliplot}.
It should be noted that for the triple systems %
	HD~112381, HD~121336, HD~138138, and HD~146331
we are not sensitive to planets which would be orbiting in between the two tight components of these hierarchical triple systems.
Except for HD~217379 and HD~285281 that have an estimated age of $\sim 145$ Myr and 
$\sim 45$ Myr respectively, all remaining targets in the sample have an
age less than ~22 Myr as illustrated in Fig. \ref{fig:sample_histogram}.

\section{Observational strategy}\label{sec:obs-strategy}

The observations were carried out with SPHERE at VLT \citep{beuzit_sphere:_2008} in IRDIFS mode 
which simultaneously acquires data in dual-band imaging through the $H2H3$ filters
  \citep[
  $\lambda_{H2} = 1.593 \pm 0.055 \mu \mathrm{m}$;
  $\lambda_{H3} = 1.667 \pm 0.056 \mu m$,][]{vigan_photometric_2010} with IRDIS \citep{dohlen_infra-red_2008}, and integral field spectroscopy in $Y$ -- $J$ 
 (0.95--1.35 $\mu$ m, $R_\lambda\sim$  54)
 with the IFS \citep{claudi_sphere_2008}. Both instruments are situated
  behind the 185 mas diameter apodized Lyot coronagraph \citep{carbillet_apodized_2011, guerri_apodized_2011} which is masking the brighter star
  in $H$ band. 
  The observations were all carried out in pupil-tracking (i.e. where the field of view rotates), in order 
  	to apply Angular Differential Imaging \citep[ADI, ][]{schneider_high_2003, liu_substructure_2004, marois_angular_2006}
  as described in Sect. \ref{sec:data-reduction}.
  
The same observing sequence is carried out for each target, which starts with calibrations followed by science
observations. The calibration sub-sequence is composed of long exposure sky frames taken by offsetting the stars out of the field of view
in order to carry out bad pixel correction and to estimate the background flux.
We rely on unsaturated non-coronagraphic exposures with both stars in the field-of-view to calibrate the photometry, the PSF profile, and the
 astrometric configuration of the targeted binary system.
The last part of the calibration sequence are the star centring frames where satellite spots
are generated through a sinusoidal pattern on the deformable mirror \citep{jovanovic_artificial_2015, rickman_spectral_2020} in order to determine
precisely the position of the primary star behind the coronagraph.
Typically the total observing time dedicated to each target is an hour, which besides the 15
 minutes spent on telescope slewing and target acquisition is split in $80/20\%$ 
 between the science observation and calibrations. 

A certain number of targets have been carried out through an ESO filler programme, which implies bad weather conditions and sub-optimal field
 rotation. In certain cases these targets have been re-observed
 in better conditions and with larger field rotation (Table \ref{tab:obs}).

The observational strategy exploits the large field of view of IRDIS to simultaneously 
acquire both stars of all selected binaries. In principle this doubles the total number of stars probed in
comparison to a single star survey. However the sensitivity around the second star is lower because the star is not behind a coronagraph, and the field of view gets smaller as the secondary star gets closer to the edge of the detector.

Second epoch follow-up observations were carried out on all sub-stellar candidates in order to verify if they are 
	co-moving and thus physically bound. For these observations the star centring satellite spots were kept
during the whole observing sequence to improve the astrometric fitting in case of a confirmed planetary companion 
detection.

\section{Data reduction}\label{sec:data-reduction}

The IRDIS data reduction is based on GRAPHIC \citep{hagelberg_geneva_2016}, with modifications in order to have an end-to-end reduction for 
SPHERE data
\citep{cheetham_direct_2018}.
 The pipeline cosmetics pre-processing involves sky subtraction, flat fielding, bad pixel correction, filter splitting, 
 distortion correction (based on \citet{maire_sphere_2016})
 and individual frame registration. 
For the PSF subtraction we use ADI combined with
algorithms based on principal component analysis \citep[PCA; ][]{amara_pynpoint:_2012, soummer_detection_2012} applied 
on $2 \times FWHM$ wide concentric annular sections. In order to minimise companion self-subtraction we exclude frames where the
field rotation is less than $1.25 \times FWHM$ (the detailed procedure is described in \citet{cheetham_direct_2018}).
The final images are then derotated in Fourier space and median combined to produce the final image.

Additionally a Spectral Differential Imaging \citep[SDI,][]{sparks_imaging_2002} reduction is also used where the $H3$ filter image is spatially and flux rescaled by the wavelength ratio and subtracted 
from the simultaneous $H2$ image. The resulting frames are then run through the same PCA
 algorithm. This Angular and Spectral Differential Imaging (ASDI) data reduction product thus adds a third potential companion detection
  image to the two ADI reductions of the $H2$ and $H3$ filters.

The IFS data reduction uses the SPHERE Data Reduction and Handling 
pipeline \citep[DRH,][]{pavlov_sphere_2008} to produce calibrated data. This calibration includes the background 
subtraction, bad pixel correction, the wavelength calibration, correction for the spectral cross-talk \citep{mesa_performance_2015} and the flat fielding. The wavelength-dependent centring of the frames is performed using the satellite spots. 
The extracted IFS data cube consists of temporal sequences of 39 monochromatic images with a format of $290 \times 290$ pixels each,
which are then processed using the PCA-based pipeline PynPoint \citep{stolker_pynpoint:_2019}.

Companions close to the secondary star may not be uncovered with a data reduction centred on
 the primary star because of the typical azimuthal smearing of bright field components 
 from the ADI processing. More importantly ADI (and ASDI) rely on the fact that the target PSF 
 is not rotating, the data reduction can thus only be optimised for one central PSF at a time.
The only changes in the data reduction for the secondary component is that the centring is done
 by fitting a 
2D Gaussian to the secondary PSF instead of using the satellite spots. This comes from the fact that 
the secondary component is not behind a coronagraph so that the satellite spots disappear in the stellar halo. 
For similar mass binaries the PSF of the secondary is slightly saturated but still usable for a Gaussian fit. 

Each telescope pointing thus results in at least 5 data products, two IRDIS reductions for each star in 
the field and one IFS reduction for the brightest star in the field. In the case of triple systems 
we only reduced the two brightest components as the third star was in every case too close to one of the brighter stars.

Contrast curves are then computed by first estimating the noise as a function of the
separation to the central star by measuring the standard deviation in concentric $\lambda/D$
wide annuli. These contrast curves are then calibrated for throughput by injecting fake-planets based on a non-saturated PSF of the star before PCA subtraction. Following the method 
proposed by \citet{mawet_fundamental_2014} the injection is done in order to 
keep a constant false positive rate of
$2.9 \times 10^{-7}$ to correct for small sample statistics. 
The procedure is repeated 10 times with a varying azimuth.
Systematic effects such as detector defects are not taken into account which thus results 
in a more conservative detection limit.

\begin{figure}
	\centering
	\includegraphics[width=9cm,clip]{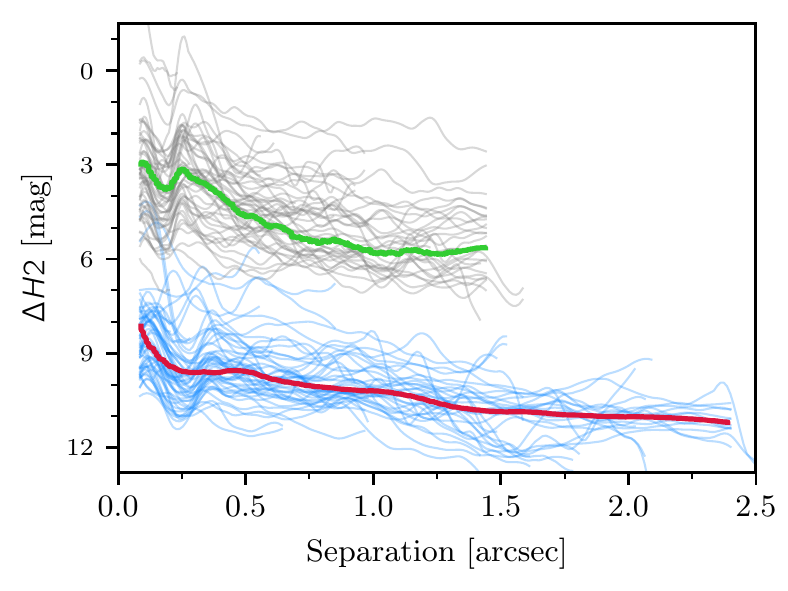}
	\caption{All IRDIS H2 contrast curves for the primary and secondary stars, blue and grey respectively, along with their median contrast curves in red and green respectively.}
	\label{fig:contrast}
\end{figure}

All the IRDIS $H2$ contrast curves for the primary and secondary stars are given in Fig.~\ref{fig:contrast} with an arbitrary cut-off at $\sim1.5$ and $\sim2.5$ arcseconds for readability, along with the median contrast curves of the primaries and of the secondaries.
$H3$ yielded a similar sensitivity while ASDI and IFS 
contrast curves where not used as they were deemed not robust enough for a statistical analysis, 
even though ASDI and ADI-IFS data reduction was done.

\section{Observational results}\label{sec:obs-results}

Except for HIP~1910 \citep{chauvin_adaptive_2003}, 
HD~165189 \citep{herschel_catalogue_1874}, HD~285281 \citep{mcalister_iccd_1990, daemgen_sub-stellar_2015} which were already identified as co-moving binaries, and
HD~138138 \citep{hartkopf_iccd_1996, kohler_multiplicity_1998} which is a known triple system, all
the other targets 
in our sample were suggested to be physically bound prior to our observations
only based on a statistical approach showing that
the probability of the secondary to be a background object was low \citep{chauvin_adaptive_2003, daemgen_sub-stellar_2015,
	elliott_search_2015, fabricius_tycho_2002, gaia_collaboration_gaia_2018, hartkopf_iccd_1996, herschel_catalogue_1874, 
	janson_multiplicity_2013, kouwenhoven_primordial_2005, kouwenhoven_primordial_2006, kohler_multiplicity_1998, mcalister_iccd_1990, tokovinin_msc_1997}.
As all our targets have now been observed at least once by our programme we are able to verify through astrometry whether the 
systems in our sample are indeed gravitationally bound. The more than five years time baseline between the binary discovery observations and our
observations is sufficient for an unambiguous determination of the binaries' physical link.

In order to measure the position angle (PA), separation, and magnitude difference between the primary and secondary star we have used the photometric calibration frames taken at the
 beginning of every sequence. In this data all the PSFs are non-coronagraphic and unsaturated ensuring the best photometric and
  astrometric accuracy. The PSF measurements are carried out using the DAOPHOT FIND algorithm \citep{stetson_daophot:_1987} as implemented in 
Astropy \citep{astropy_collaboration_astropy:_2013} where peaks over a given threshold are searched and a two-component Gaussian is then fitted to that peak. The resulting
 FWHM  fit in x and y is used to determine the roundness of the PSF.
From these measurements we determine the astrometric position as well as the brightness
of the companion with respect to the primary star. Given that all observations are taken in pupil stabilised mode we need to take
into account the parallactic angle correction when deriving the position angle (PA).
This angle correction is calculated by GRAPHIC \citep{hagelberg_geneva_2016}. The resulting PA measurement 
 when including PSF fitting (estimated at 0.4 pixel), true north determination \citep{maire_sphere_2016}, and instrument
  pupil offset results in
 a 0.3 degree uncertainty.
The separation measurement once corrected for the anamorphic distortion between x and y axis is mostly dependent on the PSF
 fitting accuracy resulting in an estimated 0.005\arcsec uncertainty on the separation.
For the magnitude measurement we have used the peak value of the Gaussian fit on the primary and companion star.
The magnitude of the companion star is then derived by taking the 2MASS $H$ band magnitude \citep{skrutskie_two_2006} of
the primary and applying the magnitude difference. This quick first order approach is thus not taking into account the 
difference between the 2MASS-$H$ and IRDIS-$H2$ filters, which thus yields an overall uncertainty of 0.5 mag on
the companion star magnitude.

\begin{figure}
	\centering
	\includegraphics[width=9cm,clip]{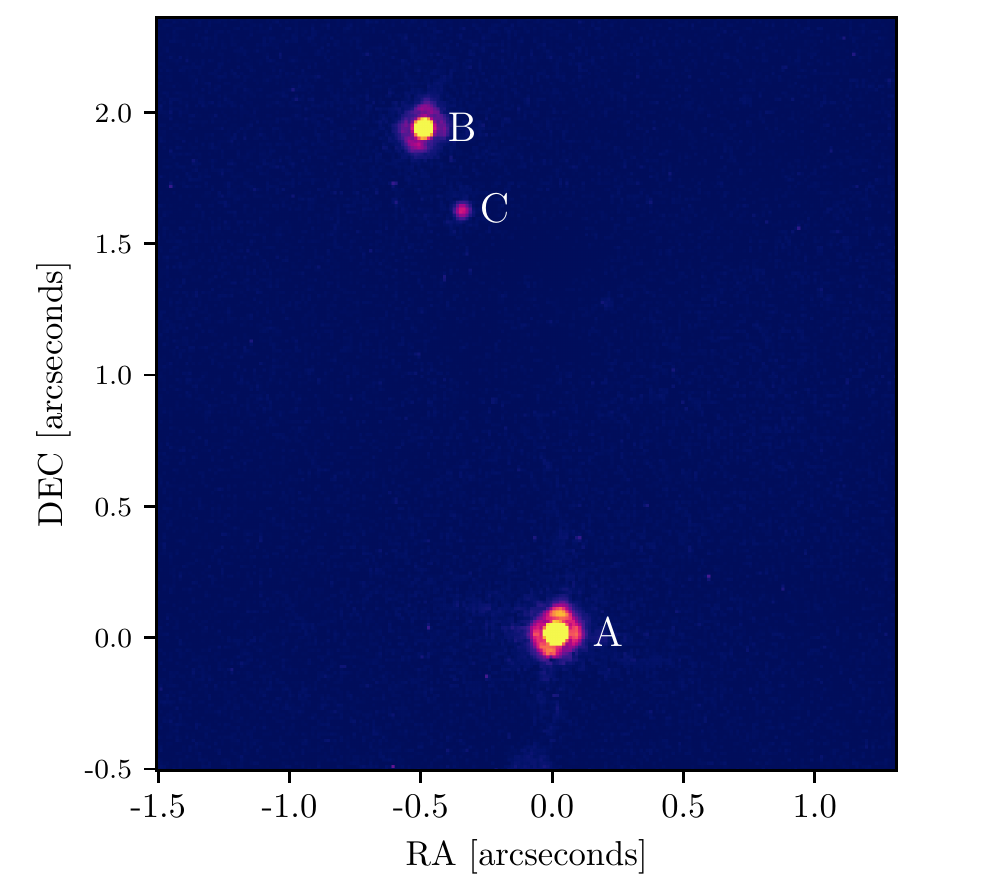}
	\caption{Binary HD~121336~AB along with the third newly discovered component C.}%
	\label{fig:hd121336}
\end{figure}

A third faint stellar component was detected in the HD~121336~AB binary (Fig. \ref{fig:hd121336}). Given the faintness and close proximity to the secondary it probably 
was beyond reach of previous instruments so that only a single epoch is available for the moment.
Using BT-Settl models with Solar metallicity \citep{allard_bt-settl_2014} we estimate the mass of component C to be around 
0.95 M$_\sun$. 
Given that component C is only at $0.346\arcsec$ of B (with a PA of $205.3\deg$ and $\Delta H$=2.3 with respect to B), we can assume that B-C form a binary orbiting A, if it is bound.

Table \ref{tab:astro_error} summarises the uncertainties on the measured stellar
companions' PAs, separations, and magnitudes as well as those reported in the literature
we used to source previous measurements. 
The measurements from this and previous work for the binaries are listed in Table \ref{tab:vibes_sample}.

To determine if the secondary component is bound or not we plot the positions from all epochs (orange crosses in Fig. \ref{fig:astrometry_1} to \ref{fig:astrometry_3}) and compare them to the relative motion
a background object would exhibit (blue lines). The line starts at 
the companion coordinates of the first observed position and then shows 
the astrometric track an object would follow with respect to the primary star if it had no proper motion. 
The green crosses along this line represent the positions a background object would have at the given
observing dates.
The wobbles are caused by the parallactic motion.
The proper motion we used is based on Gaia \citep{collaboration_gaia_2016, lindegren_gaia_2018} and Hipparcos \citep{perryman_hipparcos_1997} and it should be noted that all
the parallax measurements were graded as best quality, which means that the double star nature of the
targets did not interfere with the observations.

Figures \ref{fig:orbit_1} to \ref{fig:orbit_3} show the positions of the secondary star at all epochs. 
The primary star is represented by the blue cross at the origin of the polar plot, while the red dots
represent the secondary positions. The orange error bars on the secondary positions are mostly invisible
due to the precision of the measurements. A green curve is also plotted to illustrate the
 counter-clockwise movement the secondary would have if it was on a face-on circular orbit with the
 observed time baseline.
The apparent movement caused by the proper motion is not included in this plot as it would be too small
to be visible.
Orbital motion is clearly visible in the $\sim150$ years binary HIP~1910~AB.

From this analysis we can astrometrically confirm that 21 previously known visible binaries and three triple systems are indeed  physically bound.
The only exception are HD~165189 which was 
already known to be bound and HD~121336~AB where the binary is proven to be bound for the first time 
through astrometry but where a third component is also detected for the first time.

\section{Statistical analysis}\label{sec:analysis}

We are able to place upper limits on the occurrence rates of brown dwarfs and giant planets in binary systems. We must also rely on model predictions for the underlying population distributions in order to constrain these frequencies.
As wide-orbit exoplanets and brown dwarfs in stellar binaries have yet to be studied in depth on the theoretical side, we use population synthesis models developed for single stars, and assume that circumprimary sub-stellar companions follow similar distributions in mass and separation. Our model population consists of two sub-populations, based on the GI \citep{boss_evolution_1998} and core accretion \citep[CA, ][]{mizuno_formation_1980,lin_tidal_1986,pollack_formation_1996} formation scenarios. The GI synthetic population comes from the disc fragmentation simulations first presented in \citet{forgan_towards_2013} and later updated by \citet{forgan_towards_2018}. The synthetic CA population was obtained from the latest version of the Bern models of planetary formation and evolution \citep{mordasini_planetary_2018}, and corresponds to population NG76 from the new generation planetary population synthesis (NGPPS) series \citep{emsenhuber_new_2020, emsenhuber_new_2020-1}. Combined together, these two population synthesis models form the full population model used in our statistical analysis. The CA formation channel predicts lower-mass and closer-separation planets, while the GI mechanism predicts more massive brown dwarfs on wider orbits \citep[see e.g.][]{vigan_sphere_2020}.
The CA and GI populations serve as a first order approach as they are based on single star models
	 which do not take into account an outer 
perturber. CA populations should be affected by truncation of the disc \citep{artymowicz_dynamics_1994}, an increase of the random velocities of the planetesimals, and a possible destabilisation of the planetary system through N-body interactions with the binary, including Kozai \citep{lidov_evolution_1962, kozai_secular_1962}. The outer stellar perturber could stabilise the protoplanetary disc which would inhibit GI planet formation \citep{forgan_stellar_2009}  but it could also trigger spiral density waves which would spark GI \citep{batygin_formation_2011, rafikov_planet_2013}.

In order to constrain the statistical properties of our observed sample, we first converted the obtained detection limits for each target into detection probability maps in terms of companion masses and semi-major axes. We used the COND-2003 evolutionary tracks \citep{baraffe_evolutionary_2003} to convert observed contrasts into companion mass using the respective magnitude, distance and age of each target. Given the large range in the levels of contrast obtained, we found these models to be the only ones covering the full span of achieved detection limits. We then followed the Monte-Carlo procedure described in \citet{vigan_sphere_2020} to perform the de-projection of the limits from projected separation onto semi-major axis. We adopted for this the Beta distribution from \citet{bowler_population-level_2020} to describe the eccentricity distribution of the overall population of wide sub-stellar companions.

\begin{figure*}
	\centering
	\includegraphics[width=\textwidth]{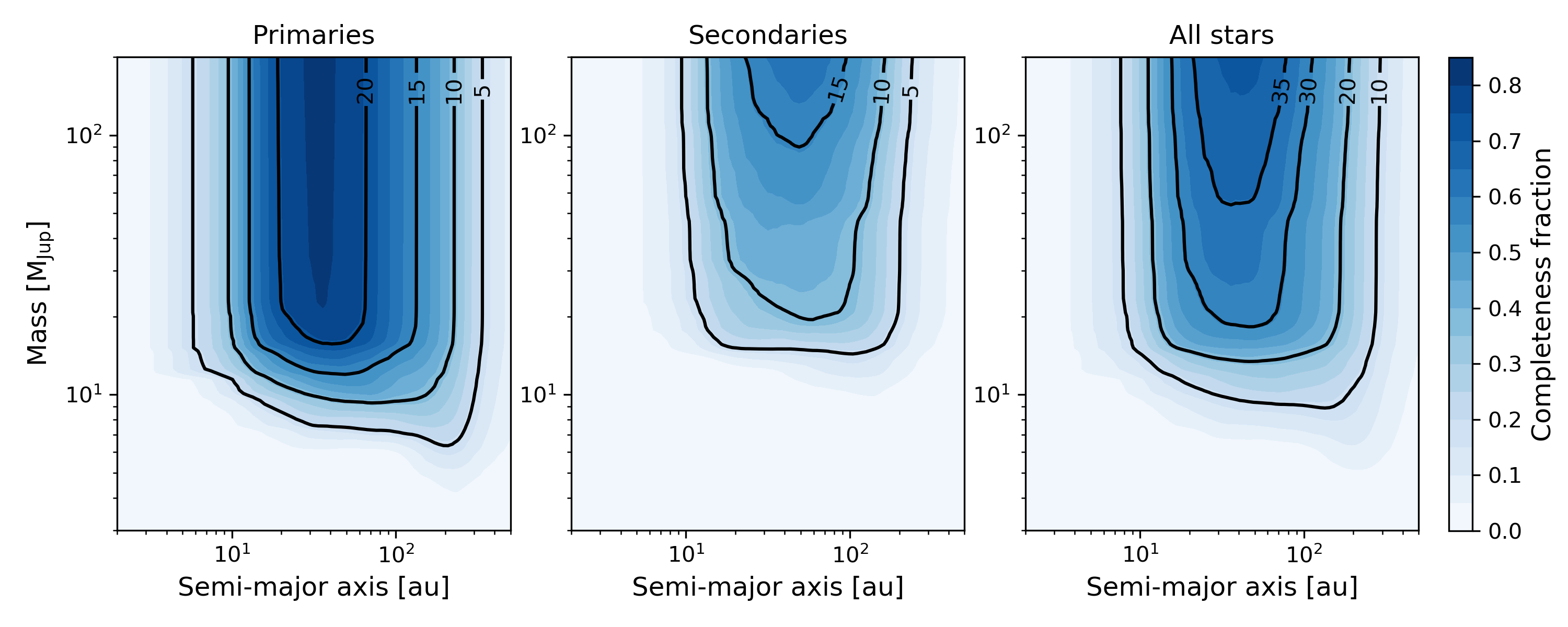}
	\caption{Average completeness of the survey considering the primary stars only (left panel), secondaries only (centre) and all stars together (right). The colourmap is the same in all panels and indicates the average completeness level as a function of mass and semi-major axis for each sub-sample. The black contours show the number of stars from each subset around which companions of given masses and separations are detectable, and are therefore out of 27 stars for the left and middle panels, and out of 54 targets for the right panel.}
	\label{fig:completeness}
\end{figure*}

The resulting survey completeness is presented in Fig.~\ref{fig:completeness}, showing the average 2-dimensional completeness around the stellar primaries (left panel), stellar secondaries (middle panel) and considering all binary components together (right panel). These provide the average probability of detecting an object of given mass and semi-major axis. The poorer contrasts achieved around the secondaries result in a significant loss in the depth of the mass limits reached around these stars compared to the primaries. In addition, some detection limits were cut at very short projected separations due to the small angular separation of the stellar binary system. As the limits around the most distant stars in the sample start at wider physical separations than the cuts made on some of these targets, there is no region in the parameter space with a completeness fraction of 1. 

From the derived completeness maps and adopted model populations, we constrained the frequencies of sub-stellar companions in visual stellar binaries using the statistical code from \citet{fontanive_constraining_2018}. This Markov Chain Monte Carlo (MCMC) sampling tool was built using the {\sc emcee} \citep{foreman-mackey_emcee_2013} python algorithm, and aims at placing statistical constraints on stellar and sub-stellar populations (occurrence rates, mass and separation distributions of companions) based on observed surveys \citep[e.g.][]{fontanive_constraining_2018, fontanive_high_2019, vigan_sphere_2020}.
We performed separately the same analyses on the 27 stellar primaries, the 27 secondaries, and the combined 54 binary components, in order to place constraints on the occurrence rates of sub-stellar companions around primaries, secondaries and either component of wide binary systems, respectively.
In all cases, we fitted the relative frequencies associated with each part of the model population and derive the total fraction of sub-stellar companions as the sum of the two parts, as was done in \citet{vigan_sphere_2020}. 

\begin{figure*}
	\centering
	\includegraphics[width=\textwidth]{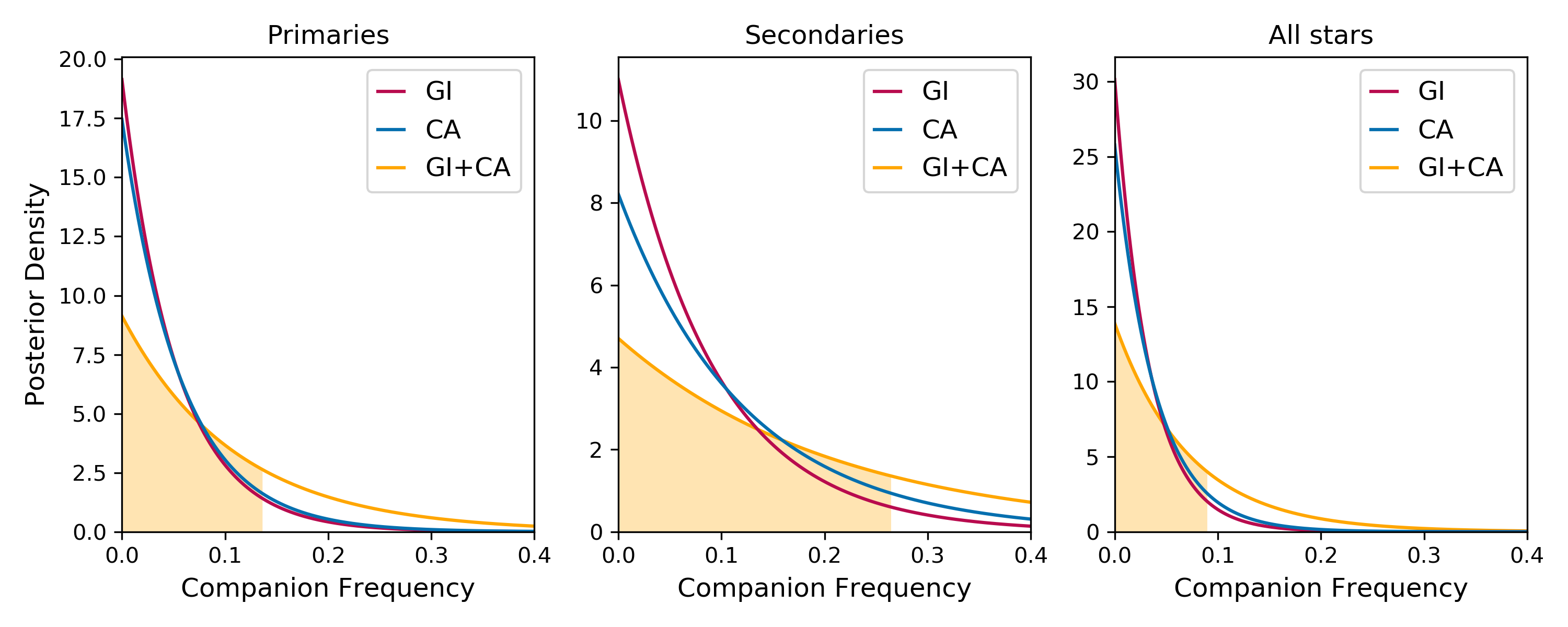}
	\caption{Posterior probability density functions of the frequencies of sub-stellar companions with masses in the range 10--75 M$_\mathrm{Jup}$ and semi-major axes between 10--200 au around the primary stars (left), secondary stars (centre) and all binary components (right). Each plot shows the posteriors obtained from the MCMC analysis for the relative frequencies of the two parts of the model population (GI in red and CA in blue), as well as for the total frequency for the full model (GI+CA in yellow). The shaded yellow areas mark the intervals of 68\% confidence derived on the total frequencies of sub-stellar companions in the considered ranges.}
	\label{fig:posteriors}
\end{figure*}

Based on the average completeness maps presented in Fig.~\ref{fig:completeness}, we chose to focus on the region of parameter space ranging from 10--200 au in semi-major axis, and from 10--75 M$_\mathrm{Jup}$ in mass. We used uniform priors between 0 and 1 for the two companion fractions. In each simulation, the MCMC code was run with 2000 walkers taking 5000 steps each. We found that convergence of the chains was reached after a couple hundred steps and discarded the initial 500 steps as the 'burn-in' phase. The results from the three analyses performed are presented in Fig.~\ref{fig:posteriors}, for the primaries (left), secondaries (centre) and all stars (right). The posterior distributions of the companion fractions associated with the GI and CA formation models are shown in red and blue, respectively. The yellow curves are inferred from the sum of the GI and CA frequencies at each step in the MCMC, and thus provide the total fraction of sub-stellar companions between 10--200 au and 10--75 M$_\mathrm{Jup}$.

We obtained upper limits for all cases. %
In general, roughly comparable limits were placed on both parts of the model, since the lack of detected companions does not allow us to scale the relative fractions between the GI and CA populations. The posteriors of the CA part (blue) are slightly wider due to the lower completeness levels achieved in the regions of the parameter space where most CA planets are predicted. This effect is emphasised for the secondaries (middle panel) around which the completeness reached is strongly decreased at the lowest masses and separations. The much lower completeness fraction reached around the secondaries in the probed parameter space (see Fig.~\ref{fig:completeness}) is also responsible for the overall looser constraints derived for this subset.
The shaded yellow areas in Fig.~\ref{fig:posteriors} indicate the 68\% confidence intervals for the full sub-stellar companion occurrences rates. The resulting values (1-$\sigma$ level) are: $<$13.7\% for the frequency of sub-stellar companions around primaries in visual binaries, $<$26.5\% for the fraction of sub-stellar companions around secondaries in visual binaries, and an occurrence rate of $<$9.0\% for giant planets and brown dwarfs around either component of visual binaries. Sample size is clearly a driving factor in the obtained results, with the tightest constraints derived the full sample of 54 stars. Of course, these results are only valid under the assumption that the GI and CA theories are the only possible formation channels for brown dwarfs and giant planets in the explored mass and separation ranges.

\section{Discussion}\label{sec:discussion}

\citet{vigan_sphere_2020} recently presented the first statistical results from the SPHERE GTO F150 (Desidera et al., submitted) single star sub-sample from the SHINE \citep[SpHere INfrared survey for Exoplanets,][]{chauvin_discovery_2017} survey. This sub-sample is based on the first half of SHINE targets observed excluding bad condition observations and newly discovered stellar binaries.
Our results appear to be broadly consistent with the frequencies of sub-stellar companions reported in \citet{vigan_sphere_2020} for single stars, although a direct comparison between the VIBES and SHINE F150 sample is not trivial. Indeed, our targets cover a wide range of spectral types, while the statistical analyses performed in \citet{vigan_sphere_2020} considered three different spectral bins, with different associated population models. The most relevant comparison to our results is arguably the constraints they obtained for FGK stars using the same GI and CA synthetic populations as in the present work, for which they derived a companion fraction of $5.7^{+3.8}_{-2.8}$\% between 1--75 M$_\mathrm{Jup}$ and 5--300 au. This is in agreement with our obtained value of $<$9\% for our full sample of 54 stars over the ranges 10--75 M$_\mathrm{Jup}$ and 10--200 au, noting that this fraction would be enhanced over the wider parameter space probed in the SHINE survey. Since our sample of secondary stars is somewhat biased towards lower stellar masses, we also compare the results from our analysis on secondaries to the occurrence rate measured by \citet{vigan_sphere_2020} for M dwarfs. Our loose constraints of $<$26.5\% is also consistent with the frequency of $12.6
^{+12.9}_{-7.1}$\% and $4.7^{+2.0}_{-1.5}$\%
from the SHINE \citep{vigan_sphere_2020} and GPIES \citep[5--80 M$_\mathrm{Jup}$, 10--100 au,][]{nielsen_gemini_2019} surveys respectively. Although we note again that the two analyses are not directly comparable and that slightly different companion mass and semi-major axis ranges were explored.

Several studies have suggested that systems with giant planets on very tight orbits may be more likely to host a wide stellar companion \citep{eggenberger_statistical_2004, desidera_properties_2007, mugrauer_multiplicity_2007}. In the Friends of hot Jupiters campaign, \citet{ngo_friends_2016} found that the occurrence of 50--2000 au binary companions is strongly increased when hot Jupiters are present, and concluded that binarity plays a role in the formation or evolution of such giant planets. More recently, \citet{fontanive_high_2019} searched for stellar companions to stars hosting planets and brown dwarfs companions with masses $>$7 M$_\mathrm{Jup}$ within 1 au and measured a further inflated binary fraction of $\sim$80\% on separations in the range 20--10,000 au. These results confirmed the trends observed for lower-mass planets, and suggested that the effects of stellar multiplicity on the presence of massive close-in companions may be enhanced for higher-mass planets and brown dwarfs.
The study conducted by \citet{fontanive_high_2019} is of particular relevant for imaging programmes as the masses of the inner companions probed in that survey (7--60 M$_\mathrm{Jup}$) are comparable to the typical masses of sub-stellar companions detectable with direct imaging. This mass range is indeed very similar to the mass interval explored in the statistical analyses presented in Sect. \ref{sec:analysis} (10--75 M$_\mathrm{Jup}$).

It has been proposed that these massive planets and brown dwarfs detected on short periods originally formed at wider orbital separations (i.e. where the directly-imaged population lies) and that their inward migration was induced or facilitated by the gravitational influence of a stellar companion \citep{ngo_friends_2016,fontanive_high_2019}. Although the precise stages in the system's evolution at which these processes would take place are not clear, we would expect in this case a number of giant planets and brown dwarf companions to reside at large separations in extremely young binary star systems. This makes targets like the ones probed in the VIBES survey potentially promising samples for the direct detection of sub-stellar companions.

However, even if binarity is indeed required to obtain massive planets on hot Jupiter orbits, such systems remain sparse and this scenario must hence be a rather rare event (there are less than 40 known systems with masses $>$7 M$_\mathrm{Jup}$ within 1 au despite the magnified sensitivity to bigger and more massive companions with the transit and radial velocity detection methods; see \citealp{fontanive_high_2019}). The low occurrence of such a mechanism could be due to the migration processes involved being somewhat inefficient, or be a direct consequence of low formation rates of wide giant planets in binaries. Therefore, the positive correlation between the existence of massive short-period planets and stellar multiplicity may not necessarily imply a higher frequency of massive planets in binaries. The null detection from our survey appears to confirm this idea, although larger sample sizes will be needed to more robustly measure the frequency of distant sub-stellar companions among visual binaries compared to single stars.

The SPOTS survey \citep{thalmann_spots:_2014, bonavita_spots:_2016, asensio-torres_spots:_2018}
which searched for companions in the $\sim30$--300~au range around tight binaries found upper limits on the
frequency of
circumbinary planets (1--15 M$_\text{Jup}$) and brown dwarfs (16--70 M$_\text{Jup}$) of 10 and 6\%
respectively, at a 95\% confidence level. Contrary to VIBES, the SPOTS survey sensitivity was only
marginally impacted by the binary nature of the targets yielding tighter upper limits but still no
planet detection.
Circumbinary discs are thought not to be perturbed by the binary star beyond two to four times the 
binary separation \citep{holman_long-term_1999} so that the formation history would be very different from
the intra-binary planet formation where the secondary star acts as an external perturber on the disc
and on the planet orbit \citep{ngo_friends_2016, fontanive_high_2019}. The results of the SPOTS survey even
though probing a different type of binary planet population is compatible with our results.

\section{Conclusion}

We presented the 'VIsual Binary Exoplanet survey with Sphere' (VIBES). 
We observed a total of 23 binaries and 4 triple systems with SPHERE in IRDIFS mode. The main results
of the survey are as follows.

\begin{itemize}
\item We have searched for sub-stellar companions around 46 stars which were part of a binary system,
around 4 stars in triples which were orbiting another binary star, and around 4 binaries in triples which were orbiting another star. No sub-stellar companion was detected as all companion candidates were followed up and turned out
to be background objects.

\item We derived (1-$\sigma$ level) upper limits of $<$13.7\% for the frequency of sub-stellar companions
around primaries in visual binaries, $<$26.5\% for the fraction of sub-stellar companions around secondaries
in visual binaries, and an occurrence rate of $<$9.0\% for giant planets and brown dwarfs (10--75 M$_\mathrm{Jup}$) around either
component of visual binaries, in the 10-200 au range.

\item We confirmed 20 binaries and 2 triple systems to be co-moving and thus physically bound.

\item One new 0.95 M$_\sun$ star was detected in the previously known binary system HD~121336.
\end{itemize}

The upper limits we have derived for planets in binaries are compatible with frequencies obtained in
single stars direct imaging surveys such as SHINE 
 \citep[$12.6
^{+12.9}_{-7.1}$\%,][]{vigan_sphere_2020} and GPIES \citep[$4.7^{+2.0}_{-1.5}$\%,][]{nielsen_gemini_2019}.
A direct comparison is however not possible given the difference in parameter ranges probed by these three surveys.

Studies of planets in binaries has increased in recent years specially through the search of stellar companions 
to transiting and RV planets, revealing the short period planets population. To probe the wider orbit planets
direct imaging remains the primary technique,
however current direct imaging techniques have a limited performance in the search of planets within 
close binaries.
Dual star coronagraphs \citep{cady_dual-mask_2011, aleksanyan_multiple-star_2017, kuhn_high-contrast_2018} 
and multi-star wavefront control techniques \citep{sirbu_techniques_2017} offer promising prospects
to significantly increase planet sensitivity in binaries. Nonetheless statistical completeness
is limited by the varying separation range which is set by the binary separation instead of the field of view.

\begin{acknowledgements}
	
We would like to thank Markus Janson for his input on the binaries he has observed, Ken Rice and Christoph Mordasini for useful discussions on population models, and the anonymous referee for the very 
constructive comments which significantly improved the scientific quality of the article.
J.H. is supported by SNSF through the Ambizione grant \#PZ00P2\_180098.
J.H, N.E, S.Q, J.K acknowledge the financial support from the SNSF through the National
Centre for Competence in Research “PlanetS”.
This research has made use of the SIMBAD database, operated at CDS, Strasbourg, France, and of NASA's Astrophysics Data System. This research made use of Astropy, a community-developed core Python package for Astronomy \citep{astropy_collaboration_astropy:_2013}, as well as numpy \citep{walt_numpy_2011}, and pandas \citep{mckinney_pandas:_2011}.
\end{acknowledgements}
	
\bibliographystyle{bibtex/aa} %
\bibliography{bibtex/VIBES.bib}

\begin{thebibliography}{117}
\expandafter\ifx\csname natexlab\endcsname\relax\def\natexlab#1{#1}\fi

\bibitem[{Aleksanyan {et~al.}(2017)Aleksanyan, Kravets, \&
  Brasselet}]{aleksanyan_multiple-star_2017}
Aleksanyan, A., Kravets, N., \& Brasselet, E. 2017, Physical Review Letters,
  118, 203902

\bibitem[{Allard(2014)}]{allard_bt-settl_2014}
Allard, F. 2014, in Proceedings of the {International} {Astronomical} {Union},
  Vol. 299 (IAU), 271--272, 00017

\bibitem[{Amara \& Quanz(2012)}]{amara_pynpoint:_2012}
Amara, A. \& Quanz, S.~P. 2012, Monthly Notices of the Royal Astronomical
  Society, 427, 948

\bibitem[{Artymowicz \& Lubow(1994)}]{artymowicz_dynamics_1994}
Artymowicz, P. \& Lubow, S.~H. 1994, The Astrophysical Journal, 421, 651, 00930

\bibitem[{Asensio-Torres {et~al.}(2018)Asensio-Torres, Janson, Bonavita,
  Desidera, Thalmann, Kuzuhara, Henning, Marzari, Meyer, Calissendorff, \&
  Uyama}]{asensio-torres_spots:_2018}
Asensio-Torres, R., Janson, M., Bonavita, M., {et~al.} 2018, Astronomy and
  Astrophysics, 619, A43

\bibitem[{{Astropy Collaboration} {et~al.}(2013){Astropy Collaboration},
  Robitaille, Tollerud, Greenfield, Droettboom, Bray, Aldcroft, Davis,
  Ginsburg, Price-Whelan, Kerzendorf, Conley, Crighton, Barbary, Muna,
  Ferguson, Grollier, Parikh, Nair, Unther, Deil, Woillez, Conseil, Kramer,
  Turner, Singer, Fox, Weaver, Zabalza, Edwards, Azalee~Bostroem, Burke, Casey,
  Crawford, Dencheva, Ely, Jenness, Labrie, Lim, Pierfederici, Pontzen, Ptak,
  Refsdal, Servillat, \& Streicher}]{astropy_collaboration_astropy:_2013}
{Astropy Collaboration}, Robitaille, T.~P., Tollerud, E.~J., {et~al.} 2013,
  Astronomy and Astrophysics, 558, 33

\bibitem[{Baraffe {et~al.}(2003)Baraffe, Chabrier, Barman, Allard, \&
  Hauschildt}]{baraffe_evolutionary_2003}
Baraffe, I., Chabrier, G., Barman, T.~S., Allard, F., \& Hauschildt, P.~H.
  2003, Astronomy and Astrophysics, 402, 701

\bibitem[{Batygin {et~al.}(2011)Batygin, Morbidelli, \&
  Tsiganis}]{batygin_formation_2011}
Batygin, K., Morbidelli, A., \& Tsiganis, K. 2011, Astronomy and Astrophysics,
  533, A7

\bibitem[{Bell {et~al.}(2015)Bell, Mamajek, \&
  Naylor}]{bell_self-consistent_2015}
Bell, C. P.~M., Mamajek, E.~E., \& Naylor, T. 2015, Monthly Notices of the
  Royal Astronomical Society, 454, 593

\bibitem[{Beuzit {et~al.}(2008)Beuzit, Feldt, Dohlen, Mouillet, Puget, Wildi,
  Abe, Antichi, Baruffolo, Baudoz, Boccaletti, Carbillet, Charton, Claudi,
  Downing, Fabron, Feautrier, Fedrigo, Fusco, Gach, Gratton, Henning, Hubin,
  Joos, Kasper, Langlois, Lenzen, Moutou, Pavlov, Petit, Pragt, Rabou, Rigal,
  Roelfsema, Rousset, Saisse, Schmid, Stadler, Thalmann, Turatto, Udry, Vakili,
  \& Waters}]{beuzit_sphere:_2008}
Beuzit, J.-L., Feldt, M., Dohlen, K., {et~al.} 2008, in , 701418, 00665

\bibitem[{Biller {et~al.}(2013)Biller, Liu, Wahhaj, Nielsen, Hayward, Males,
  Skemer, Close, Chun, Ftaclas, Clarke, Thatte, Shkolnik, Reid, Hartung, Boss,
  Lin, Alencar, de~Gouveia Dal~Pino, Gregorio-Hetem, \&
  Toomey}]{biller_gemini/nici_2013}
Biller, B.~A., Liu, M.~C., Wahhaj, Z., {et~al.} 2013, The Astrophysical
  Journal, 777, 160

\bibitem[{Binks \& Jeffries(2014)}]{binks_lithium_2014}
Binks, A.~S. \& Jeffries, R.~D. 2014, Monthly Notices of the Royal Astronomical
  Society, 438, L11

\bibitem[{Bonavita {et~al.}(2016)Bonavita, Desidera, Thalmann, Janson, Vigan,
  Chauvin, \& Lannier}]{bonavita_spots:_2016}
Bonavita, M., Desidera, S., Thalmann, C., {et~al.} 2016, Astronomy and
  Astrophysics, 593, A38

\bibitem[{Boss(1998)}]{boss_evolution_1998}
Boss, A.~P. 1998, The Astrophysical Journal, 503, 923, 00328

\bibitem[{Bowler {et~al.}(2020)Bowler, Blunt, \&
  Nielsen}]{bowler_population-level_2020}
Bowler, B.~P., Blunt, S.~C., \& Nielsen, E.~L. 2020, The Astronomical Journal,
  159, 63

\bibitem[{Cady {et~al.}(2011)Cady, McElwain, Kasdin, \&
  Thalmann}]{cady_dual-mask_2011}
Cady, E., McElwain, M., Kasdin, N.~J., \& Thalmann, C. 2011, Publications of
  the Astronomical Society of the Pacific, 123, 333

\bibitem[{Carbillet {et~al.}(2011)Carbillet, Bendjoya, Abe, Guerri, Boccaletti,
  Daban, Dohlen, Ferrari, Robbe-Dubois, Douet, \&
  Vakili}]{carbillet_apodized_2011}
Carbillet, M., Bendjoya, P., Abe, L., {et~al.} 2011, Experimental Astronomy,
  30, 39

\bibitem[{Chauvin {et~al.}(2011)Chauvin, Beust, Lagrange, \&
  Eggenberger}]{chauvin_planetary_2011}
Chauvin, G., Beust, H., Lagrange, A.-M., \& Eggenberger, A. 2011, Astronomy and
  Astrophysics, 528, A8

\bibitem[{Chauvin {et~al.}(2017)Chauvin, Desidera, Lagrange, Vigan, Gratton,
  Langlois, Bonnefoy, Beuzit, Feldt, Mouillet, Meyer, Cheetham, Biller,
  Boccaletti, D'Orazi, Galicher, Hagelberg, Maire, Mesa, Olofsson, Samland,
  Schmidt, Sissa, Bonavita, Charnay, Cudel, Daemgen, Delorme, Janin-Potiron,
  Janson, Keppler, Le~Coroller, Ligi, Marleau, Messina, Mollière, Mordasini,
  Müller, Peretti, Perrot, Rodet, Rouan, Zurlo, Dominik, Henning, Menard,
  Schmid, Turatto, Udry, Vakili, Abe, Antichi, Baruffolo, Baudoz, Baudrand,
  Blanchard, Bazzon, Buey, Carbillet, Carle, Charton, Cascone, Claudi,
  Costille, Deboulbe, De~Caprio, Dohlen, Fantinel, Feautrier, Fusco, Gigan,
  Giro, Gisler, Gluck, Hubin, Hugot, Jaquet, Kasper, Madec, Magnard, Martinez,
  Maurel, Le~Mignant, Möller-Nilsson, Llored, Moulin, Origné, Pavlov, Perret,
  Petit, Pragt, Puget, Rabou, Ramos, Rigal, Rochat, Roelfsema, Rousset, Roux,
  Salasnich, Sauvage, Sevin, Soenke, Stadler, Suarez, Weber, Wildi, Antoniucci,
  Augereau, Baudino, Brandner, Engler, Girard, Gry, Kral, Kopytova, Lagadec,
  Milli, Moutou, Schlieder, Szulágyi, Thalmann, \&
  Wahhaj}]{chauvin_discovery_2017}
Chauvin, G., Desidera, S., Lagrange, A.-M., {et~al.} 2017, Astronomy and
  Astrophysics, 605, L9, 00002

\bibitem[{Chauvin {et~al.}(2003)Chauvin, Thomson, Dumas, Beuzit, Lowrance,
  Fusco, Lagrange, Zuckerman, \& Mouillet}]{chauvin_adaptive_2003}
Chauvin, G., Thomson, M., Dumas, C., {et~al.} 2003, Astronomy and Astrophysics,
  404, 157

\bibitem[{Chauvin {et~al.}(2015)Chauvin, Vigan, Bonnefoy, Desidera, Bonavita,
  Mesa, Boccaletti, Buenzli, Carson, Delorme, Hagelberg, Montagnier, Mordasini,
  Quanz, Segransan, Thalmann, Beuzit, Biller, Covino, Feldt, Girard, Gratton,
  Henning, Kasper, Lagrange, Messina, Meyer, Mouillet, Moutou, Reggiani,
  Schlieder, \& Zurlo}]{chauvin_vlt/naco_2015}
Chauvin, G., Vigan, A., Bonnefoy, M., {et~al.} 2015, Astronomy and
  Astrophysics, 573, A127

\bibitem[{Cheetham {et~al.}(2018)Cheetham, Ségransan, Peretti, Delisle,
  Hagelberg, Beuzit, Forveille, Marmier, Udry, \& Wildi}]{cheetham_direct_2018}
Cheetham, A., Ségransan, D., Peretti, S., {et~al.} 2018, Astronomy and
  Astrophysics, 614, A16

\bibitem[{Claudi {et~al.}(2008)Claudi, Turatto, Gratton, Antichi, Bonavita,
  Bruno, Cascone, De~Caprio, Desidera, \& Giro}]{claudi_sphere_2008}
Claudi, R.~U., Turatto, M., Gratton, R.~G., {et~al.} 2008, in Astronomical
  {Telescopes} and {Instrumentation}: {Synergies} {Between} {Ground} and
  {Space}, 70143E--70143E

\bibitem[{Collaboration {et~al.}(2016)Collaboration, Prusti, de~Bruijne, Brown,
  Vallenari, Babusiaux, Bailer-Jones, Bastian, Biermann, Evans, Eyer, Jansen,
  Jordi, Klioner, Lammers, Lindegren, Luri, Mignard, Milligan, Panem,
  Poinsignon, Pourbaix, Randich, Sarri, Sartoretti, Siddiqui, Soubiran,
  Valette, van Leeuwen, Walton, Aerts, Arenou, Cropper, Drimmel, Høg, Katz,
  Lattanzi, O'Mullane, Grebel, Holland, Huc, Passot, Bramante, Cacciari,
  Castañeda, Chaoul, Cheek, De~Angeli, Fabricius, Guerra, Hernández,
  Jean-Antoine-Piccolo, Masana, Messineo, Mowlavi, Nienartowicz,
  Ordóñez-Blanco, Panuzzo, Portell, Richards, Riello, Seabroke, Tanga,
  Thévenin, Torra, Els, Gracia-Abril, Comoretto, Garcia-Reinaldos, Lock,
  Mercier, Altmann, Andrae, Astraatmadja, Bellas-Velidis, Benson, Berthier,
  Blomme, Busso, Carry, Cellino, Clementini, Cowell, Creevey, Cuypers,
  Davidson, De~Ridder, de~Torres, Delchambre, Dell'Oro, Ducourant, Frémat,
  García-Torres, Gosset, Halbwachs, Hambly, Harrison, Hauser, Hestroffer,
  Hodgkin, Huckle, Hutton, Jasniewicz, Jordan, Kontizas, Korn, Lanzafame,
  Manteiga, Moitinho, Muinonen, Osinde, Pancino, Pauwels, Petit, Recio-Blanco,
  Robin, Sarro, Siopis, Smith, Smith, Sozzetti, Thuillot, van Reeven, Viala,
  Abbas, Abreu~Aramburu, Accart, Aguado, Allan, Allasia, Altavilla, Álvarez,
  Alves, Anderson, Andrei, Anglada~Varela, Antiche, Antoja, Antón, Arcay,
  Atzei, Ayache, Bach, Baker, Balaguer-Núñez, Barache, Barata, Barbier,
  Barblan, Baroni, Barrado~y Navascués, Barros, Barstow, Becciani, Bellazzini,
  Bellei, Bello~García, Belokurov, Bendjoya, Berihuete, Bianchi, Bienaymé,
  Billebaud, Blagorodnova, Blanco-Cuaresma, Boch, Bombrun, Borrachero,
  Bouquillon, Bourda, Bouy, Bragaglia, Breddels, Brouillet, Brüsemeister,
  Bucciarelli, Budnik, Burgess, Burgon, Burlacu, Busonero, Buzzi, Caffau,
  Cambras, Campbell, Cancelliere, Cantat-Gaudin, Carlucci, Carrasco,
  Castellani, Charlot, Charnas, Charvet, Chassat, Chiavassa, Clotet, Cocozza,
  Collins, Collins, Costigan, Crifo, Cross, Crosta, Crowley, Dafonte, Damerdji,
  Dapergolas, David, David, De~Cat, de~Felice, de~Laverny, De~Luise, De~March,
  de~Martino, de~Souza, Debosscher, del Pozo, Delbo, Delgado, Delgado,
  di~Marco, Di~Matteo, Diakite, Distefano, Dolding, Dos~Anjos, Drazinos,
  Durán, Dzigan, Ecale, Edvardsson, Enke, Erdmann, Escolar, Espina, Evans,
  Eynard~Bontemps, Fabre, Fabrizio, Faigler, Falcão, Farràs~Casas, Faye,
  Federici, Fedorets, Fernández-Hernández, Fernique, Fienga, Figueras,
  Filippi, Findeisen, Fonti, Fouesneau, Fraile, Fraser, Fuchs, Furnell, Gai,
  Galleti, Galluccio, Garabato, García-Sedano, Garé, Garofalo, Garralda,
  Gavras, Gerssen, Geyer, Gilmore, Girona, Giuffrida, Gomes, González-Marcos,
  González-Núñez, González-Vidal, Granvik, Guerrier, Guillout, Guiraud,
  Gúrpide, Gutiérrez-Sánchez, Guy, Haigron, Hatzidimitriou, Haywood, Heiter,
  Helmi, Hobbs, Hofmann, Holl, Holland, Hunt, Hypki, Icardi, Irwin, Jevardat~de
  Fombelle, Jofré, Jonker, Jorissen, Julbe, Karampelas, Kochoska, Kohley,
  Kolenberg, Kontizas, Koposov, Kordopatis, Koubsky, Kowalczyk, Krone-Martins,
  Kudryashova, Kull, Bachchan, Lacoste-Seris, Lanza, Lavigne,
  Le~Poncin-Lafitte, Lebreton, Lebzelter, Leccia, Leclerc, Lecoeur-Taibi,
  Lemaitre, Lenhardt, Leroux, Liao, Licata, Lindstrøm, Lister, Livanou, Lobel,
  Löffler, López, Lopez-Lozano, Lorenz, Loureiro, MacDonald,
  Magalhães~Fernandes, Managau, Mann, Mantelet, Marchal, Marchant, Marconi,
  Marie, Marinoni, Marrese, Marschalkó, Marshall, Martín-Fleitas, Martino,
  Mary, Matijevič, Mazeh, McMillan, Messina, Mestre, Michalik, Millar,
  Miranda, Molina, Molinaro, Molinaro, Molnár, Moniez, Montegriffo, Monteiro,
  Mor, Mora, Morbidelli, Morel, Morgenthaler, Morley, Morris, Mulone, Muraveva,
  Musella, Narbonne, Nelemans, Nicastro, Noval, Ordénovic, Ordieres-Meré,
  Osborne, Pagani, Pagano, Pailler, Palacin, Palaversa, Parsons, Paulsen,
  Pecoraro, Pedrosa, Pentikäinen, Pereira, Pichon, Piersimoni, Pineau, Plachy,
  Plum, Poujoulet, Prša, Pulone, Ragaini, Rago, Rambaux, Ramos-Lerate,
  Ranalli, Rauw, Read, Regibo, Renk, Reylé, Ribeiro, Rimoldini, Ripepi, Riva,
  Rixon, Roelens, Romero-Gómez, Rowell, Royer, Rudolph, Ruiz-Dern, Sadowski,
  Sagristà~Sellés, Sahlmann, Salgado, Salguero, Sarasso, Savietto, Schnorhk,
  Schultheis, Sciacca, Segol, Segovia, Segransan, Serpell, Shih, Smareglia,
  Smart, Smith, Solano, Solitro, Sordo, Soria~Nieto, Souchay, Spagna, Spoto,
  Stampa, Steele, Steidelmüller, Stephenson, Stoev, Suess, Süveges, Surdej,
  Szabados, Szegedi-Elek, Tapiador, Taris, Tauran, Taylor, Teixeira, Terrett,
  Tingley, Trager, Turon, Ulla, Utrilla, Valentini, van Elteren, Van~Hemelryck,
  van Leeuwen, Varadi, Vecchiato, Veljanoski, Via, Vicente, Vogt, Voss,
  Votruba, Voutsinas, Walmsley, Weiler, Weingrill, Werner, Wevers, Whitehead,
  Wyrzykowski, Yoldas, Žerjal, Zucker, Zurbach, Zwitter, Alecu, Allen,
  Allende~Prieto, Amorim, Anglada-Escudé, Arsenijevic, Azaz, Balm, Beck,
  Bernstein, Bigot, Bijaoui, Blasco, Bonfigli, Bono, Boudreault, Bressan,
  Brown, Brunet, Bunclark, Buonanno, Butkevich, Carret, Carrion, Chemin,
  Chéreau, Corcione, Darmigny, de~Boer, de~Teodoro, de~Zeeuw, Delle~Luche,
  Domingues, Dubath, Fodor, Frézouls, Fries, Fustes, Fyfe, Gallardo, Gallegos,
  Gardiol, Gebran, Gomboc, Gómez, Grux, Gueguen, Heyrovsky, Hoar, Iannicola,
  Isasi~Parache, Janotto, Joliet, Jonckheere, Keil, Kim, Klagyivik, Klar,
  Knude, Kochukhov, Kolka, Kos, Kutka, Lainey, LeBouquin, Liu, Loreggia,
  Makarov, Marseille, Martayan, Martinez-Rubi, Massart, Meynadier, Mignot,
  Munari, Nguyen, Nordlander, Ocvirk, O'Flaherty, Olias~Sanz, Ortiz, Osorio,
  Oszkiewicz, Ouzounis, Palmer, Park, Pasquato, Peltzer, Peralta, Péturaud,
  Pieniluoma, Pigozzi, Poels, Prat, Prod'homme, Raison, Rebordao, Risquez,
  Rocca-Volmerange, Rosen, Ruiz-Fuertes, Russo, Sembay, Serraller~Vizcaino,
  Short, Siebert, Silva, Sinachopoulos, Slezak, Soffel, Sosnowska, Straižys,
  ter Linden, Terrell, Theil, Tiede, Troisi, Tsalmantza, Tur, Vaccari, Vachier,
  Valles, Van~Hamme, Veltz, Virtanen, Wallut, Wichmann, Wilkinson, Ziaeepour,
  \& Zschocke}]{collaboration_gaia_2016}
Collaboration, G., Prusti, T., de~Bruijne, J. H.~J., {et~al.} 2016, Astronomy
  and Astrophysics, 595, A1

\bibitem[{Correia {et~al.}(2008)Correia, Udry, Mayor, Eggenberger, Naef,
  Beuzit, Perrier, Queloz, Sivan, Pepe, Santos, \&
  Ségransan}]{correia_elodie_2008}
Correia, A. C.~M., Udry, S., Mayor, M., {et~al.} 2008, Astronomy and
  Astrophysics, 479, 271, 00047

\bibitem[{Daemgen {et~al.}(2015)Daemgen, Bonavita, Jayawardhana, Lafrenière,
  \& Janson}]{daemgen_sub-stellar_2015}
Daemgen, S., Bonavita, M., Jayawardhana, R., Lafrenière, D., \& Janson, M.
  2015, The Astrophysical Journal, 799, 155

\bibitem[{De~Rosa {et~al.}(2014)De~Rosa, Patience, Wilson, Schneider,
  Wiktorowicz, Vigan, Marois, Song, Macintosh, Graham, Doyon, Bessell, Thomas,
  \& Lai}]{de_rosa_vast_2014}
De~Rosa, R.~J., Patience, J., Wilson, P.~A., {et~al.} 2014, Monthly Notices of
  the Royal Astronomical Society, 437, 1216

\bibitem[{de~Zeeuw {et~al.}(1999)de~Zeeuw, Hoogerwerf, de~Bruijne, Brown, \&
  Blaauw}]{de_zeeuw_hipparcos_1999}
de~Zeeuw, P.~T., Hoogerwerf, R., de~Bruijne, J. H.~J., Brown, A. G.~A., \&
  Blaauw, A. 1999, The Astronomical Journal, 117, 354

\bibitem[{Desidera \& Barbieri(2007)}]{desidera_properties_2007}
Desidera, S. \& Barbieri, M. 2007, Astronomy and Astrophysics, 462, 345

\bibitem[{Desidera {et~al.}(2015)Desidera, Covino, Messina, Carson, Hagelberg,
  Schlieder, Biazzo, Alcalá, Chauvin, Vigan, Beuzit, Bonavita, Bonnefoy,
  Delorme, D'Orazi, Esposito, Feldt, Girardi, Gratton, Henning, Lagrange,
  Lanzafame, Launhardt, Marmier, Melo, Meyer, Mouillet, Moutou, Segransan,
  Udry, \& Zaidi}]{desidera_vlt/naco_2015}
Desidera, S., Covino, E., Messina, S., {et~al.} 2015, Astronomy and
  Astrophysics, 573, A126, 00000

\bibitem[{Dohlen {et~al.}(2008)Dohlen, Langlois, Saisse, Hill, Origne, Jacquet,
  Fabron, Blanc, Llored, Carle, Moutou, Vigan, Boccaletti, Carbillet, Mouillet,
  \& Beuzit}]{dohlen_infra-red_2008}
Dohlen, K., Langlois, M., Saisse, M., {et~al.} 2008, in , 118, 00000

\bibitem[{Dressing \& Charbonneau(2013)}]{dressing_occurrence_2013}
Dressing, C.~D. \& Charbonneau, D. 2013, The Astrophysical Journal, 767, 95,
  00397

\bibitem[{Dvorak(1982)}]{dvorak_planetenbahnen_1982}
Dvorak, R. 1982, Oesterreichische Akademie Wissenschaften Mathematisch
  naturwissenschaftliche Klasse Sitzungsberichte Abteilung, 191, 423

\bibitem[{Dvorak(1984)}]{dvorak_numerical_1984}
Dvorak, R. 1984, Celestial Mechanics, 34, 369

\bibitem[{Eggenberger {et~al.}(2004)Eggenberger, Udry, \&
  Mayor}]{eggenberger_statistical_2004}
Eggenberger, A., Udry, S., \& Mayor, M. 2004, Astronomy and Astrophysics, 417,
  353

\bibitem[{Elliott {et~al.}(2015)Elliott, Huélamo, Bouy, Bayo, Melo, Torres,
  Sterzik, Quast, Chauvin, \& Barrado}]{elliott_search_2015}
Elliott, P., Huélamo, N., Bouy, H., {et~al.} 2015, Astronomy and Astrophysics,
  580, A88

\bibitem[{Emsenhuber {et~al.}(2020{\natexlab{a}})Emsenhuber, Mordasini, Burn,
  Alibert, Benz, \& Asphaug}]{emsenhuber_new_2020}
Emsenhuber, A., Mordasini, C., Burn, R., {et~al.} 2020{\natexlab{a}}, arXiv
  e-prints, 2007, arXiv:2007.05561

\bibitem[{Emsenhuber {et~al.}(2020{\natexlab{b}})Emsenhuber, Mordasini, Burn,
  Alibert, Benz, \& Asphaug}]{emsenhuber_new_2020-1}
Emsenhuber, A., Mordasini, C., Burn, R., {et~al.} 2020{\natexlab{b}}, arXiv
  e-prints, 2007, arXiv:2007.05562

\bibitem[{Evans {et~al.}(2018)Evans, Southworth, Smalley, Jørgensen, Dominik,
  Andersen, Bozza, Bramich, Burgdorf, Ciceri, D'Ago, Figuera~Jaimes, Gu, Hinse,
  Henning, Hundertmark, Kains, Kerins, Korhonen, Kokotanekova, Kuffmeier,
  Longa-Peña, Mancini, MacKenzie, Popovas, Rabus, Rahvar, Sajadian, Snodgrass,
  Skottfelt, Surdej, Tronsgaard, Unda-Sanzana, von Essen, Wang, \&
  Wertz}]{evans_high-resolution_2018}
Evans, D.~F., Southworth, J., Smalley, B., {et~al.} 2018, Astronomy and
  Astrophysics, 610, A20

\bibitem[{Fabricius {et~al.}(2002)Fabricius, Høg, Makarov, Mason, Wycoff, \&
  Urban}]{fabricius_tycho_2002}
Fabricius, C., Høg, E., Makarov, V.~V., {et~al.} 2002, Astronomy and
  Astrophysics, 384, 180

\bibitem[{Fontanive {et~al.}(2018)Fontanive, Biller, Bonavita, \&
  Allers}]{fontanive_constraining_2018}
Fontanive, C., Biller, B., Bonavita, M., \& Allers, K. 2018, Monthly Notices of
  the Royal Astronomical Society, 479, 2702

\bibitem[{Fontanive {et~al.}(2019)Fontanive, Rice, Bonavita, Lopez, {Mužić},
  ~, \& Biller}]{fontanive_high_2019}
Fontanive, C., Rice, K., Bonavita, M., {et~al.} 2019, Monthly Notices of the
  Royal Astronomical Society, 485, 4967

\bibitem[{Foreman-Mackey {et~al.}(2013)Foreman-Mackey, Hogg, Lang, \&
  Goodman}]{foreman-mackey_emcee_2013}
Foreman-Mackey, D., Hogg, D.~W., Lang, D., \& Goodman, J. 2013, Publications of
  the Astronomical Society of the Pacific, 125, 306

\bibitem[{Forgan \& Rice(2009)}]{forgan_stellar_2009}
Forgan, D. \& Rice, K. 2009, Monthly Notices of the Royal Astronomical Society,
  400, 2022, publisher: Oxford Academic

\bibitem[{Forgan \& Rice(2013)}]{forgan_towards_2013}
Forgan, D. \& Rice, K. 2013, Monthly Notices of the Royal Astronomical Society,
  432, 3168

\bibitem[{Forgan {et~al.}(2018)Forgan, Hall, Meru, \&
  Rice}]{forgan_towards_2018}
Forgan, D.~H., Hall, C., Meru, F., \& Rice, W. K.~M. 2018, Monthly Notices of
  the Royal Astronomical Society, 474, 5036

\bibitem[{Fortney {et~al.}(2008)Fortney, Marley, Saumon, \&
  Lodders}]{fortney_synthetic_2008}
Fortney, J.~J., Marley, M.~S., Saumon, D., \& Lodders, K. 2008, The
  Astrophysical Journal, 683, 1104

\bibitem[{Fressin {et~al.}(2013)Fressin, Torres, Charbonneau, Bryson,
  Christiansen, Dressing, Jenkins, Walkowicz, \& Batalha}]{fressin_false_2013}
Fressin, F., Torres, G., Charbonneau, D., {et~al.} 2013, The Astrophysical
  Journal, 766, 81, 00562

\bibitem[{Gagné {et~al.}(2018)Gagné, Mamajek, Malo, Riedel, Rodriguez,
  Lafrenière, Faherty, Roy-Loubier, Pueyo, Robin, \&
  Doyon}]{gagne_banyan_2018}
Gagné, J., Mamajek, E.~E., Malo, L., {et~al.} 2018, The Astrophysical Journal,
  856, 23

\bibitem[{{Gaia Collaboration} {et~al.}(2018){Gaia Collaboration}, Brown,
  Vallenari, Prusti, de~Bruijne, Babusiaux, Bailer-Jones, Biermann, Evans,
  Eyer, Jansen, Jordi, Klioner, Lammers, Lindegren, Luri, Mignard, Panem,
  Pourbaix, Randich, Sartoretti, Siddiqui, Soubiran, van Leeuwen, Walton,
  Arenou, Bastian, Cropper, Drimmel, Katz, Lattanzi, Bakker, Cacciari,
  Castañeda, Chaoul, Cheek, De~Angeli, Fabricius, Guerra, Holl, Masana,
  Messineo, Mowlavi, Nienartowicz, Panuzzo, Portell, Riello, Seabroke, Tanga,
  Thévenin, Gracia-Abril, Comoretto, Garcia-Reinaldos, Teyssier, Altmann,
  Andrae, Audard, Bellas-Velidis, Benson, Berthier, Blomme, Burgess, Busso,
  Carry, Cellino, Clementini, Clotet, Creevey, Davidson, De~Ridder, Delchambre,
  Dell'Oro, Ducourant, Fernández-Hernández, Fouesneau, Frémat, Galluccio,
  García-Torres, González-Núñez, González-Vidal, Gosset, Guy, Halbwachs,
  Hambly, Harrison, Hernández, Hestroffer, Hodgkin, Hutton, Jasniewicz,
  Jean-Antoine-Piccolo, Jordan, Korn, Krone-Martins, Lanzafame, Lebzelter,
  Löffler, Manteiga, Marrese, Martín-Fleitas, Moitinho, Mora, Muinonen,
  Osinde, Pancino, Pauwels, Petit, Recio-Blanco, Richards, Rimoldini, Robin,
  Sarro, Siopis, Smith, Sozzetti, Süveges, Torra, van Reeven, Abbas,
  Abreu~Aramburu, Accart, Aerts, Altavilla, Álvarez, Alvarez, Alves, Anderson,
  Andrei, Anglada~Varela, Antiche, Antoja, Arcay, Astraatmadja, Bach, Baker,
  Balaguer-Núñez, Balm, Barache, Barata, Barbato, Barblan, Barklem, Barrado,
  Barros, Barstow, Bartholomé~Muñoz, Bassilana, Becciani, Bellazzini,
  Berihuete, Bertone, Bianchi, Bienaymé, Blanco-Cuaresma, Boch, Boeche,
  Bombrun, Borrachero, Bossini, Bouquillon, Bourda, Bragaglia, Bramante,
  Breddels, Bressan, Brouillet, Brüsemeister, Brugaletta, Bucciarelli,
  Burlacu, Busonero, Butkevich, Buzzi, Caffau, Cancelliere, Cannizzaro,
  Cantat-Gaudin, Carballo, Carlucci, Carrasco, Casamiquela, Castellani,
  Castro-Ginard, Charlot, Chemin, Chiavassa, Cocozza, Costigan, Cowell, Crifo,
  Crosta, Crowley, Cuypers, Dafonte, Damerdji, Dapergolas, David, David,
  de~Laverny, De~Luise, De~March, de~Martino, de~Souza, de~Torres, Debosscher,
  del Pozo, Delbo, Delgado, Delgado, Di~Matteo, Diakite, Diener, Distefano,
  Dolding, Drazinos, Durán, Edvardsson, Enke, Eriksson, Esquej,
  Eynard~Bontemps, Fabre, Fabrizio, Faigler, Falcão, Farràs~Casas, Federici,
  Fedorets, Fernique, Figueras, Filippi, Findeisen, Fonti, Fraile, Fraser,
  Frézouls, Gai, Galleti, Garabato, García-Sedano, Garofalo, Garralda, Gavel,
  Gavras, Gerssen, Geyer, Giacobbe, Gilmore, Girona, Giuffrida, Glass, Gomes,
  Granvik, Gueguen, Guerrier, Guiraud, Gutiérrez-Sánchez, Haigron,
  Hatzidimitriou, Hauser, Haywood, Heiter, Helmi, Heu, Hilger, Hobbs, Hofmann,
  Holland, Huckle, Hypki, Icardi, Janßen, Jevardat~de Fombelle, Jonker,
  Juhász, Julbe, Karampelas, Kewley, Klar, Kochoska, Kohley, Kolenberg,
  Kontizas, Kontizas, Koposov, Kordopatis, Kostrzewa-Rutkowska, Koubsky,
  Lambert, Lanza, Lasne, Lavigne, Le~Fustec, Le~Poncin-Lafitte, Lebreton,
  Leccia, Leclerc, Lecoeur-Taibi, Lenhardt, Leroux, Liao, Licata, Lindstrøm,
  Lister, Livanou, Lobel, López, Managau, Mann, Mantelet, Marchal, Marchant,
  Marconi, Marinoni, Marschalkó, Marshall, Martino, Marton, Mary, Massari,
  Matijevič, Mazeh, McMillan, Messina, Michalik, Millar, Molina, Molinaro,
  Molnár, Montegriffo, Mor, Morbidelli, Morel, Morris, Mulone, Muraveva,
  Musella, Nelemans, Nicastro, Noval, O'Mullane, Ordénovic, Ordóñez-Blanco,
  Osborne, Pagani, Pagano, Pailler, Palacin, Palaversa, Panahi, Pawlak,
  Piersimoni, Pineau, Plachy, Plum, Poggio, Poujoulet, Prša, Pulone, Racero,
  Ragaini, Rambaux, Ramos-Lerate, Regibo, Reylé, Riclet, Ripepi, Riva, Rivard,
  Rixon, Roegiers, Roelens, Romero-Gómez, Rowell, Royer, Ruiz-Dern, Sadowski,
  Sagristà~Sellés, Sahlmann, Salgado, Salguero, Sanna, Santana-Ros, Sarasso,
  Savietto, Schultheis, Sciacca, Segol, Segovia, Ségransan, Shih, Siltala,
  Silva, Smart, Smith, Solano, Solitro, Sordo, Soria~Nieto, Souchay, Spagna,
  Spoto, Stampa, Steele, Steidelmüller, Stephenson, Stoev, Suess, Surdej,
  Szabados, Szegedi-Elek, Tapiador, Taris, Tauran, Taylor, Teixeira, Terrett,
  Teyssandier, Thuillot, Titarenko, Torra~Clotet, Turon, Ulla, Utrilla, Uzzi,
  Vaillant, Valentini, Valette, van Elteren, Van~Hemelryck, van Leeuwen,
  Vaschetto, Vecchiato, Veljanoski, Viala, Vicente, Vogt, von Essen, Voss,
  Votruba, Voutsinas, Walmsley, Weiler, Wertz, Wevers, Wyrzykowski, Yoldas,
  Žerjal, Ziaeepour, Zorec, Zschocke, Zucker, Zurbach, \&
  Zwitter}]{gaia_collaboration_gaia_2018}
{Gaia Collaboration}, Brown, A. G.~A., Vallenari, A., {et~al.} 2018, Astronomy
  and Astrophysics, 616, A1

\bibitem[{Guerri {et~al.}(2011)Guerri, Daban, Robbe-Dubois, Douet, Abe,
  Baudrand, Carbillet, Boccaletti, Bendjoya, Gouvret, \&
  Vakili}]{guerri_apodized_2011}
Guerri, G., Daban, J.-B., Robbe-Dubois, S., {et~al.} 2011, Experimental
  Astronomy, 30, 59

\bibitem[{Hagelberg {et~al.}(2016)Hagelberg, Ségransan, Udry, \&
  Wildi}]{hagelberg_geneva_2016}
Hagelberg, J., Ségransan, D., Udry, S., \& Wildi, F. 2016, Monthly Notices of
  the Royal Astronomical Society, 455, 2178

\bibitem[{Hartkopf {et~al.}(1996)Hartkopf, Mason, McAlister, Turner, Barry,
  Franz, \& Prieto}]{hartkopf_iccd_1996}
Hartkopf, W.~I., Mason, B.~D., McAlister, H.~A., {et~al.} 1996, The
  Astronomical Journal, 111, 936

\bibitem[{Hatzes {et~al.}(2003)Hatzes, Cochran, Endl, McArthur, Paulson,
  Walker, Campbell, \& Yang}]{hatzes_planetary_2003}
Hatzes, A.~P., Cochran, W.~D., Endl, M., {et~al.} 2003, The Astrophysical
  Journal, 599, 1383, 00265

\bibitem[{Herschel {et~al.}(1874)Herschel, Main, \&
  Pritchard}]{herschel_catalogue_1874}
Herschel, J. F.~W., Main, R., \& Pritchard, C. 1874, Memoirs of the Royal
  Astronomical Society, 40, 1

\bibitem[{Holman \& Wiegert(1999)}]{holman_long-term_1999}
Holman, M.~J. \& Wiegert, P.~A. 1999, The Astronomical Journal, 117, 621

\bibitem[{Howard {et~al.}(2012)Howard, Marcy, Bryson, Jenkins, Rowe, Batalha,
  Borucki, Koch, Dunham, Gautier, Van~Cleve, Cochran, Latham, Lissauer, Torres,
  Brown, Gilliland, Buchhave, Caldwell, Christensen-Dalsgaard, Ciardi, Fressin,
  Haas, Howell, Kjeldsen, Seager, Rogers, Sasselov, Steffen, Basri,
  Charbonneau, Christiansen, Clarke, Dupree, Fabrycky, Fischer, Ford, Fortney,
  Tarter, Girouard, Holman, Johnson, Klaus, Machalek, Moorhead, Morehead,
  Ragozzine, Tenenbaum, Twicken, Quinn, Isaacson, Shporer, Lucas, Walkowicz,
  Welsh, Boss, Devore, Gould, Smith, Morris, Prsa, Morton, Still, Thompson,
  Mullally, Endl, \& MacQueen}]{howard_planet_2012}
Howard, A.~W., Marcy, G.~W., Bryson, S.~T., {et~al.} 2012, The Astrophysical
  Journal Supplement Series, 201, 15, 00592

\bibitem[{Janson {et~al.}(2013)Janson, Lafrenière, Jayawardhana, Bonavita,
  Girard, Brandeker, \& Gizis}]{janson_multiplicity_2013}
Janson, M., Lafrenière, D., Jayawardhana, R., {et~al.} 2013, The Astrophysical
  Journal, 773, 170

\bibitem[{Jovanovic {et~al.}(2015)Jovanovic, Guyon, Martinache, Pathak,
  Hagelberg, \& Kudo}]{jovanovic_artificial_2015}
Jovanovic, N., Guyon, O., Martinache, F., {et~al.} 2015, The Astrophysical
  Journal Letters, 813, L24, 00000

\bibitem[{Kenyon \& Hartmann(1995)}]{kenyon_pre-main-sequence_1995}
Kenyon, S.~J. \& Hartmann, L. 1995, The Astrophysical Journal Supplement
  Series, 101, 117

\bibitem[{Kohler \& Leinert(1998)}]{kohler_multiplicity_1998}
Kohler, R. \& Leinert, C. 1998, Astronomy and Astrophysics, 331, 977

\bibitem[{Kouwenhoven(2006)}]{kouwenhoven_primordial_2006}
Kouwenhoven, M. B.~N. 2006, Ph.D. Thesis

\bibitem[{Kouwenhoven {et~al.}(2005)Kouwenhoven, Brown, Zinnecker, Kaper, \&
  Portegies~Zwart}]{kouwenhoven_primordial_2005}
Kouwenhoven, M. B.~N., Brown, A. G.~A., Zinnecker, H., Kaper, L., \&
  Portegies~Zwart, S.~F. 2005, Astronomy and Astrophysics, 430, 137

\bibitem[{Kozai(1962)}]{kozai_secular_1962}
Kozai, Y. 1962, The Astronomical Journal, 67, 579

\bibitem[{Kraus \& Hillenbrand(2009)}]{kraus_coevality_2009}
Kraus, A.~L. \& Hillenbrand, L.~A. 2009, The Astrophysical Journal, 704, 531

\bibitem[{Kraus {et~al.}(2016)Kraus, Ireland, Huber, Mann, \&
  Dupuy}]{kraus_impact_2016}
Kraus, A.~L., Ireland, M.~J., Huber, D., Mann, A.~W., \& Dupuy, T.~J. 2016, The
  Astronomical Journal, 152, 8

\bibitem[{Kühn {et~al.}(2018)Kühn, Daemgen, Wang, Morales, Bottom, Serabyn,
  Shelton, Delorme, \& Tinyanont}]{kuhn_high-contrast_2018}
Kühn, J., Daemgen, S., Wang, J., {et~al.} 2018, 0702, 1070242, conference
  Name: Ground-based and Airborne Instrumentation for Astronomy VII ISBN:
  9781510619579 Place: eprint: arXiv:1808.00585

\bibitem[{Lidov(1962)}]{lidov_evolution_1962}
Lidov, M.~L. 1962, Planetary and Space Science, 9, 719

\bibitem[{Lin \& Papaloizou(1986)}]{lin_tidal_1986}
Lin, D. N.~C. \& Papaloizou, J. 1986, The Astrophysical Journal, 307, 395

\bibitem[{Lindegren {et~al.}(2018)Lindegren, Hernández, Bombrun, Klioner,
  Bastian, Ramos-Lerate, de~Torres, Steidelmüller, Stephenson, Hobbs, Lammers,
  Biermann, Geyer, Hilger, Michalik, Stampa, McMillan, Castañeda, Clotet,
  Comoretto, Davidson, Fabricius, Gracia, Hambly, Hutton, Mora, Portell, van
  Leeuwen, Abbas, Abreu, Altmann, Andrei, Anglada, Balaguer-Núñez, Barache,
  Becciani, Bertone, Bianchi, Bouquillon, Bourda, Brüsemeister, Bucciarelli,
  Busonero, Buzzi, Cancelliere, Carlucci, Charlot, Cheek, Crosta, Crowley,
  de~Bruijne, de~Felice, Drimmel, Esquej, Fienga, Fraile, Gai, Garralda,
  González-Vidal, Guerra, Hauser, Hofmann, Holl, Jordan, Lattanzi, Lenhardt,
  Liao, Licata, Lister, Löffler, Marchant, Martin-Fleitas, Messineo, Mignard,
  Morbidelli, Poggio, Riva, Rowell, Salguero, Sarasso, Sciacca, Siddiqui,
  Smart, Spagna, Steele, Taris, Torra, van Elteren, van Reeven, \&
  Vecchiato}]{lindegren_gaia_2018}
Lindegren, L., Hernández, J., Bombrun, A., {et~al.} 2018, Astronomy and
  Astrophysics, 616, A2

\bibitem[{Liu(2004)}]{liu_substructure_2004}
Liu, M.~C. 2004, Science, 305, 1442

\bibitem[{Liu {et~al.}(2010)Liu, Wahhaj, Biller, Nielsen, Chun, Close, Ftaclas,
  Hartung, Hayward, Clarke, Reid, Shkolnik, Tecza, Thatte, Alencar, Artymowicz,
  Boss, Burrows, Pino, Gregorio-Hetem, Ida, Kuchner, Lin, \&
  Toomey}]{liu_gemini_2010}
Liu, M.~C., Wahhaj, Z., Biller, B.~A., {et~al.} 2010, in Adaptive {Optics}
  {Systems} {II}, Vol. 7736 (International Society for Optics and Photonics),
  77361K

\bibitem[{Maire {et~al.}(2016)Maire, Langlois, Dohlen, Lagrange, Gratton,
  Chauvin, Desidera, Girard, Milli, Vigan, Zins, Delorme, Beuzit, Claudi,
  Feldt, Mouillet, Puget, Turatto, \& Wildi}]{maire_sphere_2016}
Maire, A.-L., Langlois, M., Dohlen, K., {et~al.} 2016, Ground-based and
  Airborne Instrumentation for Astronomy VI, 9908, 990834

\bibitem[{Marley {et~al.}(2007)Marley, Fortney, Hubickyj, Bodenheimer, \&
  Lissauer}]{marley_luminosity_2007}
Marley, M.~S., Fortney, J.~J., Hubickyj, O., Bodenheimer, P., \& Lissauer,
  J.~J. 2007, The Astrophysical Journal, 655, 541

\bibitem[{Marois {et~al.}(2006)Marois, Lafrenière, Doyon, Macintosh, \&
  Nadeau}]{marois_angular_2006}
Marois, C., Lafrenière, D., Doyon, R., Macintosh, B., \& Nadeau, D. 2006, The
  Astrophysical Journal, 641, 556, 00000

\bibitem[{Martí \& Beaugé(2012)}]{marti_stellar_2012}
Martí, J.~G. \& Beaugé, C. 2012, Astronomy and Astrophysics, 544, A97

\bibitem[{Mawet {et~al.}(2014)Mawet, Milli, Wahhaj, Pelat, Absil, Delacroix,
  Boccaletti, Kasper, Kenworthy, Marois, Mennesson, \&
  Pueyo}]{mawet_fundamental_2014}
Mawet, D., Milli, J., Wahhaj, Z., {et~al.} 2014, The Astrophysical Journal,
  792, 97

\bibitem[{McAlister {et~al.}(1990)McAlister, Hartkopf, \&
  Franz}]{mcalister_iccd_1990}
McAlister, H., Hartkopf, W.~I., \& Franz, O.~G. 1990, The Astronomical Journal,
  99, 965

\bibitem[{Mckinney(2011)}]{mckinney_pandas:_2011}
Mckinney, W. 2011, Python High Performance Science Computer

\bibitem[{Mesa {et~al.}(2015)Mesa, Gratton, Zurlo, Vigan, Claudi, Alberi,
  Antichi, Baruffolo, Beuzit, Boccaletti, Bonnefoy, Costille, Desidera, Dohlen,
  Fantinel, Feldt, Fusco, Giro, Henning, Kasper, Langlois, Maire, Martinez,
  Moeller-Nilsson, Mouillet, Moutou, Pavlov, Puget, Salasnich, Sauvage, Sissa,
  Turatto, Udry, Vakili, Waters, \& Wildi}]{mesa_performance_2015}
Mesa, D., Gratton, R., Zurlo, A., {et~al.} 2015, Astronomy and Astrophysics,
  576, A121

\bibitem[{Mizuno(1980)}]{mizuno_formation_1980}
Mizuno, H. 1980, Progress of Theoretical Physics, 64, 544

\bibitem[{Mordasini(2018)}]{mordasini_planetary_2018}
Mordasini, C. 2018, Handbook of Exoplanets, 143

\bibitem[{Mugrauer {et~al.}(2007)Mugrauer, Neuhäuser, \&
  Mazeh}]{mugrauer_multiplicity_2007}
Mugrauer, M., Neuhäuser, R., \& Mazeh, T. 2007, Astronomy and Astrophysics,
  469, 755

\bibitem[{Ngo {et~al.}(2016)Ngo, Knutson, Hinkley, Bryan, Crepp, Batygin,
  Crossfield, Hansen, Howard, Johnson, Mawet, Morton, Muirhead, \&
  Wang}]{ngo_friends_2016}
Ngo, H., Knutson, H.~A., Hinkley, S., {et~al.} 2016, The Astrophysical Journal,
  827, 8, 00002

\bibitem[{Nielsen {et~al.}(2019)Nielsen, De~Rosa, Macintosh, Wang, Ruffio,
  Chiang, Marley, Saumon, Savransky, Ammons, Bailey, Barman, Blain, Bulger,
  Burrows, Chilcote, Cotten, Czekala, Doyon, Duchêne, Esposito, Fabrycky,
  Fitzgerald, Follette, Fortney, Gerard, Goodsell, Graham, Greenbaum, Hibon,
  Hinkley, Hirsch, Hom, Hung, Dawson, Ingraham, Kalas, Konopacky, Larkin, Lee,
  Lin, Maire, Marchis, Marois, Metchev, Millar-Blanchaer, Morzinski,
  Oppenheimer, Palmer, Patience, Perrin, Poyneer, Pueyo, Rafikov, Rajan,
  Rameau, Rantakyrö, Ren, Schneider, Sivaramakrishnan, Song, Soummer, Tallis,
  Thomas, Ward-Duong, \& Wolff}]{nielsen_gemini_2019}
Nielsen, E.~L., De~Rosa, R.~J., Macintosh, B., {et~al.} 2019, The Astronomical
  Journal, 158, 13

\bibitem[{Pavlov {et~al.}(2008)Pavlov, Möller-Nilsson, Feldt, Henning, Beuzit,
  \& Mouillet}]{pavlov_sphere_2008}
Pavlov, A., Möller-Nilsson, O., Feldt, M., {et~al.} 2008, Advanced Software
  and Control for Astronomy II, 7019, 701939

\bibitem[{Pecaut {et~al.}(2012)Pecaut, Mamajek, \& Bubar}]{pecaut_revised_2012}
Pecaut, M.~J., Mamajek, E.~E., \& Bubar, E.~J. 2012, The Astrophysical Journal,
  746, 154

\bibitem[{Perryman {et~al.}(1997)Perryman, Lindegren, Kovalevsky, Hog, Bastian,
  Bernacca, Creze, Donati, Grenon, Grewing, van Leeuwen, van~der Marel,
  Mignard, Murray, Le~Poole, Schrijver, Turon, Arenou, Froeschle, \&
  Petersen}]{perryman_hipparcos_1997}
Perryman, M. a.~C., Lindegren, L., Kovalevsky, J., {et~al.} 1997, Astronomy and
  Astrophysics, 500, 501

\bibitem[{Pollack {et~al.}(1996)Pollack, Hubickyj, Bodenheimer, Lissauer,
  Podolak, \& Greenzweig}]{pollack_formation_1996}
Pollack, J.~B., Hubickyj, O., Bodenheimer, P., {et~al.} 1996, Icarus, 124, 62

\bibitem[{Rafikov(2013)}]{rafikov_planet_2013}
Rafikov, R.~R. 2013, The Astrophysical Journal Letters, 765, L8, 00000

\bibitem[{Raghavan {et~al.}(2010)Raghavan, McAlister, Henry, Latham, Marcy,
  Mason, Gies, White, \& ten Brummelaar}]{raghavan_survey_2010}
Raghavan, D., McAlister, H.~A., Henry, T.~J., {et~al.} 2010, The Astrophysical
  Journal Supplement Series, 190, 1

\bibitem[{Rickman {et~al.}(2020)Rickman, Ségransan, Hagelberg, Beuzit,
  Cheetham, Delisle, Forveille, \& Udry}]{rickman_spectral_2020}
Rickman, E.~L., Ségransan, D., Hagelberg, J., {et~al.} 2020, Astronomy and
  Astrophysics, 635, A203

\bibitem[{Santerne {et~al.}(2014)Santerne, Hébrard, Deleuil, Havel, Correia,
  Almenara, Alonso, Arnold, Barros, Behrend, Bernasconi, Boisse, Bonomo,
  Bouchy, Bruno, Damiani, Díaz, Gravallon, Guillot, Labrevoir, Montagnier,
  Moutou, Rinner, Santos, Abe, Audejean, Bendjoya, Gillier, Gregorio, Martinez,
  Michelet, Montaigut, Poncy, Rivet, Rousseau, Roy, Suarez, Vanhuysse, \&
  Verilhac}]{santerne_sophie_2014}
Santerne, A., Hébrard, G., Deleuil, M., {et~al.} 2014, Astronomy and
  Astrophysics, 571, A37

\bibitem[{Schneider \& Silverstone(2003)}]{schneider_high_2003}
Schneider, G. \& Silverstone, M. 2003, in , 69, 00000

\bibitem[{Schwarz {et~al.}(2016)Schwarz, Funk, Zechner, \&
  Bazsó}]{schwarz_new_2016}
Schwarz, R., Funk, B., Zechner, R., \& Bazsó, A. 2016, Monthly Notices of the
  Royal Astronomical Society, 460, 3598, 00003

\bibitem[{Shkolnik {et~al.}(2017)Shkolnik, Allers, Kraus, Liu, \&
  Flagg}]{shkolnik_all-sky_2017}
Shkolnik, E.~L., Allers, K.~N., Kraus, A.~L., Liu, M.~C., \& Flagg, L. 2017,
  The Astronomical Journal, 154, 69

\bibitem[{Sirbu {et~al.}(2017)Sirbu, Thomas, Belikov, \&
  Bendek}]{sirbu_techniques_2017}
Sirbu, D., Thomas, S., Belikov, R., \& Bendek, E. 2017, The Astrophysical
  Journal, 849, 142

\bibitem[{Skrutskie {et~al.}(2006)Skrutskie, Cutri, Stiening, Weinberg,
  Schneider, Carpenter, Beichman, Capps, Chester, Elias, Huchra, Liebert,
  Lonsdale, Monet, Price, Seitzer, Jarrett, Kirkpatrick, Gizis, Howard, Evans,
  Fowler, Fullmer, Hurt, Light, Kopan, Marsh, McCallon, Tam, Van~Dyk, \&
  Wheelock}]{skrutskie_two_2006}
Skrutskie, M.~F., Cutri, R.~M., Stiening, R., {et~al.} 2006, The Astronomical
  Journal, 131, 1163

\bibitem[{Song {et~al.}(2012)Song, Zuckerman, \& Bessell}]{song_new_2012}
Song, I., Zuckerman, B., \& Bessell, M.~S. 2012, The Astronomical Journal, 144,
  8

\bibitem[{Soummer {et~al.}(2012)Soummer, Pueyo, \&
  Larkin}]{soummer_detection_2012}
Soummer, R., Pueyo, L., \& Larkin, J. 2012, The Astrophysical Journal Letters,
  755, L28

\bibitem[{Sparks \& Ford(2002)}]{sparks_imaging_2002}
Sparks, W.~B. \& Ford, H.~C. 2002, The Astrophysical Journal, 578, 543

\bibitem[{Stetson(1987)}]{stetson_daophot:_1987}
Stetson, P.~B. 1987, Publications of the Astronomical Society of the Pacific,
  99, 191

\bibitem[{Stolker {et~al.}(2019)Stolker, Bonse, Quanz, Amara, Cugno, Bohn, \&
  Boehle}]{stolker_pynpoint:_2019}
Stolker, T., Bonse, M.~J., Quanz, S.~P., {et~al.} 2019, Astronomy and
  Astrophysics, 621, A59

\bibitem[{Tetzlaff {et~al.}(2011)Tetzlaff, Neuhäuser, \&
  Hohle}]{tetzlaff_catalogue_2011}
Tetzlaff, N., Neuhäuser, R., \& Hohle, M.~M. 2011, Monthly Notices of the
  Royal Astronomical Society, 410, 190

\bibitem[{Thalmann {et~al.}(2014)Thalmann, Desidera, Bonavita, Janson, Usuda,
  Henning, Köhler, Carson, Boccaletti, Bergfors, Brandner, Feldt, Goto, Klahr,
  Marzari, \& Mordasini}]{thalmann_spots:_2014}
Thalmann, C., Desidera, S., Bonavita, M., {et~al.} 2014, Astronomy and
  Astrophysics, 572, A91

\bibitem[{Tokovinin(1997)}]{tokovinin_msc_1997}
Tokovinin, A.~A. 1997, Astronomy and Astrophysics Supplement Series, 124, 75

\bibitem[{Torres {et~al.}(2009)Torres, Loinard, Mioduszewski, \&
  Rodríguez}]{torres_vlba_2009}
Torres, R.~M., Loinard, L., Mioduszewski, A.~J., \& Rodríguez, L.~F. 2009, The
  Astrophysical Journal, 698, 242

\bibitem[{Udry \& Santos(2007)}]{udry_statistical_2007}
Udry, S. \& Santos, N.~C. 2007, Annual Review of Astronomy and Astrophysics,
  45, 397

\bibitem[{Vigan {et~al.}(2020)Vigan, Fontanive, Meyer, Biller, Bonavita, Feldt,
  Desidera, Marleau, Emsenhuber, Galicher, Rice, Forgan, Mordasini, Gratton,
  Le~Coroller, Maire, Cantalloube, Chauvin, Cheetham, Hagelberg, Lagrange,
  Langlois, Bonnefoy, Beuzit, Boccaletti, D'Orazi, Delorme, Dominik, Henning,
  Janson, Lagadec, Lazzoni, Ligi, Menard, Mesa, Messina, Moutou, Müller,
  Perrot, Samland, Schmid, Schmidt, Sissa, Turatto, Udry, Zurlo, Abe, Antichi,
  Asensio-Torres, Baruffolo, Baudoz, Baudrand, Bazzon, Blanchard, Bohn,
  Brown~Sevilla, Carbillet, Carle, Cascone, Charton, Claudi, Costille,
  De~Caprio, Delboulbé, Dohlen, Engler, Fantinel, Feautrier, Fusco, Gigan,
  Girard, Giro, Gisler, Gluck, Gry, Hubin, Hugot, Jaquet, Kasper, Le~Mignant,
  Llored, Madec, Magnard, Martinez, Maurel, Möller-Nilsson, Mouillet, Moulin,
  Origné, Pavlov, Perret, Petit, Pragt, Puget, Rabou, Ramos, Rickman, Rigal,
  Rochat, Roelfsema, Rousset, Roux, Salasnich, Sauvage, Sevin, Soenke, Stadler,
  Suarez, Wahhaj, Weber, \& Wildi}]{vigan_sphere_2020}
Vigan, A., Fontanive, C., Meyer, M., {et~al.} 2020, arXiv e-prints, 2007,
  arXiv:2007.06573

\bibitem[{Vigan {et~al.}(2010)Vigan, Moutou, Langlois, Allard, Boccaletti,
  Carbillet, Mouillet, \& Smith}]{vigan_photometric_2010}
Vigan, A., Moutou, C., Langlois, M., {et~al.} 2010, Monthly Notices of the
  Royal Astronomical Society, 407, 71

\bibitem[{Vigan {et~al.}(2012)Vigan, Patience, Marois, Bonavita, De~Rosa,
  Macintosh, Song, Doyon, Zuckerman, Lafrenière, \&
  Barman}]{vigan_international_2012}
Vigan, A., Patience, J., Marois, C., {et~al.} 2012, Astronomy and Astrophysics,
  544, 9

\bibitem[{Walt {et~al.}(2011)Walt, Colbert, \& Varoquaux}]{walt_numpy_2011}
Walt, S. v.~d., Colbert, S.~C., \& Varoquaux, G. 2011, Computing in Science \&
  Engineering, 13, 22

\bibitem[{Wang {et~al.}(2015)Wang, Fischer, Horch, \&
  Xie}]{wang_influence_2015}
Wang, J., Fischer, D.~A., Horch, E.~P., \& Xie, J.-W. 2015, The Astrophysical
  Journal, 806, 248

\bibitem[{Wang {et~al.}(2014)Wang, Xie, Barclay, \&
  Fischer}]{wang_influence_2014}
Wang, J., Xie, J.-W., Barclay, T., \& Fischer, D.~A. 2014, The Astrophysical
  Journal, 783, 4, 00058

\bibitem[{Wilking {et~al.}(2008)Wilking, Gagne, \& Allen}]{wilking_star_2008}
Wilking, B., Gagne, M., \& Allen, L. 2008, arXiv:0811.0005 [astro-ph], arXiv:
  0811.0005

\bibitem[{Winters {et~al.}(2019)Winters, Henry, Jao, Subasavage, Chatelain,
  Slatten, Riedel, Silverstein, \& Payne}]{winters_solar_2019}
Winters, J.~G., Henry, T.~J., Jao, W.-C., {et~al.} 2019, The Astronomical
  Journal, 157, 216

\bibitem[{Zsom {et~al.}(2011)Zsom, Sándor, \& Dullemond}]{zsom_first_2011}
Zsom, A., Sándor, Z., \& Dullemond, C.~P. 2011, Astronomy and Astrophysics,
  527, A10

\end{thebibliography}

\begin{appendix} %
\section{Additional material}	
   \begin{figure*}
   \resizebox{\hsize}{!}
            {\subfigure[HIP~1910~B]{\includegraphics[width=\hsize]{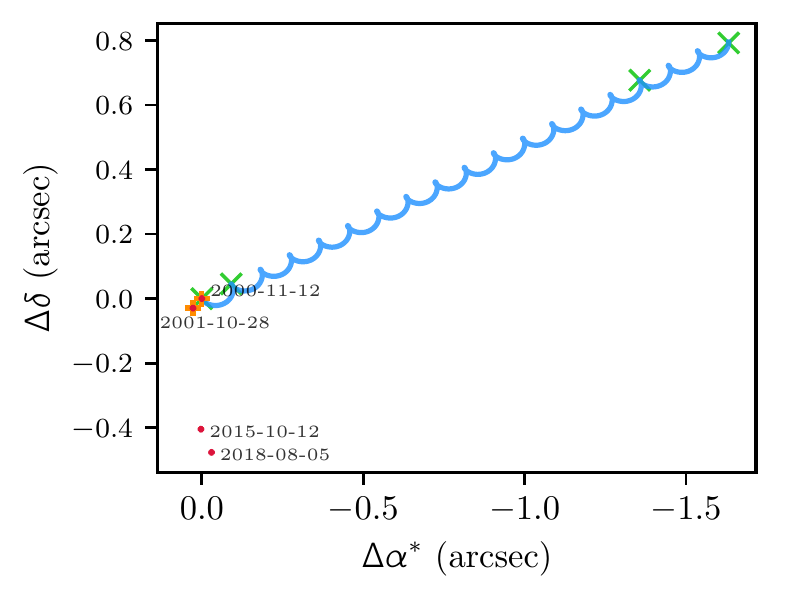}}\qquad%
            \subfigure[HD~22213~B]{\includegraphics[width=\hsize]{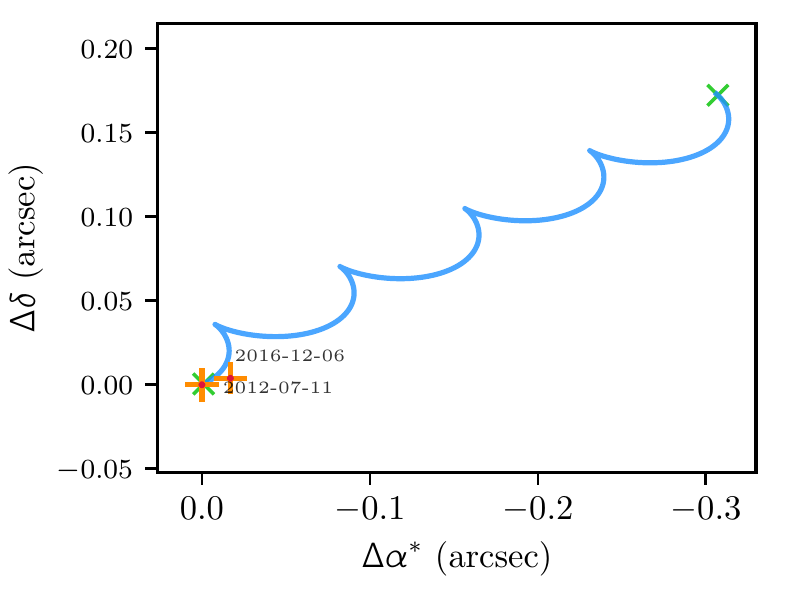}}\qquad%
            \subfigure[HD~285281~B]{\includegraphics[width=\hsize]{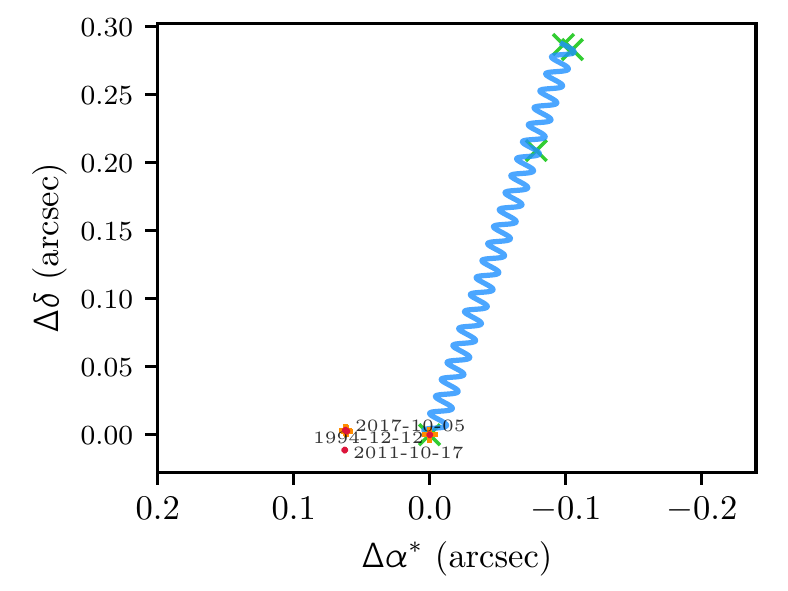}}}\\
   \resizebox{\hsize}{!}
            {\subfigure[TYC~8083-45-5~B]{\includegraphics[width=\hsize]{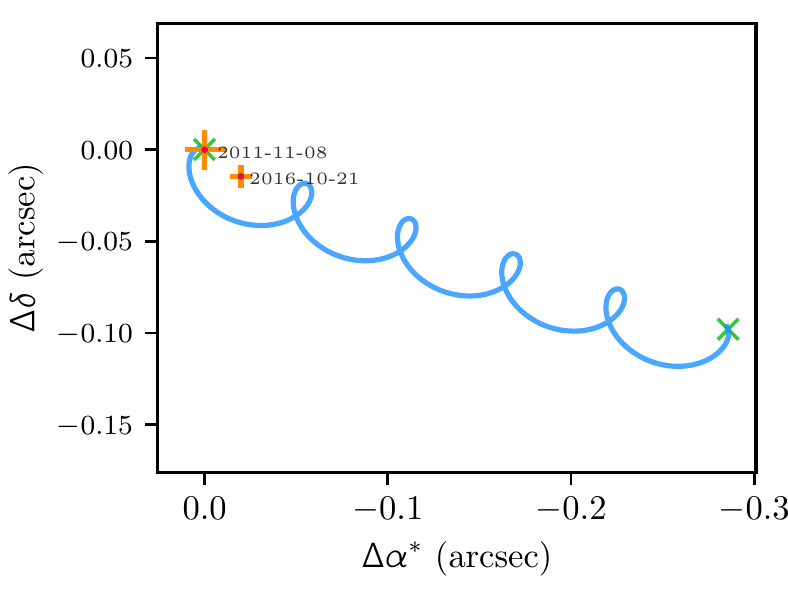}}\qquad%
            \subfigure[HD~102026~B]{\includegraphics[width=\hsize]{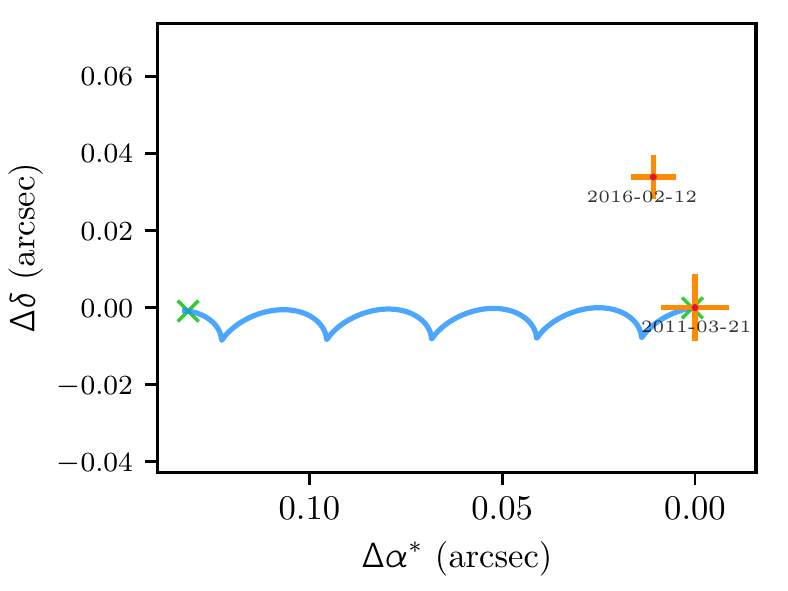}}\qquad%
            \subfigure[HD~104231~B]{\includegraphics[width=\hsize]{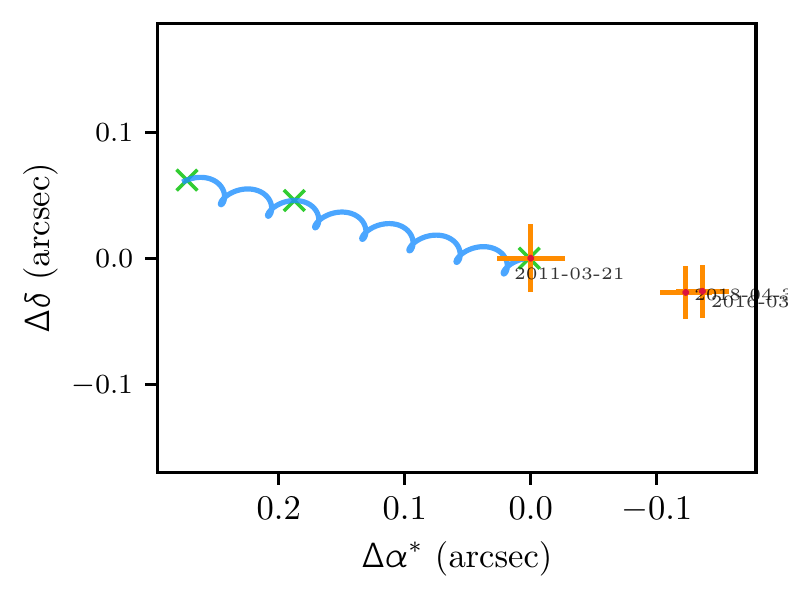}}}\\
   \resizebox{\hsize}{!}
            {\subfigure[HD~104897~B]{\includegraphics[width=\hsize]{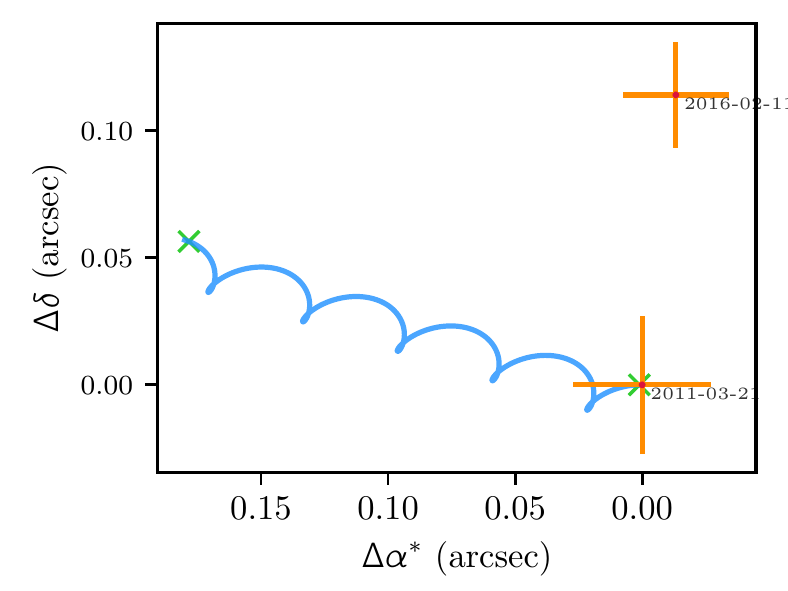}}\qquad%
            \subfigure[HD~108568~B]{\includegraphics[width=\hsize]{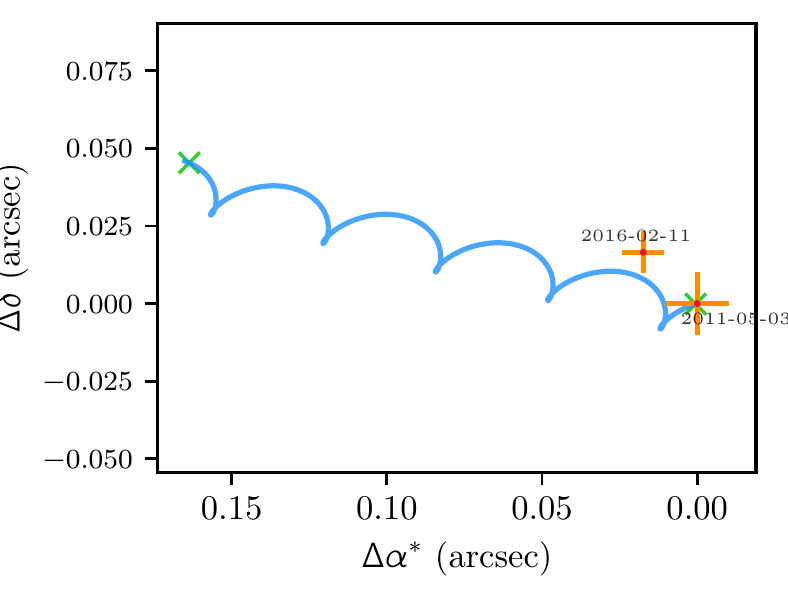}}\qquad%
            \subfigure[HD~112381~B]{\includegraphics[width=\hsize]{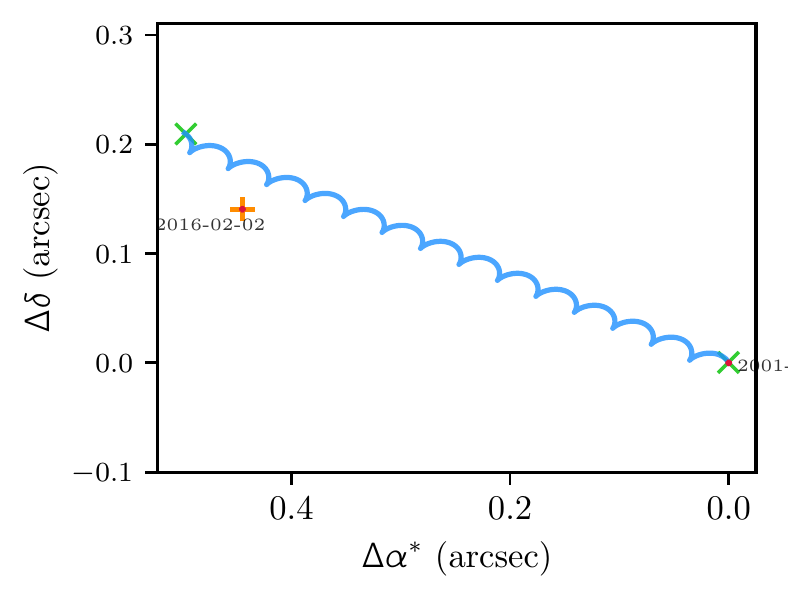}}}\\%
   \resizebox{\hsize}{!}
	{\subfigure[\Huge HD~112381~C]{\includegraphics[width=\hsize]{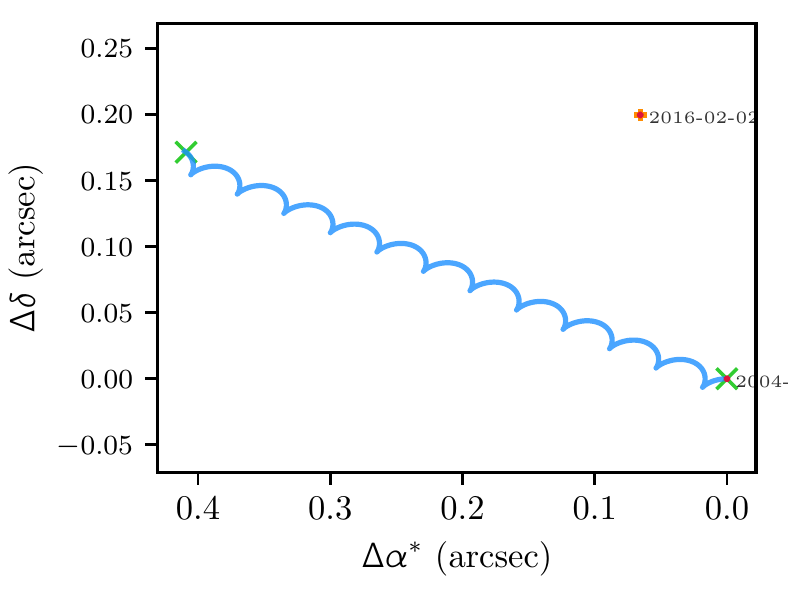}}\qquad%
	\subfigure[HD~120178~B]{\includegraphics[width=\hsize]{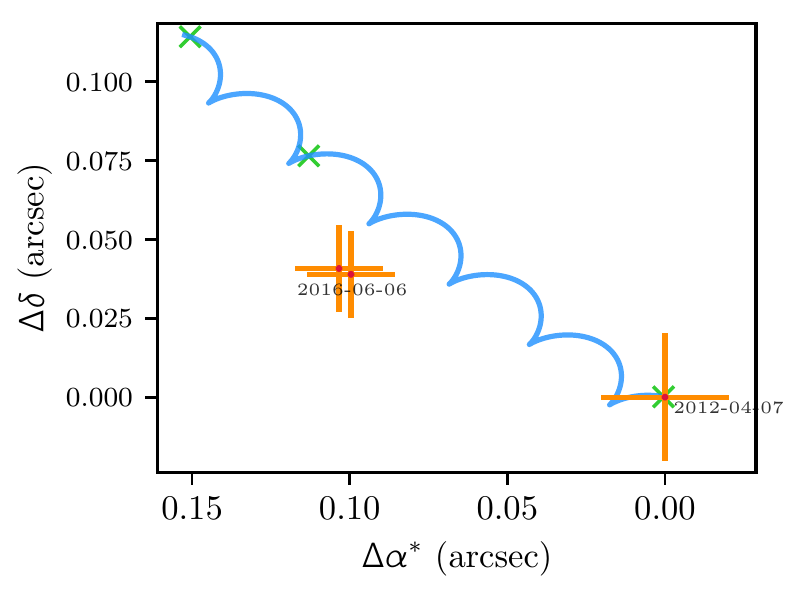}}\qquad%
			\subfigure[HD~121336~B]{\includegraphics[width=\hsize]{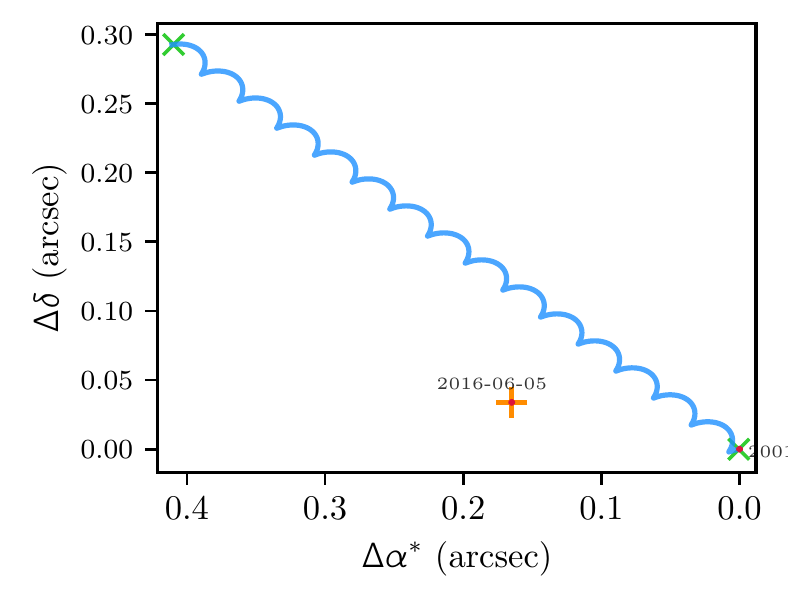}}}\\%
      \caption{Astrometric plots of the binaries. Oranges crosses represent the relative position of the companion where the origin is defined by the first observation. The blue corkscrew line is the track a background object would follow due to the primary star proper motion with the green crosses along that track representing the positions it would have at the dates of observation.}
         \label{fig:astrometry_1}
   \end{figure*}

 \begin{figure*}
 	\resizebox{\hsize}{!}
 	{\subfigure[HD~127215~B]{\includegraphics[width=\hsize]{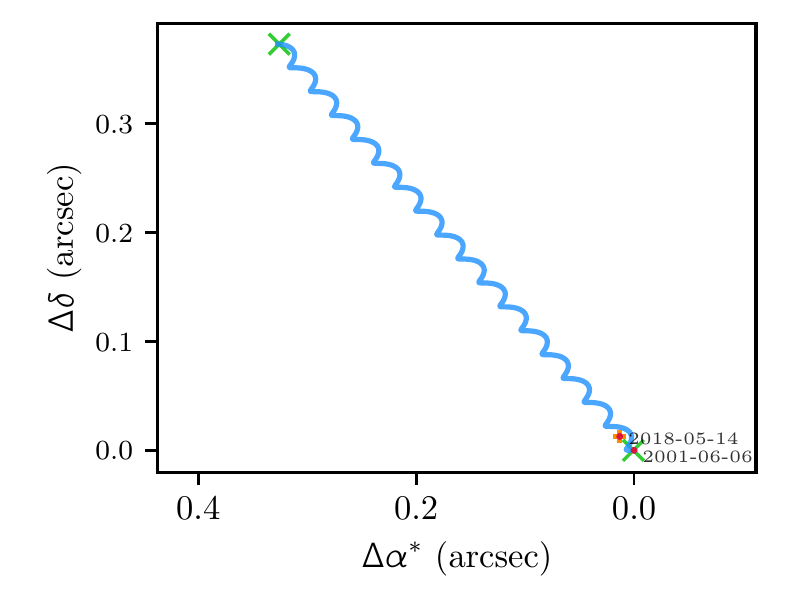}}\qquad
 		\subfigure[HD~128788~B]{\includegraphics[width=\hsize]{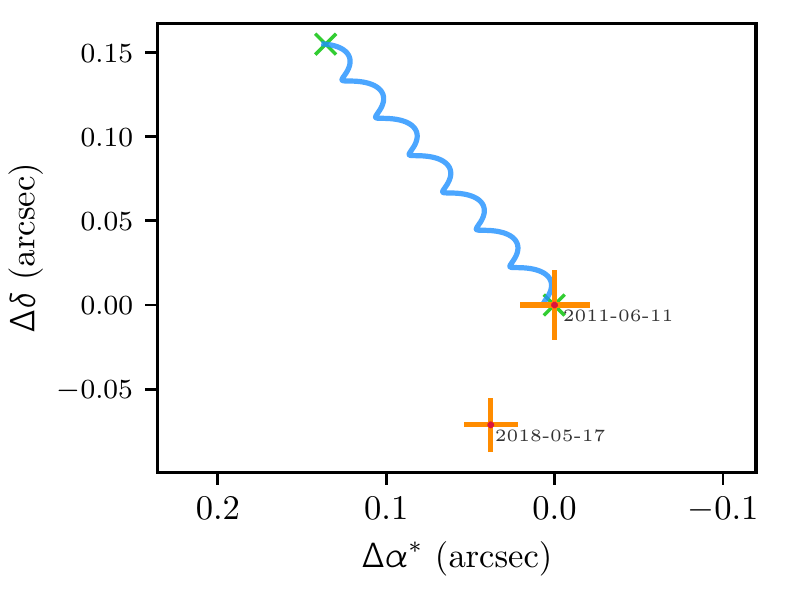}}\qquad%
 		\subfigure[HD~133954~B]{\includegraphics[width=\hsize]{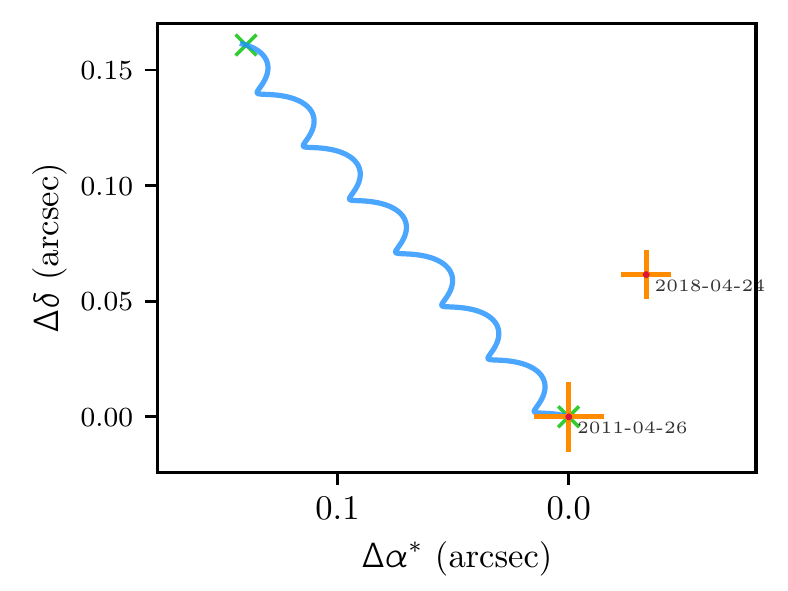}}}\\
	\resizebox{\hsize}{!}
			{\subfigure[HD~138138~B]{\includegraphics[width=\hsize]{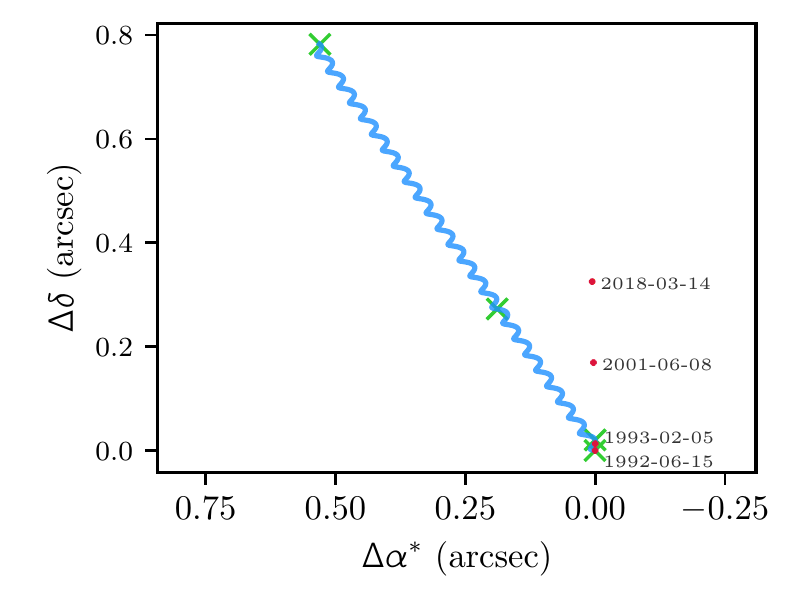}}\qquad%
			\subfigure[HD~138138~C]{\includegraphics[width=\hsize]{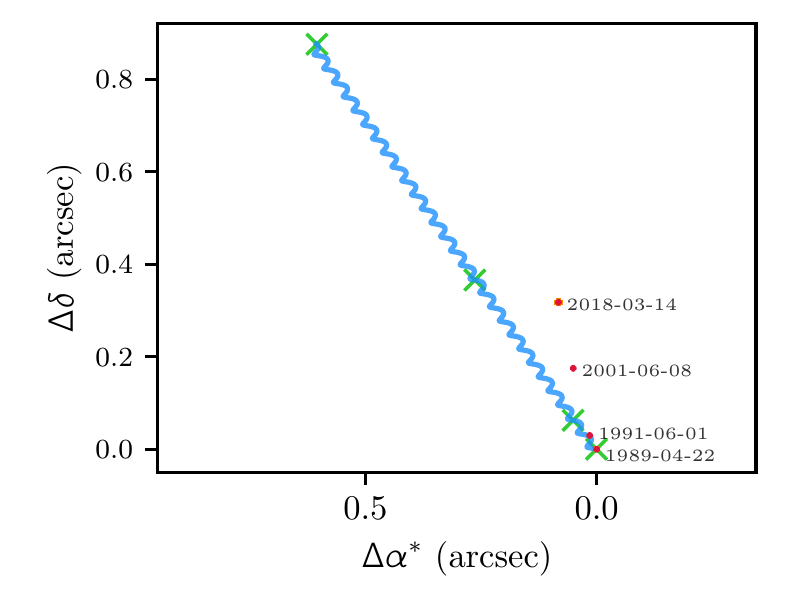}}\qquad%
			\subfigure[HD~144175~B]{\includegraphics[width=\hsize]{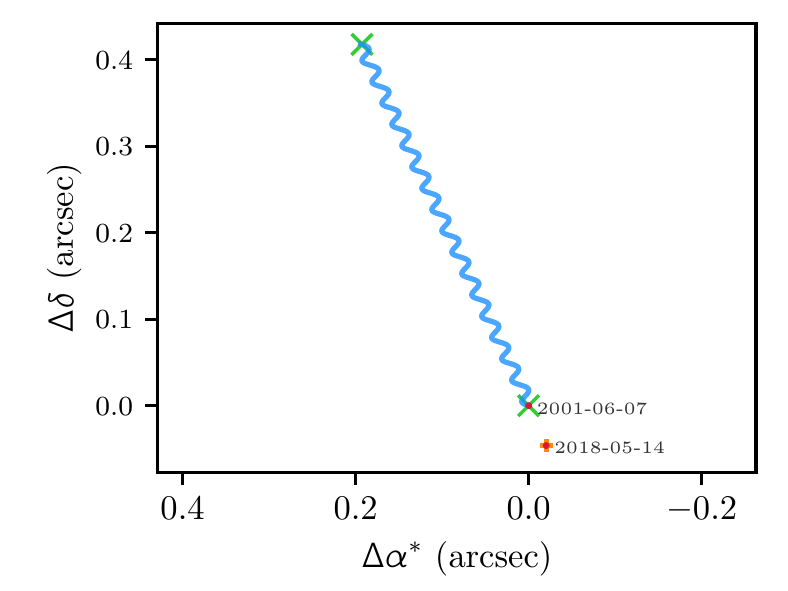}}}\\%
			\resizebox{\hsize}{!}
		{	\subfigure[HD~144823~B]{\includegraphics[width=\hsize]{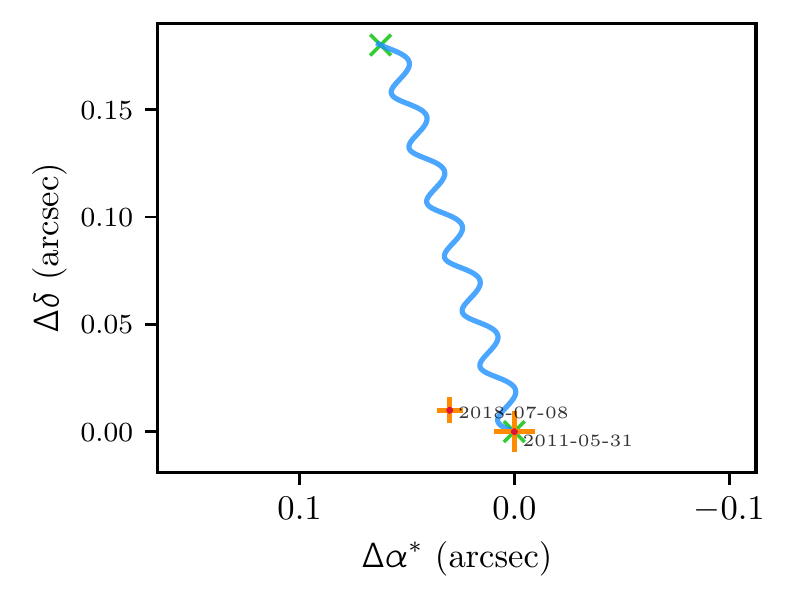}}\qquad%
			\subfigure[HD~145792~B]{\includegraphics[width=\hsize]{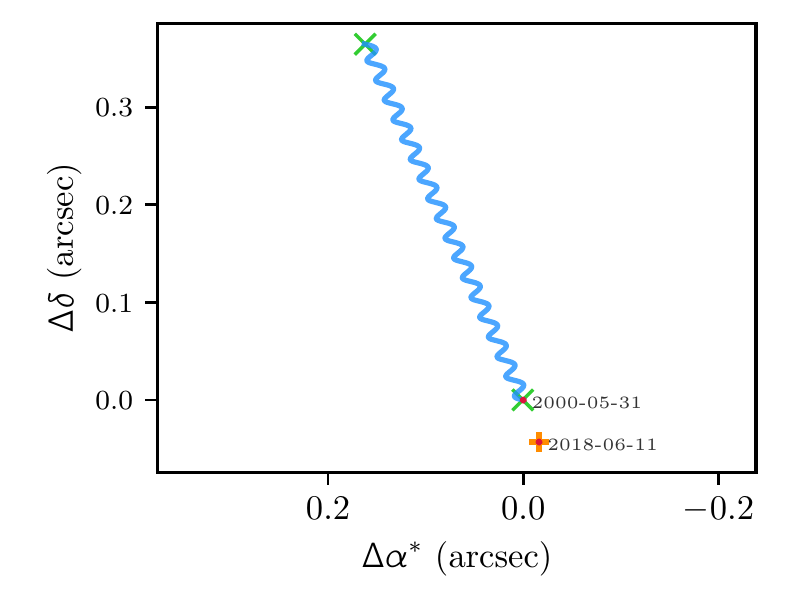}}\qquad%
			\subfigure[HD~146331~B]{\includegraphics[width=\hsize]{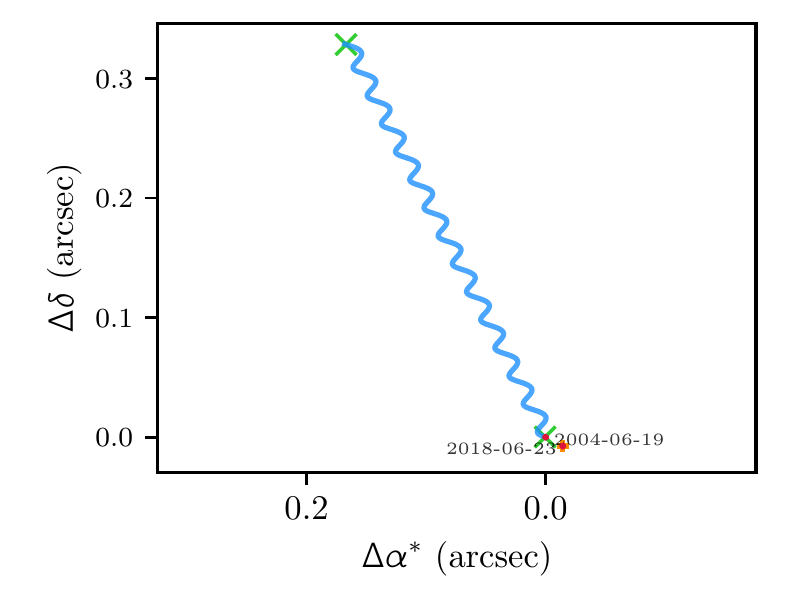}}}\\%
		\resizebox{\hsize}{!}
		{\subfigure[HD~146331~C]{\includegraphics[width=\hsize]{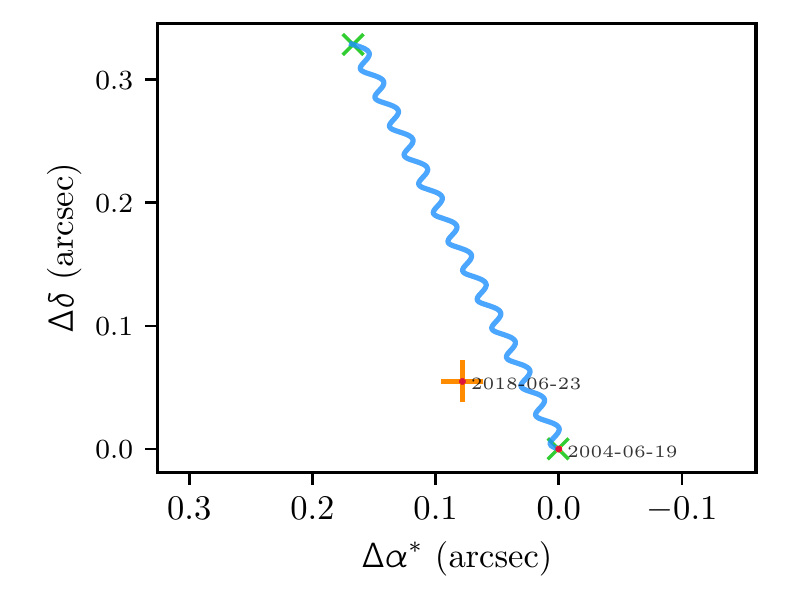}}\qquad%
		\subfigure[HD~147432~B]{\includegraphics[width=\hsize]{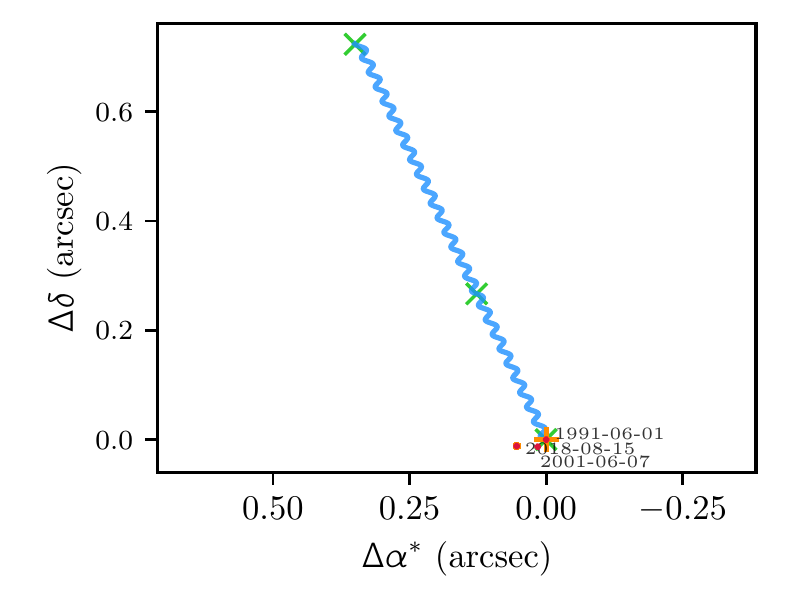}}\qquad%
		\subfigure[HD~148562~B]{\includegraphics[width=\hsize]{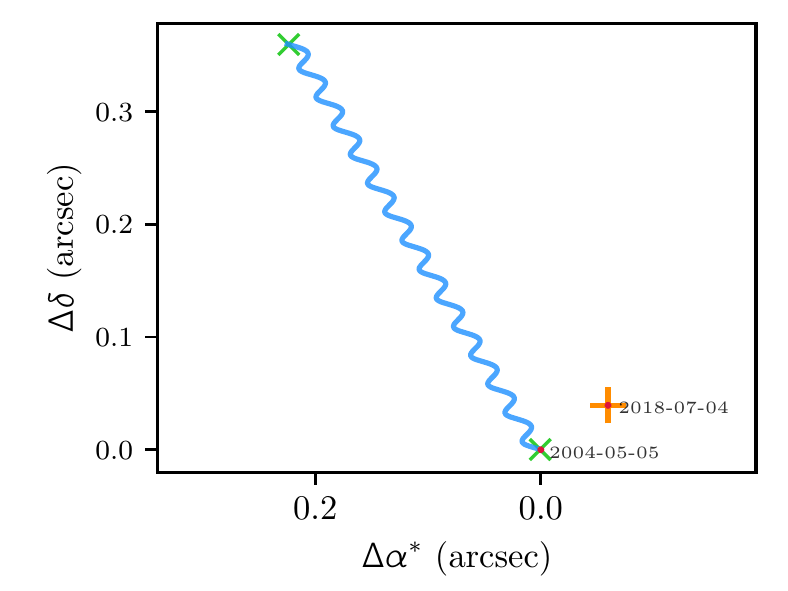}}}\\%
\caption{continued.}
\label{fig:astrometry_2}
\end{figure*}

\begin{figure*}
		\resizebox{\hsize}{!}
		{\subfigure[HD~148716~B]{\includegraphics[width=\hsize]{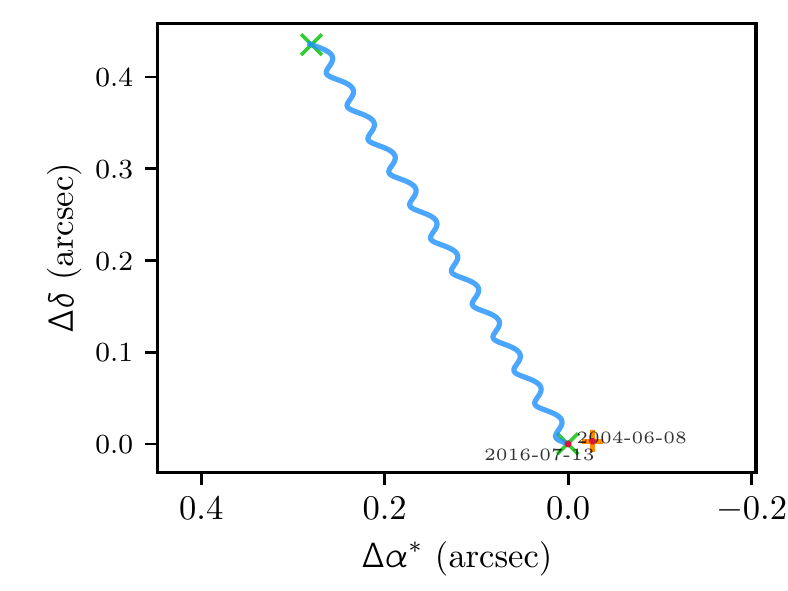}}\qquad%
			\subfigure[HD~165189~B]{\includegraphics[width=\hsize]{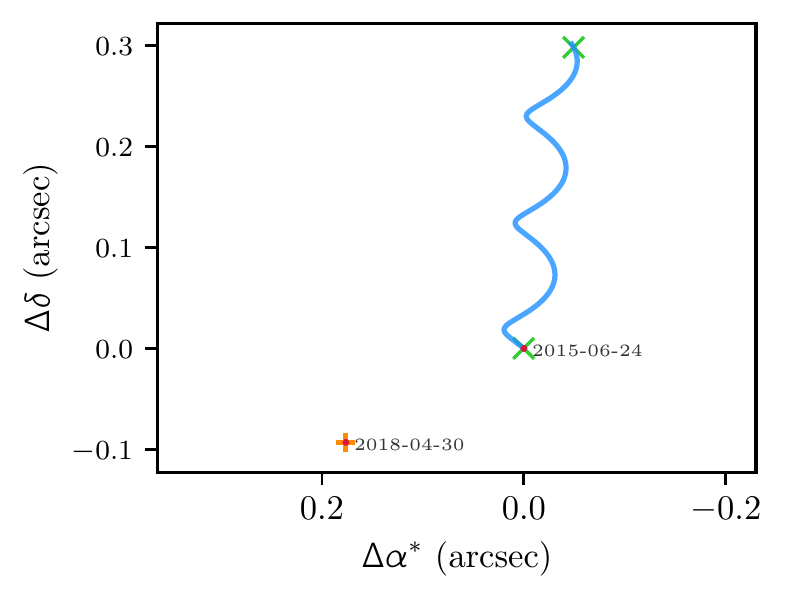}}\qquad%
			\subfigure[HD~208233~B]{\includegraphics[width=\hsize]{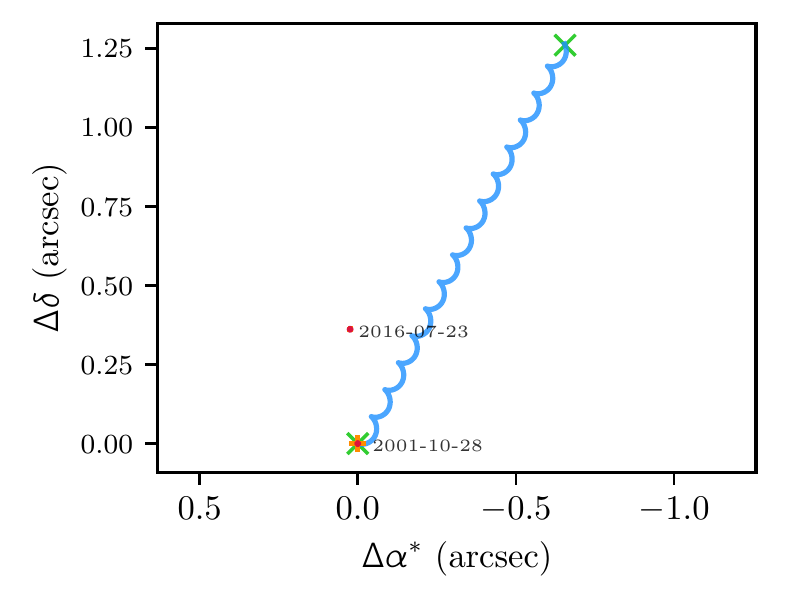}}}\\%
\resizebox{0.33\hsize}{!}
{			\subfigure[HD~217379~B]{\includegraphics[width=\hsize]{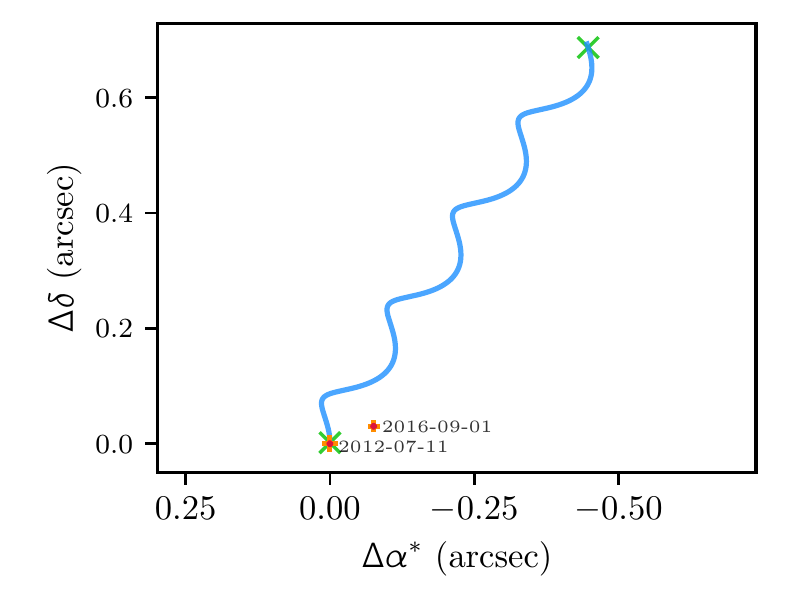}}}\\
	\caption{continued.}
	\label{fig:astrometry_3}
\end{figure*}

   \begin{figure*}
	\resizebox{\hsize}{!}
	{\subfigure[HIP~1910~B]{\includegraphics[width=\hsize]{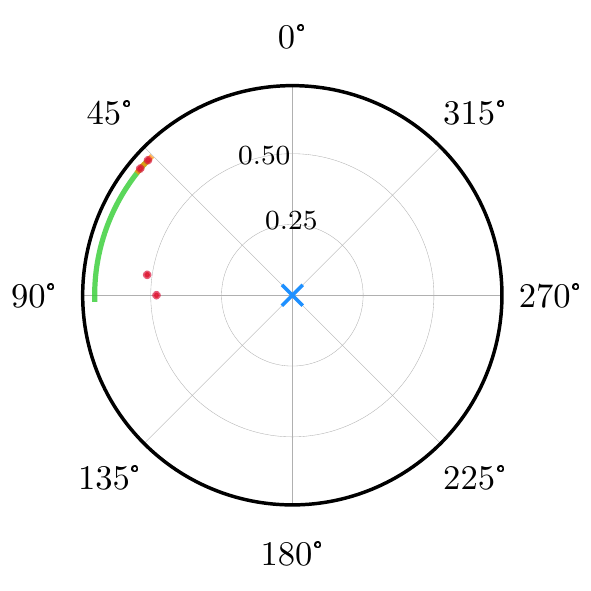}}\qquad%
		\subfigure[HD~22213~B]{\includegraphics[width=\hsize]{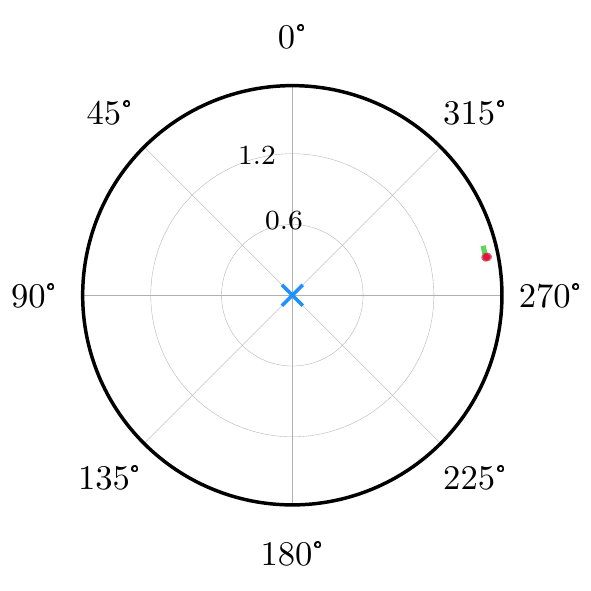}}\qquad%
		\subfigure[HD~285281~B]{\includegraphics[width=\hsize]{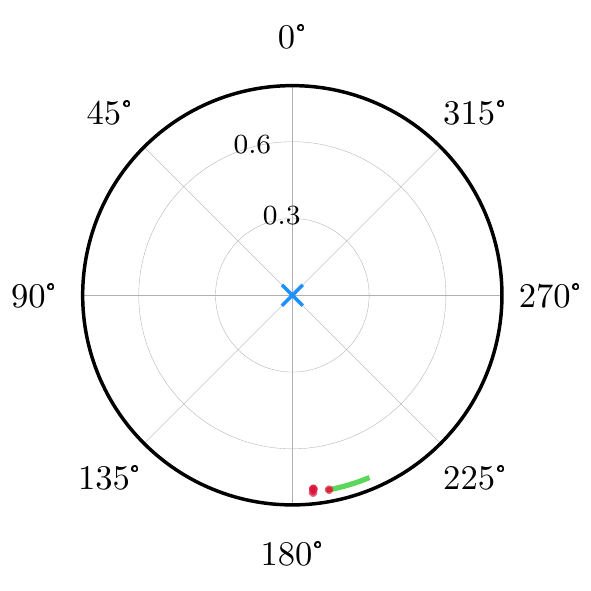}}}\\
	\resizebox{\hsize}{!}
	{\subfigure[TYC~8083-45-5~B]{\includegraphics[width=\hsize]{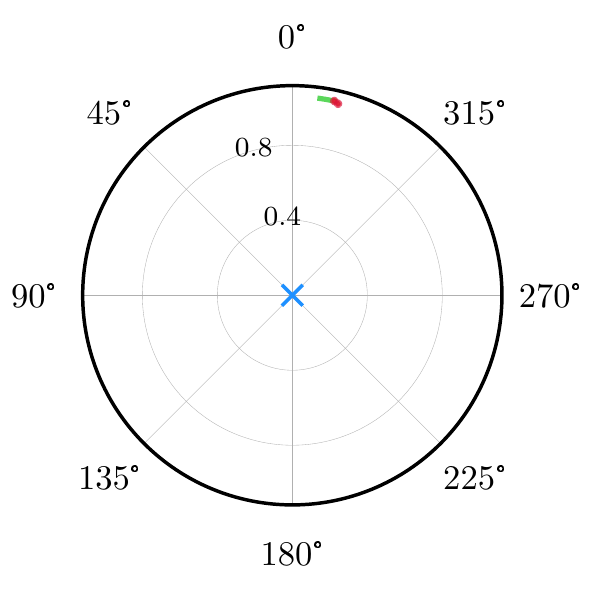}}\qquad%
		\subfigure[HD~102026~B]{\includegraphics[width=\hsize]{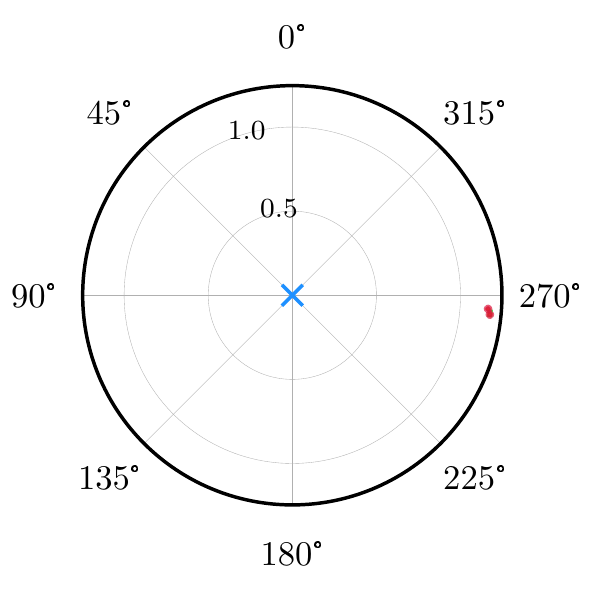}}\qquad%
		\subfigure[HD~104231~B]{\includegraphics[width=\hsize]{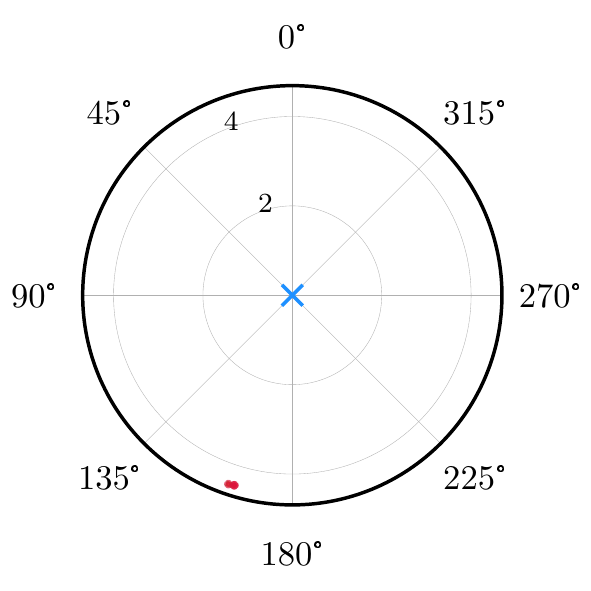}}}\\
	\resizebox{\hsize}{!}
	{\subfigure[HD~104897~B]{\includegraphics[width=\hsize]{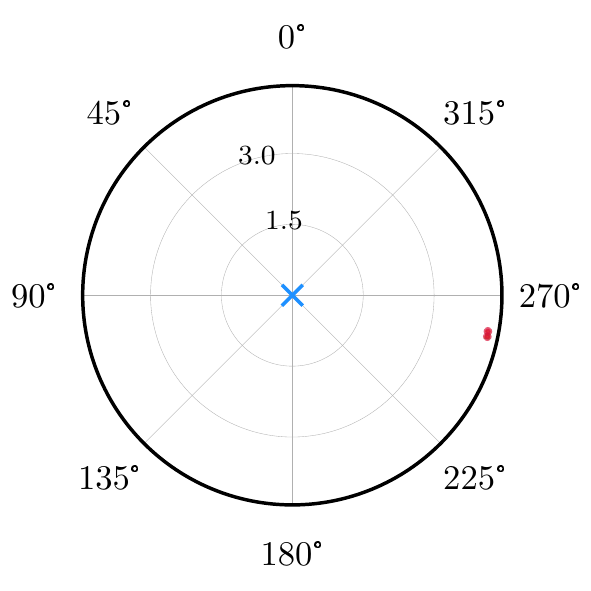}}\qquad%
		\subfigure[HD~108568~B]{\includegraphics[width=\hsize]{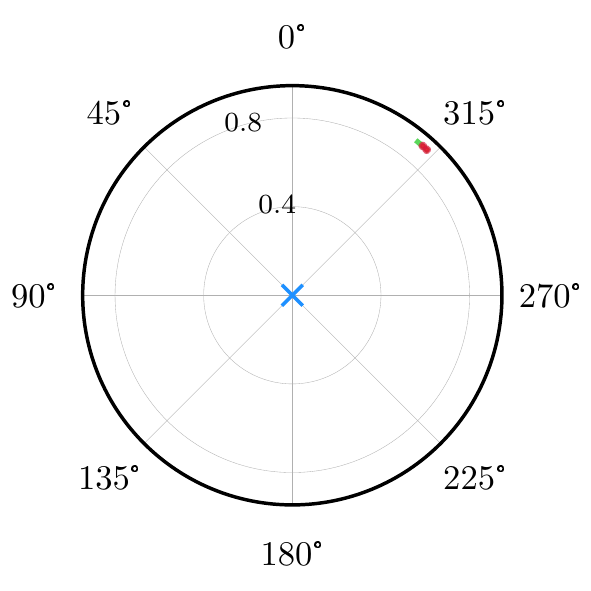}}\qquad%
		\subfigure[HD~112381~B]{\includegraphics[width=\hsize]{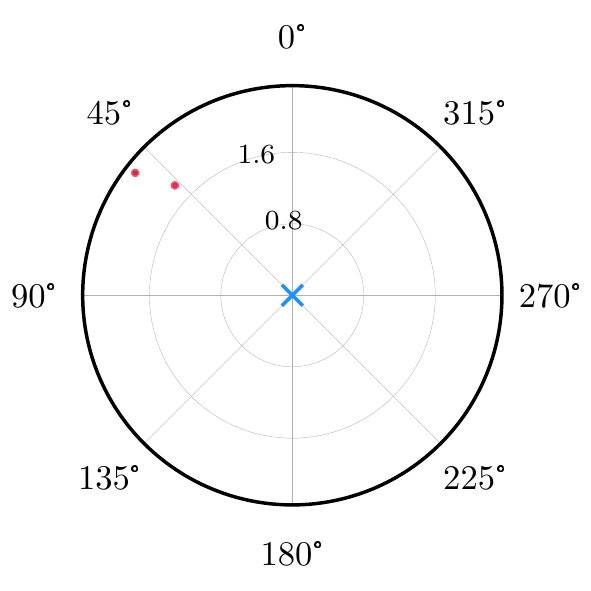}}}\\
	\caption{Position measurements of the companions from this work and literature. The blue cross in the centre represents the primary star position while the red dots show the companion position. The green arc represents the movement a face-on circular orbit would have with the observed time baseline.
		Error bars are too small to be visible in the plots.}
\label{fig:orbit_1}
\end{figure*}

\begin{figure*}
	\resizebox{\hsize}{!}
	{\subfigure[HD~112381~C]{\includegraphics[width=\hsize]{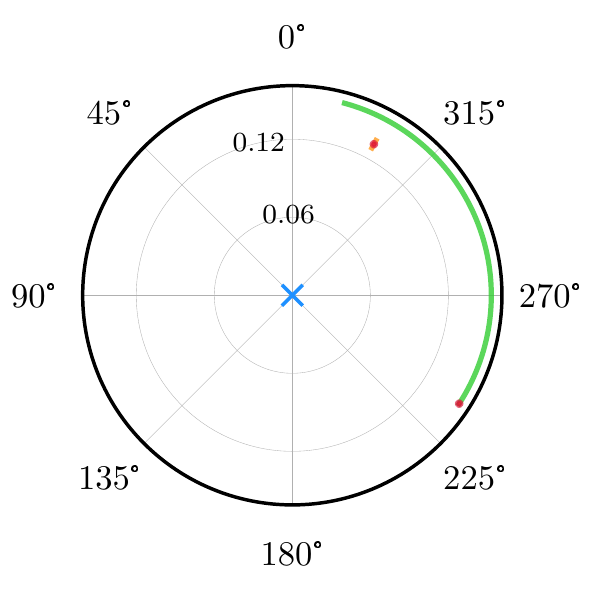}}\qquad %
		\subfigure[HD~120178~B]{\includegraphics[width=\hsize]{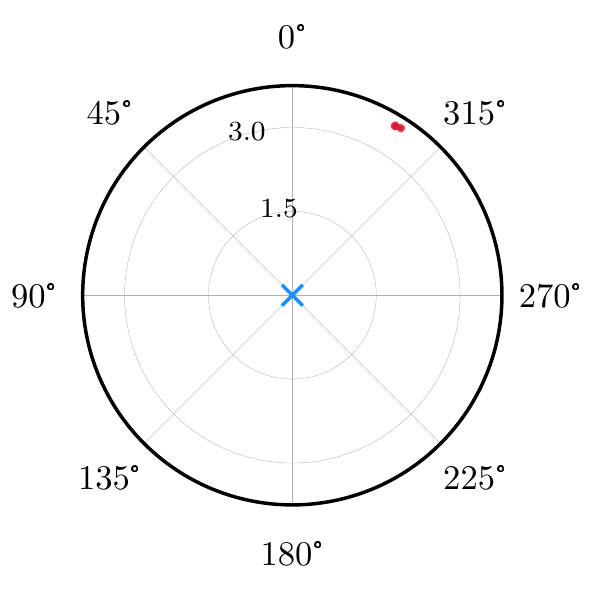}}\qquad%
		\subfigure[HD~121336~B]{\includegraphics[width=\hsize]{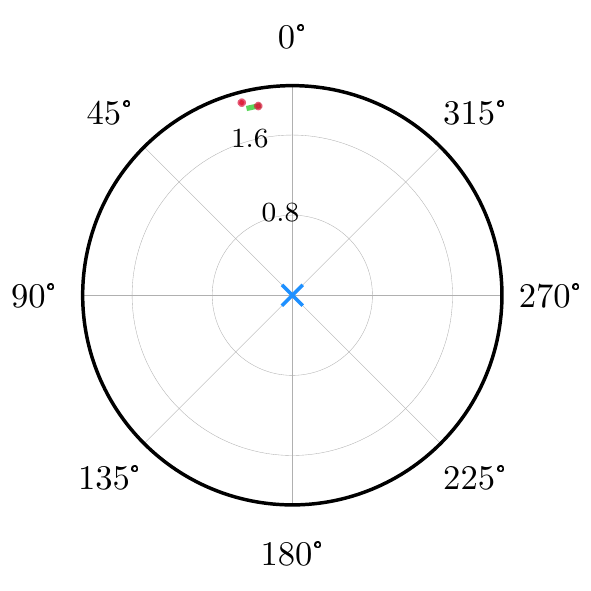}}}\\
	\resizebox{\hsize}{!}
	{\subfigure[HD~127215~B]{\includegraphics[width=\hsize]{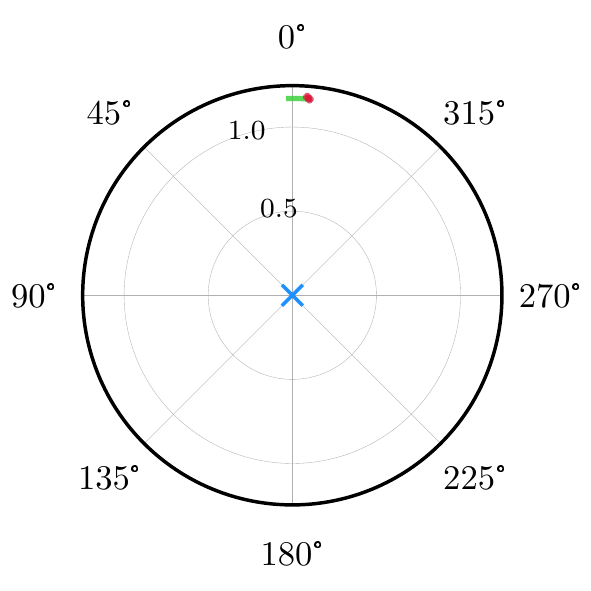}}\qquad%
	\subfigure[HD~128788~B]{\includegraphics[width=\hsize]{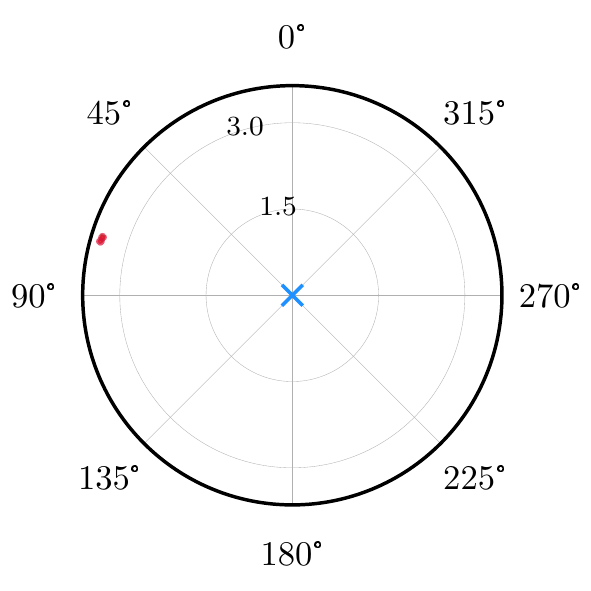}}\qquad%
		\subfigure[HD~133954~B]{\includegraphics[width=\hsize]{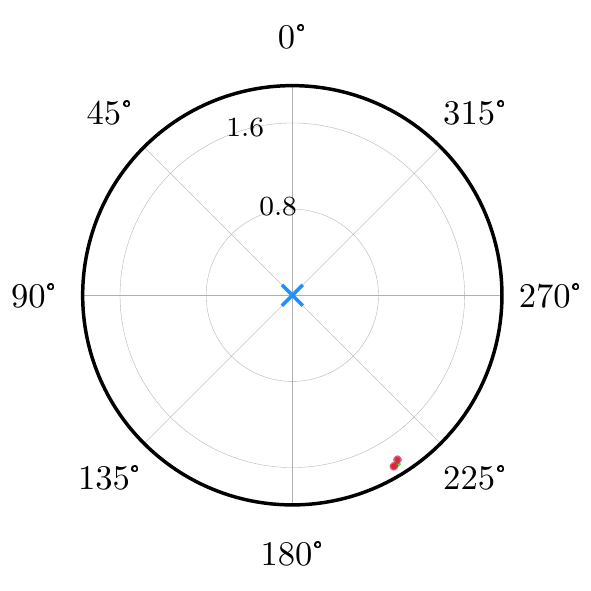}}}\\
	\resizebox{\hsize}{!}
	{\subfigure[HD~138138~B]{\includegraphics[width=\hsize]{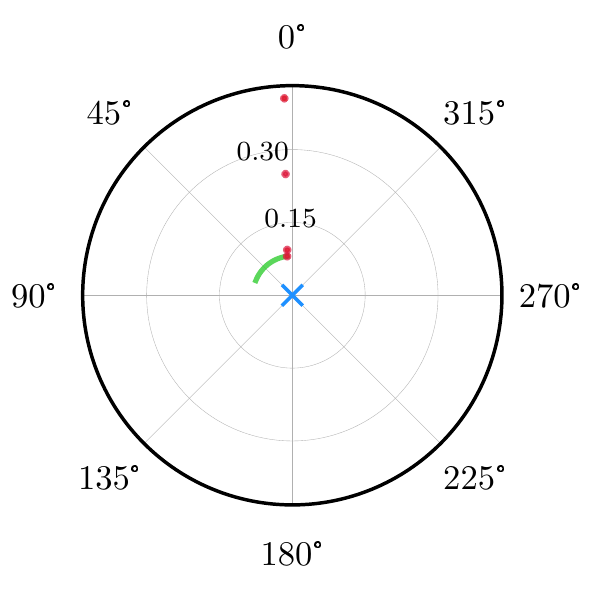}}\qquad%
	\subfigure[HD~138138~C]{\includegraphics[width=\hsize]{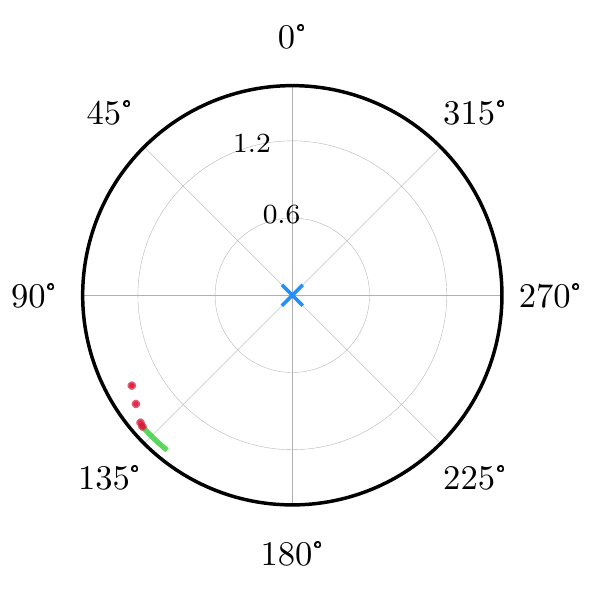}}\qquad%
	\subfigure[HD~144175~B]{\includegraphics[width=\hsize]{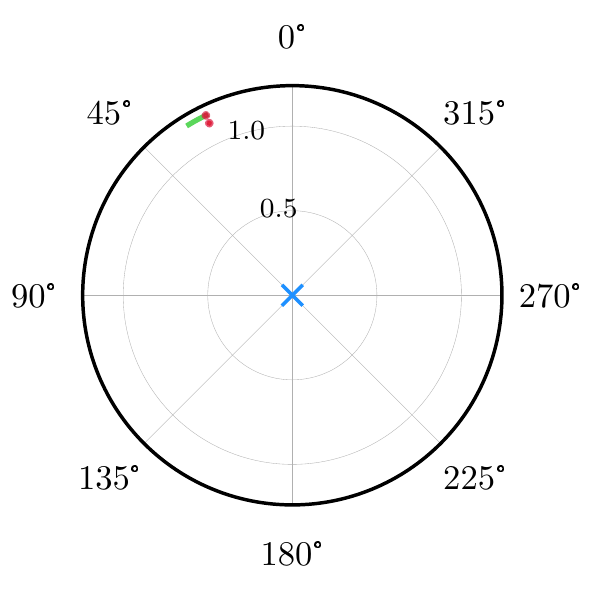}}}\\%
		\caption{continued.}
	\label{fig:orbit_2}
\end{figure*}

\begin{figure*}
\resizebox{\hsize}{!}
{\subfigure[HD~144823~B]{\includegraphics[width=\hsize]{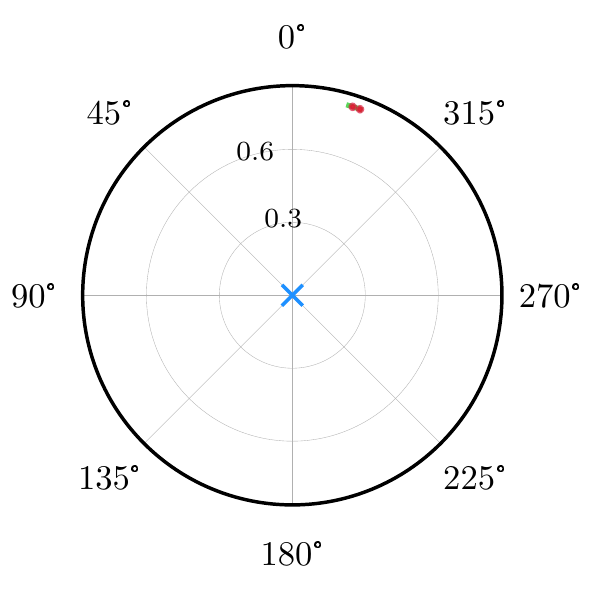}}\qquad%
	\subfigure[HD~145792~B]{\includegraphics[width=\hsize]{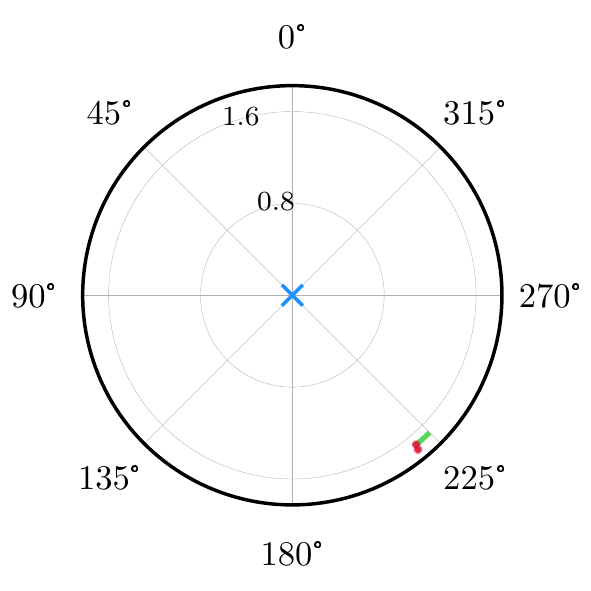}}\qquad%
	\subfigure[HD~146331~B]{\includegraphics[width=\hsize]{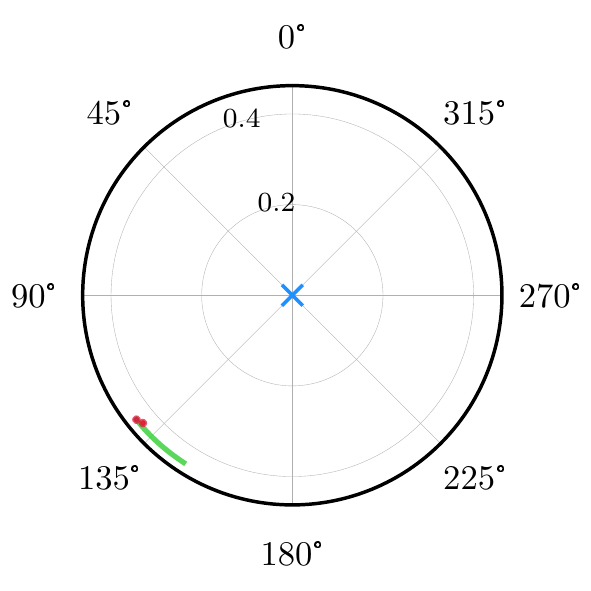}}}\\%
\resizebox{\hsize}{!}
{	\subfigure[HD~146331~C]{\includegraphics[width=\hsize]{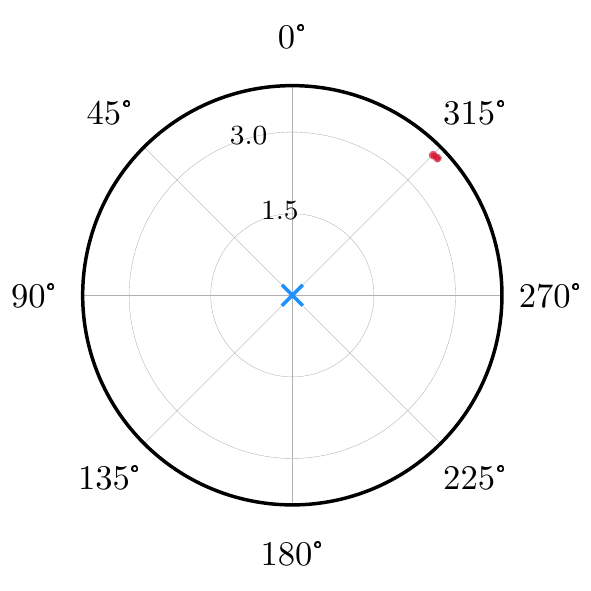}}\qquad%
	\subfigure[HD~147432~B]{\includegraphics[width=\hsize]{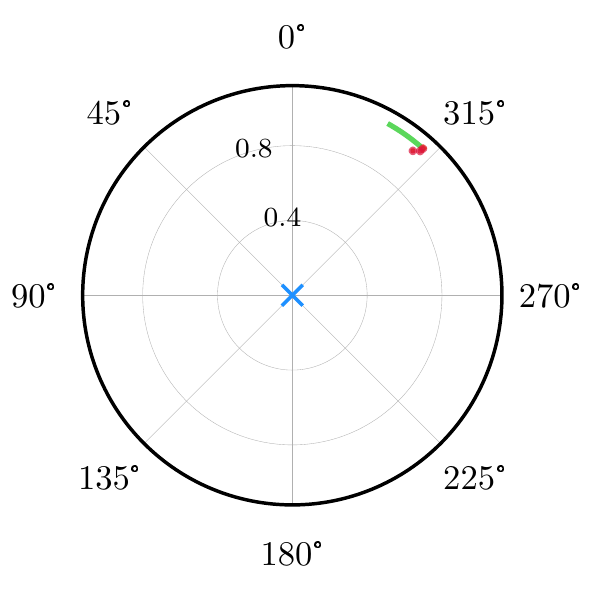}}\qquad%
	\subfigure[HD~148562~B]{\includegraphics[width=\hsize]{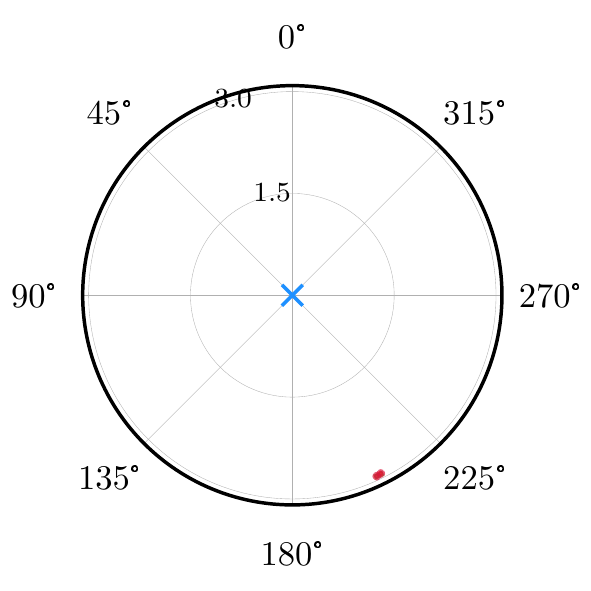}}}\\%
	\resizebox{\hsize}{!}
	{\subfigure[HD~148716~B]{\includegraphics[width=\hsize]{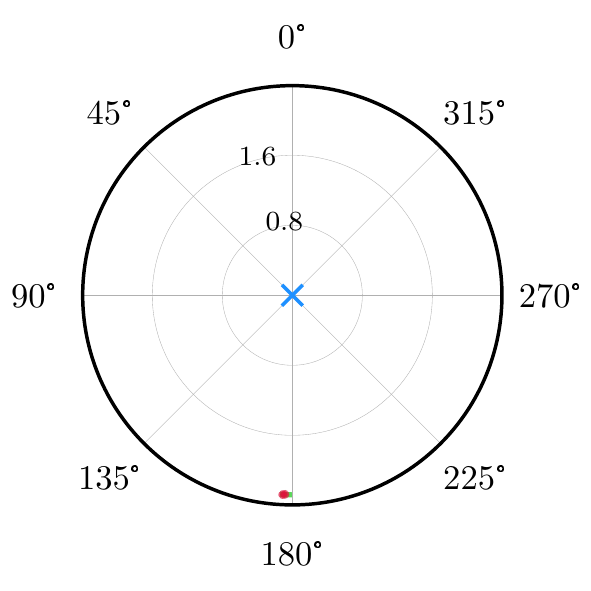}}\qquad%
	\subfigure[HD~165189~B]{\includegraphics[width=\hsize]{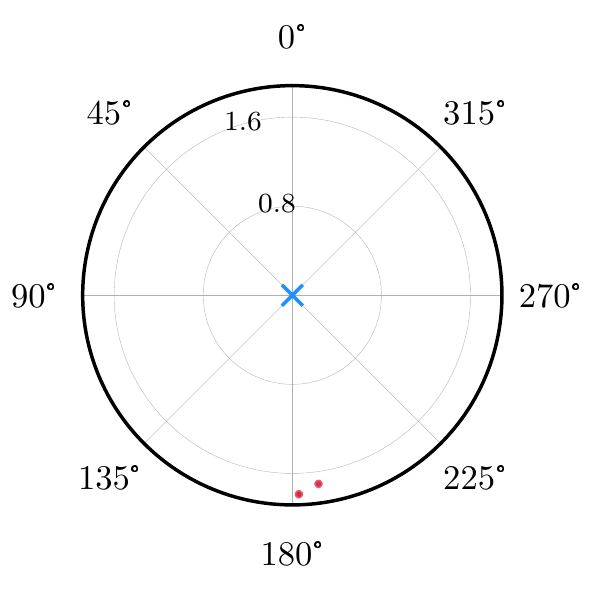}}\qquad%
		\subfigure[HD~208233~B]{\includegraphics[width=\hsize]{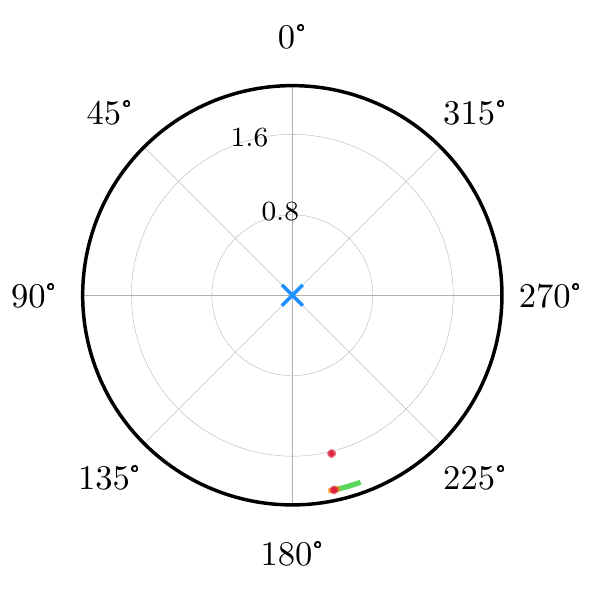}}}\\%
			\resizebox{.333\hsize}{!}
		{
		\subfigure[HD~217379~B]{\includegraphics[width=\hsize]{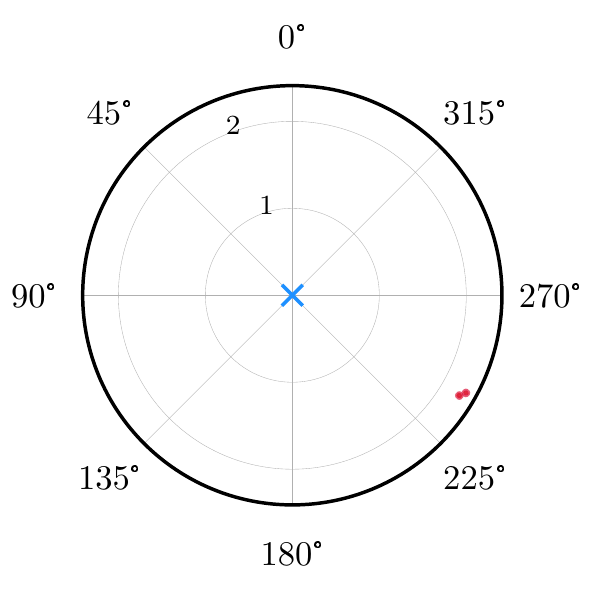}}}\\
	\caption{continued.}
	\label{fig:orbit_3}
\end{figure*}

\begin{table*} 
      \caption[]{Gaia distance, coordinates, and spectral types, and probability of association or moving group (MG) for each target using the Banyan $\Sigma$ tool.\label{tab:MG}}
      \centering
  		\renewcommand{\arraystretch}{1.3}
         \begin{tabular}{lcccccc}
            \hline 
            \hline
            \multirow{ 2}{2cm}{ID} & \multirow{ 2}{2cm}{RA} & \multirow{ 2}{2cm}{DEC} & \multirow{ 2}{2cm}{SpT} & Distance & Banyan $\Sigma$ & Banyan $\Sigma$ Prob.  \\
             & & & & (pc) & MG & (\%) \\
            \hline
            \hline
            \noalign{\smallskip}

HIP1910  & 00:24:08.9 & -62:11:04.3 & M0Ve & $44.23^{+1.09}_{-1.04}$ & THA & 58.0 \\
HD22213  & 03:34:16.3 & -12:04:07.2 & G8V & $51.57^{+0.16}_{-0.16}$ & THA & 98.9  \\
HD102026 & 11:44:09.7 & -53:44:54.3 & A1V & $219.07^{+2.26}_{-2.21}$ & FIELD & 99.9 \\
HD104231 & 12:00:09.4 & -57:07:01.9 & F5V & $102.73^{+0.47}_{-0.47}$ & LCC & 99.9  \\
HD104897 & 12:04:44.4 & -52:21:15.6 & F3V & $106.96^{+0.46}_{-0.45}$ & LCC & 97.7  \\ 
HD108568 & 12:28:40.0 & -55:27:19.3 & G1IV & $120.44^{+0.56}_{-0.55}$ & LCC & 98.9  \\
HD112381 & 12:56:58.2 & -54:35:14.4 & ApSi & $121.30^{+3.10}_{-2.95}$ & LCC & 97.6  \\
HD120178 & 13:49:09.2 & -54:13:42.3 & F5V & $133.08^{+0.64}_{-0.63}$ & LCC / UCL & 42.1 / 49.5 \\ 
HD121336 & 13:56:19.9 & -54:07:56.7 & A1V & $139.86^{+1.78}_{-1.74}$ & UCL & 63.7  \\
HD127215 & 14:31:14.3 & -40:39:21.3 & A1V & $145.99^{+1.76}_{-1.72}$ & UCL & 99.6  \\
HD128788 & 14:40:05.0 & -40:54:02.3 & A5V & $144.62^{+4.38}_{-4.13}$ & UCL & 99.8  \\
HD133954 & 15:08:42.5 & -44:29:04.3 & A2/3III & $147.77^{+1.37}_{-1.34}$ & UCL & 99.8  \\
HD138138 & 15:31:17.2 & -33:49:11.4 & A2/3V & $123.87^{+1.42}_{-1.39}$ & UCL & 98.7  \\
HD144175 & 16:05:19.1 & -23:40:08.8 & B9V & $143.72^{+2.06}_{-2.00}$ & USCO & 99.9 \\
HD144118 & 16:05:46.2 & -39:50:35.9 & A5V & $129.03^{+0.96}_{-0.95}$ & UCL & 99.5  \\
HD144823 & 16:08:43.6 & -25:22:36.5 & F3V & $160.38^{+1.18}_{-1.16}$ & USCO & 97.1  \\
HD145792 & 16:13:45.4 & -24:25:19.5 & B6IV & $143.68^{+2.76}_{-2.66}$ & USCO & 99.6  \\
HD146331 & 16:16:50.6 & -25:51:46.7 & B9V & $156.03^{+1.65}_{-1.61}$ & USCO & 99.2  \\
HD147432 & 16:22:51.7 & -23:07:07.5 & A1III/IV & $131.41^{+3.13}_{-2.99}$ & USCO & 98.9 \\
HD148562 & 16:29:54.5 & -24:58:46.0 & A2V & $130.36^{+1.37}_{-1.35}$ & ROPH & 63.2 \\
HD148716 & 16:31:11.0 & -29:59:52.4 & F3V & $118.15^{+1.09}_{-1.07}$ & UCL & 93.7  \\
HD165189 & 18:06:49.8 & -43:25:30.8 & A6V & $44.56^{+0.32}_{-0.32}$ & BPMG & 94.0  \\
HD199143 & 20:55:47.7 & -17:06:51.0 & F8V & $45.66^{+0.08}_{-0.08}$ & BPMG & 97.1  \\ 
HD208233 & 21:57:51.4 & -68:12:50.1 & G9IV & $52.38^{+0.37}_{-0.37}$ & THA & 99.1  \\
HD217379 & 23:00:27.9 & -26:18:42.7 & K6.5V & $30.54^{+0.05}_{-0.05}$ & ABD & 59.5  \\ 
HD285281 & 04:00:31.0 & +19:35:20.9 & K1 & $135.34^{+1.19}_{-1.17}$ & TAU & 99.2  \\ 
TYC8083-45-5 & 04:48:00.6 & -50:41:25.6 & K7Ve & $59.60^{+0.27}_{-0.27}$ & THA & 98.2 \\
 
            \hline
            \hline
            \noalign{\smallskip}
\end{tabular}
\end{table*}

 \begin{table*} 
      \caption[]{Log of observations with the atmospheric conditions for each run.\label{tab:obs}}
      \centering
         \label{t_Settings}
  		\renewcommand{\arraystretch}{1.3}
         \begin{tabular}{lccccccccc}
            \hline 
            \hline
            \multirow{ 3}{2cm}{HD ID} & \multirow{ 3}{2cm}{HIP ID} & \multirow{ 3}{*}{Date} &  Field & Exposure & \multicolumn{3}{c}{Observing conditions on average} \\\cline{6-8} 
            & & & rotation & time, IRDIS$^1$ &  Airmass & Seeing & Coherence time \\
             & & & [$^\circ$]& [sec / min] &  & [$''$] & [ms]  \\
            \hline
            \hline
            \noalign{\smallskip}
 -        & 1910 & 2015-10-13 & 12 & $8\times1\times124=992$ / 16.5 &1.43--1.54&  $1.18$ & $1.4$ \\
 -        & 1910 & 2018-08-05 & 14 & $8\times26\times8=1664$ / 27.7 &1.26--1.27&  $0.97$ & $3.6$ \\
 22213    & -        & 2016-12-06 & 42 & $0.8\times16\times98=1312$ / 21.9 &1.03--1.04&  $0.65$ & $6.5$ \\
 285281   & -        & 2017-10-05 & 20 & $0.8\times18\times72=1085$ / 18.1 &1.40--1.42&  $1.07$ & $4.8$ \\
 285281   & -        & 2017-12-08 & 19 & $0.8\times18\times80=1206$ / 20.1 &1.41--1.46&  $0.53$ & $7.1$ \\
 TYC8083-45-5 & -  &  2016-10-21 & 21 & $2\times17\times48=1632$ / 27.2 &1.12--1.13&  $1.13$ & $2.7$ \\
 TYC8083-45-5 & -  &  2017-10-09 & 21 & $2\times17\times48=1632$ / 27.2 &1.11--1.12&  $0.61$ & $5.2$ \\
 102026   & 57238  &  2016-02-12 & 22 & $32\times1\times80=2560$ / 42.7 &1.15--1.16&  $0.81$ & $4.4$ \\
 104231   & 58528  &  2016-03-18 & 20 & $8\times1\times256=2048$ / 34.1 &1.19--1.19&  $1.08$ & $2.6$ \\
 104231   & 58528  &  2018-04-30 & 14 & $4\times28\times11=1232$ / 20.5 &1.19--1.19&  $1.14$ & $5.6$ \\
 104897   & 58899  &  2016-02-11 & 23 & $8\times1\times256=2048$ / 34.1 &1.13--1.14&  $1.36$ & $2.7$ \\
 108568   & 60885  &  2016-02-11 & 20 & $16\times1\times144=2304$ / 38.4 &1.17--1.19&  $1.07$ & $3.3$ \\
 112381   & 63204  &  2016-02-02 & 21 & $1\times1\times831=831$ / 13.9 &1.16--1.16&  $1.73$ & $2.0$ \\
 120178   & 67428  &  2016-06-06 & 20 & $8\times16\times17=2176$ / 36.3  &1.15--1.15&  $0.97$ & $1.8$ \\
 120178   & 67428  &  2018-03-25 & 15 & $32\times27\times2=1728$ / 28.8  &1.15--1.16&  $0.85$ & $5.1$ \\
 121336   & 68080  &  2016-06-05 & 20 & $16\times8\times16=2048$ / 34.1 &1.15--1.15&  $0.76$ & $2.8$ \\
 127215   & 70998  &  2018-05-14 & 32 & $16\times8\times16=2048$ / 34.1 &1.04--1.05&  $0.95$ & $2.2$ \\
 128788   & 71708  &  2018-05-17 & 32 & $4\times15\times32=1920$ / 32.0 &1.04--1.05&  $0.31$ & $11.0$ \\
 133954   & 74104  &  2018-04-24 & 27 & $16\times9\times16=2304$ / 38.4 &1.06--1.07&  $0.48$ & $9.7$ \\
 138138   & 76001  &  2018-03-14 & 59 & $8\times16\times32=4096$ / 68.3 &1.01--1.02&  $1.07$ & $3.6$ \\
 144175   & 78809  &  2018-05-14 & 49 & $4\times30\times32=3840$ / 64.0 &1.00--1.01&  $1.00$ & $1.8$ \\
 144118   & 78853  &  2018-05-15 & 34 & $4\times30\times16=1920$ / 32.0 &1.04--1.04&  $0.67$ & $4.1$ \\
 144118   & 78853  &  2019-05-26 & 32 & $8\times15\times16=1920$ / 32.0 &1.04--1.04&  $1.09$ & $2.2$ \\
 144823   & 79097  &  2018-07-08 & 6 & $8\times15\times16=1920$ / 32.0 &1.02--1.03&  $0.50$ & $6.1$ \\
 145792   & 79530  &  2018-06-11 & 2 & $4\times13\times32=1664$ / 27.7 &1.01--1.02&  $1.11$ & $1.8$ \\
 146331   & 79771  &  2018-06-23 & 7 & $2\times15\times48=1440$ / 24.0 &1.02--1.03&  $0.70$ & $3.2$ \\
 147432   & 80238  &  2018-08-15 & 4 & $1\times52\times32=1664$ / 27.7 &1.02--1.04&  $0.66$ & $5.9$ \\
 148562   & 80799  &  2018-07-04 & 7 & $2\times15\times48=1440$ / 24.0 &1.01--1.02&  $1.55$ & $1.7$ \\
 148716   & 80896  &  2016-07-13 & 47 & $4\times15\times32=1920$ / 32.0 &1.01--1.02&  $0.77$ & $2.8$ \\
 165189   &   -    &  2016-04-30 & 28 & $2\times32\times16=1024$ / 17.1 &1.06--1.06&  $0.80$ & $4.0$ \\
 165189   &   -    &  2018-04-30 & 23 & $4\times26\times12=1248$ / 20.8 &1.06--1.06&  $0.53$ & $6.3$ \\
 199143   & 103311 &  2019-08-26 & 46 & $2\times15\times48=1448$ / 24.0 &1.01--1.02&  $0.95$ & $4.0$ \\
 208233   & 108422  &  2016-07-23 & 14 & $16\times9\times16=2304$ / 38.4 &1.38--1.38&  $0.67$ & $4.5$ \\
 217379   & 113597  &  2016-09-01 & 12 & $2\times16\times48=1536$ / 25.6 &1.01--1.02&  $0.66 $ & $6.3$ \\
            \hline
            \hline
            \noalign{\smallskip}
\end{tabular}

\begin{flushleft} {\bf Notes.} $^{(1)}$ The exposure time of the IRDIS observation is given in a format ``DIT $\times$ NDIT $\times$ NEXP = Total exposure time in seconds / Total exposure time in minutes'', where DIT is the detector integration time, NDIT is a number of integrations per exposure, NEXP is a number of exposures (files). The total exposure time of the IFS data is either the same as for the IRDIS data or slightly shorter depending on target.  

\end{flushleft}
\end{table*}

\longtab[3]{\begin{longtable}{ l l l l l l l l l l l l} %
\caption{Target sample along with astrometric measurements from literature and from this work.         The subscript A always stands for the primary star while B stands for either the secondary or tertiary        component. Uncertainties reported in Table \ref{tab:astro_error}.        }\label{tab:vibes_sample}\\
\hline\hline\noalign{\smallskip}
HD & Pair & $M_I$ & $M_{II}$ & $K_I$ & $K_{II}$ & $H_I$ & $H_{II}$ & Sep & PA & Date & Ref. \\
  &   & $M_\sun$ & $M_{II}$ & (mag) & (mag) & (mag) & (mag) & ($\arcsec$) & (deg) &   &   \\
\hline\noalign{\smallskip}
\endfirsthead
\caption{continued.}\\ 
\hline\hline\noalign{\smallskip}
HD & Pair & $M_I$ & $M_{II}$ & $K_I$ & $K_{II}$ & $H_I$ & $H_{II}$ & Sep & PA & Date & Ref. \\
  &   & $M_\sun$ & $M_\sun$ & (mag) & (mag) & (mag) & (mag) & ($\arcsec$) & (deg) &   &   \\
\hline\noalign{\smallskip}
\endhead
\hline
\endfoot
\object{HD102026} & AB & $1.47$ & $0.14$ & -- & -- & -- & -- & $1.18$ & $264.40$ & 2011-03-21 & 10 \\
 & AB & -- & -- & $8.33$ & -- & $8.35$ & $12.62$ & $1.17$ & $266.00$ & 2016-02-12 & \textbf{1} \\
\object{HD104231} & AB & $1.33$ & $0.30$ & -- & -- & -- & -- & $4.46$ & $161.30$ & 2011-03-21 & 10 \\
 & AB & -- & -- & $7.42$ & -- & $7.48$ & $9.90$ & $4.44$ & $163.07$ & 2016-03-18 & \textbf{1} \\
 & AB & -- & -- & $7.42$ & -- & $7.48$ & $10.26$ & $4.45$ & $162.91$ & 2018-04-30 & \textbf{1} \\
\object{HD104897} & AB & $1.49$ & $0.44$ & $7.35$ & -- & -- & -- & $4.22$ & $258.00$ & 2011-03-21 & 10 \\
 & AB & -- & -- & -- & -- & $7.42$ & $10.49$ & $4.21$ & $259.56$ & 2016-02-11 & \textbf{1} \\
\object{HD108568} & AB & $2.47$ & $0.24$ & $7.29$ & -- & -- & -- & $0.89$ & $317.30$ & 2011-05-03 & 10 \\
 & AB & -- & -- & -- & -- & $7.41$ & $11.11$ & $0.89$ & $318.84$ & 2016-02-11 & \textbf{1} \\
\object{HD112381} & AB & -- & -- & $6.47$ & $7.36$ & -- & -- & $1.80$ & $46.90$ & 2001-12-31 & 11 \\
 & AB & -- & -- & $6.09$ & -- & $6.22$ & $7.27$ & $2.23$ & $52.08$ & 2016-02-02 & \textbf{1} \\
 & AC & $2.05$ & $1.06$ & $6.78$ & $8.40$ & -- & -- & $0.15$ & $237.00$ & 2004-04-07 & 12 \\
 & AC & -- & -- & -- & -- & $6.22$ & $7.81$ & $0.13$ & $331.62$ & 2016-02-02 & \textbf{1} \\
\object{HD120178} & AB & $1.44$ & $0.11$ & $7.63$ & -- & -- & -- & $3.56$ & $327.00$ & 2012-04-07 & 10 \\
 & AB & -- & -- & -- & -- & $7.71$ & $12.10$ & $3.54$ & $328.69$ & 2016-06-06 & \textbf{1} \\
 & AB & -- & -- & -- & -- & $7.71$ & $11.74$ & $3.54$ & $328.76$ & 2018-03-25 & \textbf{1} \\
\object{HD121336} & AB & $2.91$ & $1.92$ & $6.28$ & $7.19$ & -- & -- & $1.92$ & $10.20$ & 2001-06-05 & 12 \\
 & AB & -- & -- & -- & -- & $6.00$ & $7.00$ & $1.99$ & $14.71$ & 2016-06-05 & \textbf{1} \\
 & AC & -- & -- & $5.94$ & -- & $6.00$ & $9.31$ & $1.65$ & $12.56$ & 2016-06-05 & \textbf{1} \\
\object{HD127215} & AB & $2.54$ & $0.48$ & $7.06$ & $10.83$ & -- & -- & $1.17$ & $355.00$ & 2001-06-06 & 12 \\
 & AB & -- & -- & $7.05$ & -- & $7.09$ & $11.08$ & $1.18$ & $355.69$ & 2018-05-14 & \textbf{1} \\
\object{HD128788} & AB & $1.45$ & $0.44$ & $7.80$ & -- & -- & -- & $3.45$ & $73.00$ & 2011-06-11 & 10 \\
 & AB & -- & -- & -- & -- & -- & -- & $3.47$ & $74.31$ & 2018-05-17 & \textbf{1} \\
\object{HD133954} & AB & $2.45$ & $0.49$ & $7.67$ & -- & -- & -- & $1.85$ & $210.70$ & 2011-04-26 & 10 \\
 & AB & -- & -- & -- & -- & $7.70$ & $10.52$ & $1.81$ & $212.60$ & 2018-04-24 & \textbf{1} \\
\object{HD138138} & AB & -- & -- & -- & -- & -- & -- & $0.08$ & $7.60$ & 1992-06-15 & 7 \\
 & AB & -- & -- & -- & -- & -- & -- & $0.09$ & $6.40$ & 1993-02-05 & 7 \\
 & AB & $1.54$ & $1.36$ & $7.60$ & $7.80$ & -- & -- & $0.25$ & $3.20$ & 2001-06-08 & 12 \\
 & AB & -- & -- & $6.43$ & -- & $6.49$ & $6.72$ & $0.41$ & $2.33$ & 2018-03-14 & \textbf{1} \\
 & AC & -- & -- & -- & -- & -- & -- & $1.55$ & $131.20$ & 1989-04-22 & 14 \\
 & AC & -- & -- & -- & -- & -- & -- & $1.54$ & $130.00$ & 1991 & 15 \\
 & AC & $1.54$ & $1.36$ & $7.60$ & $8.20$ & -- & -- & $1.48$ & $124.80$ & 2001-06-08 & 12 \\
 & AC & -- & -- & -- & -- & $6.49$ & $7.20$ & $1.43$ & $119.37$ & 2018-03-14 & \textbf{1} \\
\object{HD144118} & AB & $1.82$ & $1.14$ & $7.50$ & $8.45$ & -- & -- & $1.99$ & $270.39$ & 2001-06-08 & 12 \\
 & AB & -- & -- & -- & -- & $7.15$ & $8.29$ & $2.01$ & $271.01$ & 2018-05-15 & \textbf{1} \\
\object{HD144175} & AB & $2.03$ & $0.30$ & $7.51$ & $10.26$ & -- & -- & $1.18$ & $25.67$ & 2001-06-07 & 12 \\
 & AB & -- & -- & $7.35$ & -- & $7.37$ & $10.28$ & $1.13$ & $25.76$ & 2018-05-14 & \textbf{1} \\
\object{HD144823} & AB & $1.99$ & $0.75$ & $7.25$ & -- & -- & -- & $0.81$ & $340.00$ & 2011-05-31 & 10 \\
 & AB & -- & -- & -- & -- & $7.33$ & $10.44$ & $0.81$ & $342.23$ & 2018-07-08 & \textbf{1} \\
\object{HD145792} & AB & $3.73$ & $1.58$ & $6.60$ & $8.34$ & -- & -- & $1.69$ & $219.66$ & 2000-05-31 & 12 \\
 & AB & -- & -- & $7.11$ & -- & $6.12$ & $8.16$ & $1.73$ & $219.16$ & 2018-06-11 & \textbf{1} \\
\object{HD146331} & AB & $2.14$ & $0.19$ & $7.10$ & $11.42$ & -- & -- & $0.44$ & $128.59$ & 2004-06-19 & 12 \\
 & AB & -- & -- & $7.10$ & -- & $7.19$ & $11.64$ & $0.43$ & $130.54$ & 2018-06-23 & \textbf{1} \\
 & AC & $2.14$ & $0.19$ & $7.10$ & $10.89$ & -- & -- & $3.67$ & $313.38$ & 2004-06-19 & 12 \\
 & AC & -- & -- & -- & -- & $7.19$ & $11.12$ & $3.65$ & $314.85$ & 2018-06-23 & \textbf{1} \\
\object{HD147432} & AB & -- & -- & -- & -- & -- & -- & $1.05$ & $318.30$ & 1991 & 5 \\
 & AB & $1.94$ & $1.67$ & $7.34$ & $7.49$ & -- & -- & $1.03$ & $318.46$ & 2001-06-07 & 12 \\
 & AB & -- & -- & $6.62$ & -- & $6.69$ & $6.91$ & $1.01$ & $320.15$ & 2018-08-15 & \textbf{1} \\
\object{HD148562} & AB & $1.86$ & $0.34$ & $7.45$ & $9.80$ & -- & -- & $2.94$ & $205.02$ & 2004-05-05 & 12 \\
 & AB & -- & -- & $7.31$ & -- & $7.41$ & $10.92$ & $2.93$ & $206.40$ & 2018-07-04 & \textbf{1} \\
\object{HD148716} & AB & $1.81$ & $0.24$ & $7.44$ & $10.33$ & -- & -- & $2.28$ & $177.23$ & 2004-06-08 & 12 \\
 & AB & -- & -- & $7.45$ & -- & $7.55$ & $11.28$ & $2.28$ & $177.89$ & 2016-07-13 & \textbf{1} \\
\object{HD165189} & AB & -- & -- & $4.40$ & -- & -- & -- & $1.80$ & $249.00$ & 1874 & 8 \\
 & AB & -- & -- & -- & -- & -- & -- & $1.71$ & $187.91$ & 2015-06-24 & 6 \\
 & AB & -- & -- & -- & -- & -- & -- & $1.79$ & $181.89$ & 2018-04-30 & \textbf{1} \\
\object{HD199143} & AB & $1.19$ & $0.35$ & $8.12$ & -- & -- & -- & $1.08$ & $325.00$ & 2001-05-31 & 9 \\
 & AB & -- & -- & -- & -- & $5.95$ & $8.19$ & $0.84$ & $321.71$ & 2019-08-26 & \textbf{1} \\
\object{HD208233} & AB & $1.03$ & $0.28$ & $6.83$ & $9.78$ & -- & -- & $1.98$ & $192.10$ & 2001-10-28 & 2 \\
 & AB & -- & -- & $6.83$ & -- & $6.86$ & $9.82$ & $1.62$ & $193.94$ & 2016-07-23 & \textbf{1} \\
\object{HD217379} & AB & -- & -- & $6.75$ & $7.38$ & -- & -- & $2.24$ & $239.02$ & 2012-07-11 & 4 \\
 & AB & -- & -- & -- & -- & $6.45$ & $7.31$ & $2.29$ & $240.64$ & 2016-09-01 & \textbf{1} \\
\object{HD22213} & AB & $0.94$ & $0.54$ & $6.97$ & $5.10$ & -- & -- & $1.67$ & $281.10$ & 2012-07-11 & 4 \\
 & AB & -- & -- & -- & -- & $6.95$ & $8.84$ & $1.69$ & $281.12$ & 2016-12-06 & \textbf{1} \\
\object{HD285281} & AB & -- & -- & $7.61$ & -- & -- & -- & $0.77$ & $190.70$ & 1994-12-12 & 13 \\
 & AB & $1.11$ & $0.71$ & $7.61$ & -- & -- & -- & $0.78$ & $186.00$ & 2011-10-17 & 3 \\
 & AB & -- & -- & -- & -- & $7.75$ & $8.99$ & $0.76$ & $186.22$ & 2017-10-05 & \textbf{1} \\
 & AB & -- & -- & -- & -- & $7.75$ & $9.45$ & $0.76$ & $186.15$ & 2017-12-08 & \textbf{1} \\
\object{HIP1910} & AB & $0.71$ & $0.30$ & $7.69$ & $9.22$ & -- & -- & $0.70$ & $46.90$ & 2000-11-12 & 2 \\
 & AB & -- & -- & $7.69$ & -- & -- & -- & $0.70$ & $50.20$ & 2001-10-28 & 2 \\
 & AB & -- & -- & $7.69$ & -- & $7.71$ & $9.37$ & $0.52$ & $81.98$ & 2015-10-12 & \textbf{1} \\
 & AB & -- & -- & $7.69$ & -- & $7.71$ & $9.32$ & $0.48$ & $89.96$ & 2018-08-05 & \textbf{1} \\
\object{TYC8083-45-5} & AB & $0.70$ & $0.35$ & $8.25$ & $7.10$ & -- & -- & $1.06$ & $347.80$ & 2011-11-08 & 4 \\
 & AB & -- & -- & -- & -- & $8.08$ & $9.24$ & $1.05$ & $346.58$ & 2016-10-21 & \textbf{1} \\

		\footnotetext{
			References: (1)  this work; (2 )\citet{chauvin_adaptive_2003}; (3) \citet{daemgen_sub-stellar_2015}; (4) \citet{elliott_search_2015};
			(5) \citet{fabricius_tycho_2002}; (6) \citet{gaia_collaboration_gaia_2018}; (7) \citet{hartkopf_iccd_1996}; (8) \citet{herschel_catalogue_1874}; (9) \citet{janson_multiplicity_2013}; (10) \citet{kouwenhoven_primordial_2005}
			(11) \citet{kouwenhoven_primordial_2006}; (12) \citet{kohler_multiplicity_1998}; (13) \citet{mcalister_iccd_1990}; (14) \citet{tokovinin_msc_1997}\newline
		}
	\end{longtable}
}

\end{appendix}

\end{document}